\documentclass[a4paper,11pt,preprint,aps,prd,showpacs,amsmath,amssymb,nofootinbib,superscriptaddress,showkeys,floatfix]{revtex4-1}

\usepackage[utf8x]{inputenc}
\usepackage{graphicx}
\usepackage{hyperref}
\usepackage{natbib}
\usepackage{bm}
\usepackage{multirow}
\usepackage{epstopdf}

\allowdisplaybreaks

\begin{document}

\preprint{TUM-EFT 45/14}

\title{Quarkonium Hybrids with Nonrelativistic Effective Field Theories}
\author{Matthias Berwein}
\email{matthias.berwein@mytum.de}
\affiliation{Physik-Department, Technische Universit\"at M\"unchen, \\ James-Franck-Str.~1, 85748 Garching, Germany}
\affiliation{Excellence Cluster Universe, Technische Universit\"at M\"unchen, \\ Boltzmannstr.~2, 85748, Garching, Germany}
\author{Nora Brambilla}
\email{nora.brambilla@ph.tum.de}
\affiliation{Physik-Department, Technische Universit\"at M\"unchen, \\ James-Franck-Str.~1, 85748 Garching, Germany}
\author{Jaume \surname{Tarr\'us Castell\`a}}
\email{jaume.tarrus@tum.de}
\affiliation{Physik-Department, Technische Universit\"at M\"unchen, \\ James-Franck-Str.~1, 85748 Garching, Germany}
\author{Antonio Vairo}
\email{antonio.vairo@ph.tum.de}
\affiliation{Physik-Department, Technische Universit\"at M\"unchen, \\ James-Franck-Str.~1, 85748 Garching, Germany}

\begin{abstract}
We construct a nonrelativistic effective field theory description of heavy quarkonium hybrids from QCD. We identify the symmetries of the system made of a heavy quark, a heavy antiquark, and glue in the static limit. Corrections to this limit can be obtained order by order in an expansion in the inverse of the mass $m$ of the heavy quark. At order $1/m$ in the expansion, we obtain at the level of potential Non-Relativistic QCD a system of coupled Schr\"odinger equations that describes hybrid spin-symmetry multiplets, including the mixing of different static energies into the hybrid states, an effect known as $\Lambda$-doubling in molecular physics. In the short distance, the static potentials depend on two nonperturbative parameters, the gluelump mass and the quadratic slope, which can be determined from lattice calculations. We adopt a renormalon subtraction scheme for the calculation of the perturbative part of the potential. We numerically solve the coupled Schr\"odinger equations and obtain the masses for the lowest lying spin-symmetry multiplets for $c\bar{c}$, $b\bar{c}$, and $b\bar{b}$ hybrids. The $\Lambda$-doubling effect breaks the degeneracy between opposite parity spin-symmetry multiplets and lowers the mass of the multiplets that get mixed contributions of different static energies. We compare our findings to the experimental data, direct lattice computations, sum rules calculations, and discuss the relation to the Born--Oppenheimer approximation.
\end{abstract}


\maketitle

\section{Introduction}

During the last few years, experimental observations have revealed the existence of a large number of unexpected states close to or above the open flavor thresholds in the heavy quarkonium spectrum (see, e.g., the reviews ~\cite{Brambilla:2014jmp,Brambilla:2010cs,Bodwin:2013nua}), culminating recently with the observation of a ``charmonium-pentaquark'' state at LHCb~\cite{Aaij:2015tga}. Ref.~\cite{Brambilla:2014jmp} summarizes the situation at the time of the review with two tables listing seven states with masses at open flavor thresholds and twenty states with masses above the open flavor threshold in the charmonium and bottomonium sector. Since then, several additional states have been observed. Most of these states display special features that single them out as ``exotic states''. They are indicated with the $X$, $Y$, and $Z$ labels. There is an ongoing significant amount of experimental effort to study exotic quarkonium by measuring new states, new production mechanisms, decays and transitions, and obtaining precision and high statistics data at BESIII, LHC experiments, and prospectively at Belle2 and Panda at FAIR.

This experimental effort is matched by a correspondingly intense theoretical activity. Exotic quarkonium states are interesting, because they are candidates for nonconventional hadronic states, for example, hadrons containing four quarks or an excited gluon. Since these states are at or above the strong decay threshold, heavy-light mesons and light quark degrees of freedom should be explicitly taken into account in the dynamics. At the moment there is no direct QCD approach to study these states. In fact, even if great progress has been made in the last few years, see, e.g., Refs.~\cite{Lang:2014zaa,*Prelovsek:2014swa,Moir:2013yfa,Morningstar:2013bda,*Bulava:2013dra}, the lattice QCD study of excited states in the quarkonium sector is still challenging, particularly for states close to or above thresholds.

On the other hand, many phenomenological models for exotics have been introduced and used in the meantime. A phenomenological model is based on the choice of some relevant degrees of freedom and a phenomenological Hamiltonian that dictates the dynamics of the chosen degrees of freedom. In this way, exotic states may be interpreted as quarkonium tetraquarks (whose four quark constituents can be clustered in several different ways up to a molecular description), quarkonium hybrids, or hadro-quarkonium in dependence of the model used, see, e.g., \cite{Brambilla:2014jmp,Brambilla:2010cs,Bodwin:2013nua} for a review. Sum rules are also used to verify the dominant composition of these states~\cite{Nielsen:2014mva}. Only for special states displaying exceptional features, like the $X(3872)$, which is precisely at a threshold, a kind of universal effective field theory description can be developed based on the small energy of the state and the correspondingly large scattering length~\cite{Braaten:2010mg}.

The description of exotics should, however, be obtained from QCD. One possibility is to work out a nonrelativistic effective field theory description of these states in a way similar to what has been done by potential Non-Relativistic QCD (pNRQCD) for states away from threshold~\cite{Pineda:1997bj,Brambilla:1999xf,Brambilla:2004jw,Brambilla:2000gk,*Pineda:2000sz,Brambilla:2002nu}. Also for exotics the quark mass is still a large parameter of expansion, and the first step entails a matching from QCD to Non-Relativistic QCD (NRQCD) by integrating out the hard scale of the mass~\cite{Caswell:1985ui,Bodwin:1994jh}. However, the second step, i.e., arriving at an effective field theory of the type of pNRQCD, whose matching coefficients are the interaction potentials and the leading order dynamical equation is of the Schr\"odinger type, is more difficult.

While in the case of quarkonium systems away from the threshold a dynamically generated gap exists~\cite{Brambilla:2004jw,Brambilla:2000gk,*Pineda:2000sz,Brambilla:2002nu}, allowing us to integrate out the other degrees of freedom, when we consider quarkonium systems at or above the strong decay threshold, this is no longer the case. There is no mass gap between the heavy quarkonium and the creation of a heavy-light pair or a heavy quark pair with gluonic excitations. Thus, constructing the effective field theory entails considering, besides the heavy quark operators, all gauge invariant operators containing light quarks, heavy quarks, and excited glue operators, e.g., pions, some heavy-light mesons, quarkonium hybrids, and glueballs. We discussed above how phenomenological models just pick up some of these possible degrees of freedom and attach to them some phenomenological interaction. In an effective field theory description one should identify an appropriate expansion parameter and establish a power counting weighting the operators. This is, at the moment, still difficult.

In this paper we restrict ourselves to considering a heavy quark, a heavy antiquark, and excited glue degrees of freedom, aiming at a description of heavy quarkonium hybrids under some special conditions. Heavy quarkonium hybrids (for a review see, e.g., \cite{Meyer:2015eta}) have traditionally been described in models like the flux tube model~\cite{Isgur:1983wj,*Isgur:1984bm}, the bag model~\cite{Hasenfratz:1980jv}, the constituent gluon model~\cite{Horn:1977rq}, or in the so-called Born--Oppenheimer (BO) approximation applied to QCD~\cite{Meyer:2015eta,Griffiths:1983ah,Juge:1997nc}. The adiabatic BO approximation has been the standard method to describe the interaction between electrons and nuclei in molecules bound by electromagnetism since the early days of quantum mechanics~\cite{bo,LandauLifshitz} up to now~\cite{mh}.\footnote{For an effective field theory description of the physics of the BO approximation in QED see~\cite{BOQED}.} The BO approximation assumes that the lighter electrons adjust adiabatically to the motion of the heavier nuclei. It exploits the fact that the masses of the nuclei are much larger than the electron masses and, consequently, the time scales for the dynamics of the two types of particles are very different. It entails no restriction on the strength of the coupling between the slow and the fast degrees of freedom. In concrete terms, the BO approximation provides a method to obtain the molecular energies by solving the Schr\"odinger equation for the nuclei with a potential given by the electronic static energies at fixed nuclei positions. In particular, in the case of the diatomic molecule the electronic static energies turn out to be labeled by molecular quantum numbers corresponding to the symmetries of the diatomic molecular system.

This procedure is rooted in the existence of two classes of degrees of freedom, the ``fast'' and ``slow'' ones, and in the symmetries of the diatomic molecular system. This is the reason why the same framework can be used to describe systems of different nature but with similar characteristics. This turns out to be the case for heavy quarkonium hybrids, systems formed by a heavy quark, a heavy antiquark, and excited glue. The BO approximation has been used in this case, identifying the slow and fast degrees of freedom with the heavy quark-antiquark pair and the gluons, respectively~\cite{Meyer:2015eta,Griffiths:1983ah,Juge:1997nc}. In the static limit the quark and the antiquark serve as color source and sink at distance $\bm{r}$, and the gluonic field arranges itself in configurations described by the quantum numbers fixed by the symmetry of the system. The gluonic dynamics are, however, collective and nonperturbative. Nevertheless, the gluonic static energies (that are the analog of the electronic static energies) have been extracted from the large time behavior of lattice evaluations of generalized quark-antiquark Wilson loops at fixed spatial distance with initial and final states of the appropriate symmetry~\cite{Griffiths:1983ah,Campbell:1984fe,Ford:1988ki,Perantonis:1990dy,Lacock:1996ny,Juge:1997nc,Juge:2002br,Bali:2000vr}. This method provides, in principle, these gluonic static energies, but does not provide the gluonic wave functions.

Then, relying on a kind of BO approximation, the gluonic static energies have been introduced as potentials in a Schr\"odinger-like equation~\cite{LandauLifshitz} and some level structure has been obtained ~\cite{Juge:1999ie,Juge:1997nc}. The structure of the hybrid multiplets has also been discussed in Ref.~\cite{Braaten:2014qka,*Braaten:2014ita,*Braaten:2013boa} using the BO approximation and complementary information from the lattice. These works relied on the adiabatic and single channel BO approximation, meaning respectively that only the static potential and no mixing between different static energies have been considered. To our knowledge, up to now no analytical description of quarkonium hybrids has been worked out directly from QCD by constructing an
effective field theory that realizes the physical scale hierarchy typical of the system.\footnote{This refers to the hybrid spectroscopy. Applications of NRQCD to hybrid production can be found in~\cite{Chiladze:1998ti}.} This is what we address in the present paper.
 
The paper is organized as follows. In section~\ref{ssegee} we introduce the Non-Relativistic QCD Hamiltonian and discuss the description of the heavy quarkonium hybrid systems in NRQCD in the static limit, defining the Fock states, their symmetries, and the corresponding static energies. In section~\ref{statpnr} we give the same characterization using potential NRQCD, i.e., integrating out the soft scale of the momentum transfer and in case multipole expanding.

In particular, we match the NRQCD states and Hamiltonian to the corresponding objects in pNRQCD. In this way, glueballs and gluelumps naturally emerge in pNRQCD, where the gluelumps are defined as the color singlet combination of an octet color source coupled to a gluonic field. The hybrid static potentials appear as matching coefficients of pNRQCD. The higher degree of symmetry of the lower energy EFT induces a pattern of degeneracy in the gluelump multiplets.

In section~\ref{latdat} we introduce existing lattice evaluations of the hybrid static energies and we relate them to the definitions and the discussion given in the previous sections. In section~\ref{sch} we add the first correction to the static limit, introducing operators of order $1/m$ in NRQCD and pNRQCD. This allows us to obtain the appropriate Schr\"odinger equations as dynamical equations in pNRQCD. At this time we still neglect the spin. We work out the radial Schr\"odinger equations coupling the $\Sigma_u^-$ and the $\Pi_u$ gluonic states (due to the so-called $\Lambda$-doubling term) in detail, as these will generate all the lowest mass hybrid multiplets. We characterize the hybrid multiplets by their $J^{PC}$ quantum numbers, and we discuss the relation with the pattern of hybrid multiplets obtained in the BO approximation and other approaches.

In section~\ref{htmp} we solve the Schr\"odinger equation to get the masses of the predicted hybrid multiplets. The static potentials appearing in the Schr\"odinger equation have been defined in pNRQCD in section~\ref{statpnr}, they depend in the short range on two nonperturbative parameters. We fix the first one from lattice determinations of the gluelump mass, and we extract the second from a fit to the gluonic static energies. Then, we define an appropriate renormalon-free scheme (RS) and we obtain the heavy quarkonium hybrid masses for $c\bar{c}$, $b\bar{c}$, and $b\bar{b}$ systems.

In section~\ref{compare} we compare our results for hybrid mass multiplets to the existing experimental candidates and to results obtained using the BO approximation, direct lattice computations, and QCD sum rules. Section~\ref{conc} contains the conclusions and an outlook for future development. The appendices contain detailed information about the symmetry of the static system (Appendix~\ref{sss}), the RS scheme (Appendix~\ref{rescheme}), the derivation of the radial Schr\"odinger equation (Appendix~\ref{RSEQ}), and the numerical solution of coupled Schr\"odinger equations (Appendix~\ref{appx}).

\section{Static NRQCD: Symmetries of the static system and definition of the gluonic static energies} \label{ssegee}

We are considering a bound system made by a heavy quark $Q$, a heavy antiquark $\bar{Q}$ and some gluonic excitations: this we will generically call a heavy hybrid state.\footnote{Usually the term hybrid identifies systems where $Q\bar{Q}$ is in a color octet configuration. In the present treatment the distinction between this type of hybrid and $Q\bar{Q}$ in a color singlet state plus a glueball is often irrelevant, therefore we will make it only when necessary.} Since the quark mass $m$ is much larger than the typical hadronic scale $\Lambda_\mathrm{QCD}$, we can use NRQCD~\cite{Caswell:1985ui,Bodwin:1994jh} to describe such a system. NRQCD is obtained from QCD by integrating out the hard scale of the quark mass, which corresponds to expanding in inverse powers of the mass and including the nonanalytic dependence on the quark mass inside some matching coefficients.

The Hamiltonian of NRQCD for the one-quark-one-antiquark sector of the Fock space reads
\begin{align}
H_\mathrm{NRQCD} &= H^{(0)}+\frac{1}{m_Q}H^{(1,0)}+\frac{1}{m_{\bar{Q}}}H^{(0,1)} + \dots\,, \label{HH}\\
H^{(0)} &= \int d^3x\, \frac{1}{2}\left( \bm{E}^a\cdot\bm{E}^a +\bm{B}^a\cdot\bm{B}^a \right)-\sum_{j=1}^{n_f} \int d^3\bm{x}\, \bar{q}_j \, i \bm{D}\cdot \bm{\gamma} \, q_j \,,\label{H0}\\
H^{(1,0)} &= - \frac{1}{2} \int d^3x\, \psi^\dagger \left( \bm{D}^2 + g c_F\, \bm{\sigma} \cdot \bm{B}\right) \psi\,, \label{H10}\\
H^{(0,1)} &=\frac{1}{2}\int d^3x\, \chi^\dagger \left(\bm{D}^2+ g c_F\, \bm{\sigma} \cdot \bm{B} \right) \chi\,,\label{H01}
\end{align}
where we have shown only terms up to order $1/m$ in the quark mass expansion, $\psi$ is the Pauli spinor field that annihilates the heavy quark, $\chi$ is the Pauli spinor field that creates the heavy antiquark, $q_j$ is the Dirac spinor field that annihilates a massless quark of flavor~$j$, $iD^0=i\partial_0-gA^0$, $i\bm{D}=i\bm{\nabla}+g\bm{A}$, and the matching coefficient $c_F$ is equal to one up to loop corrections of order $\alpha_\mathrm{s}$.
The physical states are constrained to satisfy the Gauss law\footnote{Since $\bm{\Pi}^a=\bm{E}^a +O(1/m^2)$ we use the chromoelectric field $\bm{E}^a$ instead of the canonical momentum $\bm{\Pi}^a$ here and in the Hamiltonian above.}
\begin{equation}
(\bm{D}\cdot \bm{E})^a \, \vert \mathrm{phys} \rangle =
g \Bigl(\psi^\dagger T^a \psi + \chi^\dagger T^a \chi + \sum_{j=1}^{n_f} 
\bar{q}_j \gamma^0 T^a q_j\Bigr)
\vert \mathrm{phys} \rangle\,.
\label{gausslaw}
\end{equation}

Even though we include the light quarks here in the Hamiltonian and in the Gauss law, we will not consider them as external dynamical sources in the rest of the paper, in the sense that we exclude excitations with nonzero isospin,\footnote{States induced by the inclusion of these light degrees of freedom have been discussed in the BO approximation in~\cite{Braaten:2014qka,*Braaten:2014ita,*Braaten:2013boa}.} transitions through light mesons, or decays into heavy-light mesons, but we still allow for them to appear in the form of sea quarks, as in light quark loops in perturbation theory or unquenched lattice calculations. The lowest gluonic excitations are stable under these conditions, since the only remaining transitions require the emission of a glueball, and this is only possible if the mass gap between initial and final state is larger than the glueball mass.

In the static limit $m_Q,\,m_{\bar{Q}} \to \infty$ we have
\begin{equation}
H_\mathrm{NRQCD} = H^{(0)}\,,
\label{ssta}
\end{equation}
which still contains the kinetic terms associated to the gluons, while the kinetic terms of the heavy quarks vanish. In the static limit the one-quark--one-antiquark sector of the Fock space is spanned by~\cite{Brambilla:2004jw,Brambilla:2000gk,*Pineda:2000sz}
\begin{equation}
\vert \underline{n}; \bm{x}_1 ,\bm{x}_2 \rangle^{(0)} = \psi^{\dagger}(\bm{x}_1) \chi (\bm{x}_2)
|n;\bm{x}_1 ,\bm{x}_2\rangle^{(0)},\qquad \forall \bm{x}_1,\bm{x}_2\,,
\label{basis0}
\end{equation}
where $|\underline{n}; \bm{x}_1 ,\bm{x}_2\rangle^{(0)} $ is a gauge-invariant eigenstate of $H^{(0)}$ (defined up to a phase and satisfying the Gauss law) with energy $E_n^{(0)}(\bm{x}_1 ,\bm{x}_2)$; $|n;\bm{x}_1 ,\bm{x}_2\rangle^{(0)}$ encodes the purely gluonic content of the state, and it is annihilated by $\chi^{\dagger}(\bm{x})$ and $\psi(\bm{x})$ for any $\bm{x}$. It transforms like $3_{\bm{x}_1}\otimes 3_{\bm{x}_2}^{\ast}$ under color $SU(3)$. The normalizations are taken as follows
\begin{equation}
^{(0)}\langle n;\bm{x}_1 ,\bm{x}_2|m;\bm{x}_1 ,\bm{x}_2\rangle^{(0)} =\delta_{nm}\,,
\end{equation}
\begin{equation}
^{(0)}\langle \underline{n}; \bm{x}_1 ,\bm{x}_2|\underline{m}; \bm{y}_1 ,\bm{y}_2\rangle^{(0)} =\delta_{nm}
\delta^{(3)} (\bm{x}_1 -\bm{y}_1)\delta^{(3)} (\bm{x}_2 -\bm{y}_2)\,.
\label{norm}
\end{equation}
Notice that since $ H^{(0)}$ does not contain any heavy fermion field, $|n;\bm{x}_1,\bm{x}_2\rangle^{(0)}$ itself is also an eigenstate of $H^{(0)}$ with energy $E_n^{(0)}(\bm{x}_1 ,\bm{x}_2)$. We have made it explicit that the positions $\bm{x}_1$ and $\bm{x}_2$ of the quark and antiquark, respectively, are good quantum numbers for the static solution $|\underline{n};\bm{x}_1 ,\bm{x}_2 \rangle^{(0)}$, while $n$ generically denotes the remaining quantum numbers.

In static NRQCD, the gluonic excitations between static quarks have the same symmetries as the diatomic molecule~\cite{LandauLifshitz}. In the center-of-mass system, these correspond to the symmetry group $D_{\infty h}$ (substituting the parity operation by CP). According to that symmetry, the mass eigenstates are classified in terms of the angular momentum along the quark-antiquark axis ($\Lambda = 0,1,2, \dots$, to which one gives the traditional names $\Sigma, \Pi, \Delta, \dots$), CP ($g$ for even or $u$ for odd), and the reflection properties with respect to a plane that passes through the quark-antiquark axis ($+$ for even or $-$ for odd). Only the $\Sigma$ states are not degenerate with respect to the reflection symmetry. See Appendix~\ref{sss} for more details.

Translational invariance implies that $E_n^{(0)}(\bm{x}_1,\bm{x}_2) = E_n^{(0)}(r)$, where $\bm{r}=\bm{x}_1-\bm{x}_2$. This means that the gluonic static energies are functions of $r$ and of the only other scale of the system in the static limit, $\Lambda_\mathrm{QCD}$. The ground-state energy $E_{\Sigma_{g}^{+}}^{(0)}(r)$ is associated to the static quark-antiquark energy, while the other gluonic static energies $E_n^{(0)}(r)$, $n\neq0$, are associated to gluonic excitations between static quarks. Following the analogy with the diatomic molecule, the $E_n^{(0)}(r)$ play the same role as the electronic static energies. However, in the present case they are nonperturbative quantities and can be obtained in lattice QCD from generalized static Wilson loops in the limit of large interaction times $T$~\cite{Griffiths:1983ah,Campbell:1984fe,Ford:1988ki,Perantonis:1990dy,Lacock:1996ny,Juge:1997nc,Juge:2002br,Bali:2000vr}

Since the static energies are eigenvalues of the static Hamiltonian, one can exploit the following relation:
\begin{equation}
^{(0)}\left\langle \underline{n};\,\bm{x}_1,\,\bm{x}_2,\,T/2\right|\left.\underline{n};\,\bm{x}_1,\,\bm{x}_2,\,-T/2\right\rangle^{(0)}=\mathcal{N}\,\exp\left[-iE_n^{(0)}(r)\,T\right]\,,
\label{timecorr}
\end{equation}
where $\mathcal{N}=\left[\delta^{(3)}(\bm{0})\right]^2$ is a normalization constant following from~\eqref{norm}. Since the static states $\left|\underline{n};\,\bm{x}_1,\,\bm{x}_2\right\rangle^{(0)}$ form a complete basis, any state $|X_n\rangle$ can be written as an expansion in them:
\begin{equation}
 |X_n\rangle=c_n\,\left|\underline{n};\,\bm{x}_1,\,\bm{x}_2\right\rangle^{(0)}+c_{n'}\left|\underline{n}';\,\bm{x}_1,\,\bm{x}_2\right\rangle^{(0)}+\dots\,.
\end{equation}
From Eq.~\eqref{timecorr}, it then follows
\begin{equation}
 \langle X_n,\,T/2|X_n,\,-T/2\rangle=\mathcal{N}|c_n|^2\exp\left[-iE_n^{(0)}(r)\,T\right]+\mathcal{N}|c_{n'}|^2\exp\left[-iE_{n'}^{(0)}(r)\,T\right]+\dots\,.
\end{equation}
For large $T$ the exponentials will be highly oscillatory, or in the Euclidean time of lattice QCD highly suppressed, so such a correlator will be dominated by the lowest static energy. This allows us to obtain the lowest static energies without knowing the static states explicitly
\begin{equation}
 E_n^{(0)}(r)=\lim_{T\to\infty}\frac{i}{T}\log\langle X_n,\,T/2|X_n,\,-T/2\rangle\,.
 \label{eelwl}
\end{equation}
The only condition that $|X_n\rangle$ has to satisfy is that it needs to have a nonvanishing overlap with the static state, $c_n\neq 0$. This can be ensured by requiring $|X_n\rangle$ to have the same quantum numbers $n$ as the static state. Doing this also allows us to not only get the ground state energy, but also the lowest static energy for any set of excited quantum numbers $n$, because, if the quantum numbers of $|X_n\rangle$ are fixed, then it can only have an overlap with static states of the same quantum numbers.

A convenient choice for these $|X_n\rangle$ states gives the static energies in terms of Wilson loops, so we define
\begin{equation}
 |X_n\rangle=\chi(\bm{x}_2)\phi(\bm{x}_2,\bm{R})T^aP_n^a(\bm{R})\phi(\bm{R},\bm{x_1})\psi^\dagger(\bm{x}_1)|\mathrm{vac}\rangle\,.
 \label{latstat}
\end{equation}
Here the strings $\phi(\bm{x}_2,\bm{x}_1)$ are Wilson lines from $\bm{x}_1$ to $\bm{x}_2$, which are defined in general as
\begin{equation}
 \phi(x_2,x_1)=\mathcal{P}\exp\left[-ig\int_{x_1}^{x_2}dx^\mu\,A_\mu(x)\right]\,,
\end{equation}
where $\mathcal{P}$ denotes the path ordering operator. By $|\mathrm{vac}\rangle$ we mean the NRQCD vacuum, and $P_n$ is some gluonic operator that generates the right quantum numbers $n$. A list of possible operators $P_n$ is given in Table~\ref{tab3}. The large time correlator of these states is given by a static Wilson loop with insertions of $P_n$ in the strings at the center-of-mass. These generalized static Wilson loops are in principle the same quantities as those that are used to obtain the gluonic static energies on the lattice, but with suitable lattice definitions for the operators $P_n$ and allowing for further manipulations like smearing. For more details see section~\ref{latdat}.

For the ground state energy $E_{\Sigma_{g}^{+}}^{(0)}(r)$ one has to insert a color-neutral gluonic operator with $J^{PC}$ quantum numbers $0^{++}$ instead of $T^aP_n^a$. For the simplest choice, i.e., the unit matrix, this then coincides with the usual static Wilson loop without insertions and gives the quark-antiquark static energy. One can also replace $T^aP_n^a$ by a color-neutral gluonic operator with excited $J^{PC}$ quantum numbers. In this case one selects the lowest mass singlet plus glueball states consistent with those $J^{PC}$ quantum numbers. It is possible to get additional information about and a characterization of these gluonic static energies by using the lower energy effective field theory called pNRQCD. 

\section{Static pNRQCD: characterization of the gluonic static energies at short distances and form of the potentials} \label{statpnr}

In this section we discuss how it is possible to obtain a model independent characterization of the gluonic static energies at short distance and a definition of the hybrid potential using a low energy effective field theory called pNRQCD~\cite{Pineda:1997bj,Brambilla:1999xf}. 

Potential NRQCD is obtained from NRQCD by integrating out the soft scale of the relative momentum transfer between the quark and the antiquark, which is of the order of the inverse quark-antiquark distance $1/r$. Therefore, the matching coefficients of pNRQCD may have a nonanalytic dependence on $r$ and correspond to the quark-antiquark potentials. For short quark-antiquark distances (i.e., in the limit $\Lambda_\mathrm{QCD} \ll 1/r$) the soft scale of the quark-antiquark momentum transfer is still perturbative, and we can call that effective theory weakly-coupled pNRQCD (see~\cite{Brambilla:2004jw} or~\cite{Pineda:2011dg} for a review). The dynamical degrees of freedom of this theory are heavy quark-antiquark pairs in a color singlet, $S$, or in a color octet configuration, $O$, and low energy (ultrasoft) gluons. To ensure that the gluons are characterized by a length that is larger than the typical quark-antiquark distance, the gluon fields are multipole expanded with respect to $\bm{r}$, which means that they only depend on the center-of-mass coordinate $\bm{R}$ and time $t$. 

In the static limit and at leading order in the multipole expansion, the pNRQCD Hamiltonian is 
\begin{equation}
 \mathcal{H}^{(0)}=\int d^3R\,d^3r\,\left(V_s(r)S^{\dagger}(\bm{r},\bm{R})S(\bm{r},\bm{R})+V_o(r)O^{a\,\dagger}(\bm{r},\bm{R})O^a(\bm{r},\bm{R})\right)+\mathcal{H}_{YM}+\mathcal{O}(r)\,.
\end{equation}
We will use the symbol $\mathcal{H}$ to distinguish pNRQCD Hamiltonians from the NRQCD symbol $H$. We assume that the theory has been quantized in an $A_0^a=0$ gauge for simplicity. The $S$ and $O$ fields depend on $\bm{r}$, the center-of-mass $\bm{R}$, and the time $t$ (which is not displayed here, because the Hamiltonian as a whole is time independent). At leading order in the multipole expansion the singlet and octet degrees of freedom decouple, but the octet is still coupled to gluons because of the Gauss law. $V_s(r) $ and $V_o(r)$ are pNRQCD matching coefficients corresponding to the static quark-antiquark potential in a singlet and in an octet color configuration, respectively. These potential terms are generated by soft gluons, which are still dynamical in NRQCD but integrated out in pNRQCD, so their effect has to be included explicitly in the Hamiltonian.

$\mathcal{H}_{YM}$ has the same form as the pure Yang-Mills plus light-quark part of the NRQCD Hamiltonian given in Eq.~\eqref{H0}, but all fields are now understood as ultrasoft. The same conditions on the light quarks as discussed in the previous chapter also apply here. The inclusion or omissions of light quarks as sea quarks seems not to critically affect the pattern of the lowest hybrid masses. This is indicated by the few existing unquenched lattice calculations of the gluelump masses~\cite{Marsh:2013xsa} and static energies~\cite{Bali:2000vr}.

In the $r\to0$ limit extra symmetries for the gluonic excitations between static quarks appear. The glue dynamics no longer involve the relative coordinate $\bm{r}$, in particular, there is no longer a special direction dictated by the quark-antiquark axis. Therefore, the glue associated with a gluonic excitation between static quarks acquires a spherical symmetry. So in the center-of-mass system gluonic excitations between static quarks are classified according to representations of $\mathrm{O}(3)\otimes\mathrm{C}$~\cite{Brambilla:1999xf}, as opposed to the $D_{\infty h}$ group in NRQCD. We will indicate these quantum numbers by $K^{PC}$, where $\bm{K}$ is the angular momentum operator of the gluons.

Accordingly, in the short distance limit the static states have to be given through glueball and gluelump operators, which we will call $G$ and $G^a$ respectively. While a gluelump itself consists of the color singlet combination of a color octet source with gluons, here we will always use the term ``gluelump operator'' to refer only to the gluonic operator, since the source will always be given by the quarkonium octet field. The glueball and gluelump operators are normalized as
\begin{equation}
 \langle0|G_{m,\,i}(\bm{R})G_{n,\,j}(\bm{R})|0\rangle=\delta_{mn}\delta_{ij}\,,\hspace{10pt}\mathrm{and}\hspace{13pt}\langle0|G^a_{m,\,i}(\bm{R})G^b_{n,\,j}(\bm{R})|0\rangle=\delta_{mn}\delta_{ij}\delta^{ab}\,.
\end{equation}
Here the operators are assumed to be real. The first indices $m$ and $n$ label different types of glueballs or gluelumps, the second indices $i$ and $j$ label the different components of the respective $K^{PC}$ representation.

We can then match the eigenstates of the static NRQCD Hamiltonian to pNRQCD through
\begin{equation}
\left|\underline{n};\bm{x}_1,\bm{x}_2\right\rangle^{(0)}\stackrel{\sim}{=}\left(S^\dagger(\bm{r},\bm{R})\,\hat{n}^{\phantom{a}}_i\,G_{n,\,i}(\bm{R})+\mathcal{O}(r)\right)\left|0\right\rangle
\label{s+gmatch}
\end{equation}
for the singlet plus glueball states and
\begin{equation}
\left|\underline{n};\bm{x}_1,\bm{x}_2\right\rangle^{(0)}\stackrel{\sim}{=}\left(O^{a\,\dagger}(\bm{r},\bm{R})\,\hat{n}^{\phantom{a}}_i\,G^a_{n,\,i}(\bm{R})+\mathcal{O}(r)\right)\left|0\right\rangle
\label{glmatch}
\end{equation}
for the gluelump states, where $\hat{n}$ is some unit projection vector that fixes the $D_{\infty h}$ quantum numbers. Higher order terms in the multipole expansion will also be operators of this form, so the states will no longer be purely singlet plus glueball or gluelump, but a combination of all of these states with the right $D_{\infty h}$ quantum numbers. We use the symbol $\stackrel{\sim}{=}$ to read ``matches to'', meaning that, although the states or operators on both sides are defined in different Fock spaces, calculating amplitudes in either theory gives the same results. In this case the matching condition is that acting with the static Hamiltonian of either theory on the respective state gives the same static energy.

Since the projection vector $\hat{n}$ does not influence the static energy at leading order in the multipole expansion, several static energies are degenerate in the short distance limit $r \ll 1/\Lambda_\mathrm{QCD}$. We can see this for the gluelump states by calculating that
\begin{align}
 H^{(0)}\left|\underline{n};\bm{x}_1,\bm{x}_2\right\rangle&\stackrel{\sim}{=}\left[\int d^3R'\,d^3r'\,V_o(r)O^{a'\,\dagger}(\bm{r}',\bm{R}')O^{a'}(\bm{r}',\bm{R}'),O^{a\,\dagger}(\bm{r},\bm{R})\right]\,\hat{n}^{\phantom{a}}_iG^a_{n,\,i}(\bm{R})\left|0\right\rangle\notag\\
 &\hspace{13pt}+O^{a\,\dagger}(\bm{r},\bm{R})\,\hat{n}^{\phantom{a}}_i\left[\mathcal{H}_{YM},G^a_{n,\,i}(\bm{R})\right]\left|0\right\rangle+\mathcal{O}(r)\notag\\
 &=\left(V_o(r)+\Lambda_H+\mathcal{O}\left(r^2\right)\right)\left(O^{a\,\dagger}(\bm{r},\bm{R})\,\hat{n}^{\phantom{a}}_iG^a_{n,\,i}(\bm{R})+\mathcal{O}(r)\right)\left|0\right\rangle\,.
\label{hypot}
\end{align}
For the singlet plus glueball states the calculation goes analogously. The glueball or gluelump mass $\Lambda_H$ is the energy eigenvalue of the states generated by $G$ or $G^a$ under the Yang-Mills Hamiltonian. It depends on $n$ but it is the same for any component of $G$ or $G^a$, so the projections have no influence on the leading order of the static energy. This approximate degeneracy for small $r$ is a direct consequence of the extension of the $D_{\infty h}$ symmetry group to $\mathrm{O}(3)\otimes\mathrm{C}$.

The glueball and gluelump masses $\Lambda_H$ are well defined as eigenvalues of the Yang-Mills Hamiltonian, however, the operators that create the corresponding eigenstates are unknown. This situation is similar to the previous section, where it is also unknown how to express the exact static NRQCD states $\left|\underline{n};\bm{x}_1,\bm{x}_2\right\rangle^{(0)}$ in terms of NRQCD fields. So one can use the same approach here to determine the values of $\Lambda_H$: one uses operators with the same quantum numbers as $G$ or $G^a$ and projects out the lowest energy eigenvalue through the large time limit.

The NRQCD states $|X_n\rangle$ defined in~\eqref{latstat} match in pNRQCD at leading order in the multipole expansion to
\begin{equation}
 \left|X_n\right\rangle\stackrel{\sim}{=}\left(Z_n(r)\,O^{a\,\dagger}(\bm{r},\bm{R})P_n^a(\bm{\hat{r}},\bm{R})+\mathcal{O}(r)\right)|0\rangle\,.
\end{equation}
The matching constant $Z_n$ accounts for effects at the scale $1/r$, which have been integrated out in pNRQCD, and so it depends on $r$ in a nonanalytic way. However, it gives a vanishing term in the large time correlator~\eqref{eelwl}, so it has no influence on the static energies. Table~\ref{tab3} shows a set of convenient gluon operators $P_n^a$ corresponding to the lowest hybrid quantum numbers. The expected pattern of degeneracies in the short distance limit also can be read off from this table:
\begin{align}
\Sigma_u^- &\sim \Pi_u\,, &
\Sigma_g^- &\sim \Pi_g^{\prime} \sim \Delta_g\,,
\notag
\\
\Sigma_g^{+\,\prime} &\sim \Pi_g\,, &
\Sigma_u^{+} &\sim \Pi_u^{\prime} \sim \Delta_u \,,
\label{dege}
\end{align} 
where a prime indicates an excited state~\cite{Brambilla:1999xf} (see also~\cite{Foster:1998wu}).

\begin{table}[t]
\centerline{
\begin{tabular}{||c|c|c||}
\hline\hline
$\Lambda_\eta^\sigma$   & $K^{PC}$ & $P^a$                                                                                                                \\ \hline
$\Sigma_u^-$            & $1^{+-}$ & $\bm{\hat{r}}\cdot\bm{B},\,\bm{\hat{r}}\cdot(\bm{D}\times \bm{E})$                                                   \\ 
$\Pi_u$                 & $1^{+-}$ & $\bm{\hat{r}}\times\bm{B},\,\bm{\hat{r}}\times(\bm{D}\times \bm{E})$                                                 \\
$\Sigma_g^{+\,\prime}$  & $1^{--}$ & $\bm{\hat{r}}\cdot\bm{E},\,\bm{\hat{r}}\cdot(\bm{D}\times \bm{B})$                                                   \\
$\Pi_g$                 & $1^{--}$ & $\bm{\hat{r}}\times\bm{E},\,\bm{\hat{r}}\times(\bm{D}\times \bm{B}) $                                                \\
$\Sigma_g^-$            & $2^{--}$ & $(\bm{\hat{r}}\cdot \bm{D})(\bm{\hat{r}}\cdot \bm{B}) $                                                              \\
$\Pi_g^{\prime}$        & $2^{--}$ & $\bm{\hat{r}}\times((\bm{\hat{r}}\cdot\bm{D}) \bm{B}+\bm{D}(\bm{\hat{r}}\cdot\bm{B}))$                               \\
$\Delta_g$              & $2^{--}$ & $(\bm{\hat{r}}\times \bm{D})^i(\bm{\hat{r}}\times\bm{B})^j+(\bm{\hat{r}}\times\bm{D})^j(\bm{\hat{r}}\times\bm{B})^i$ \\
$\Sigma_u^{+}$          & $2^{+-}$ & $(\bm{\hat{r}}\cdot \bm{D})(\bm{\hat{r}}\cdot \bm{E})$                                                               \\
$\Pi_u^{\prime}$        & $2^{+-}$ & $\bm{\hat{r}}\times((\bm{\hat{r}}\cdot\bm{D}) \bm{E}+\bm{D}(\bm{\hat{r}}\cdot\bm{E})) $                              \\
$\Delta_u$              & $2^{+-}$ & $(\bm{\hat{r}}\times \bm{D})^i(\bm{\hat{r}}\times\bm{E})^j+(\bm{\hat{r}}\times\bm{D})^j(\bm{\hat{r}}\times\bm{E})^i$ \\ \hline\hline
\end{tabular}}
\caption{Gluonic excitation operators at leading order in the multipole expansion in pNRQCD up to mass dimension 3; $\bm{\hat{r}}$ denotes the unit vector in the direction of the quark-antiquark distance~$\bm{r}$. Different projections of the same fields correspond to different $D_{\infty h}$ representations, which are degenerate in the small distance limit. The cross product with $\bm{\hat{r}}$ has two linearly independent components, which correspond to the two components of $\Lambda\geq1$ representations of $D_{\infty h}$; the same applies for the symmetric tensor operators of the $\Delta$ representations. Note that the $K^{PC}$ quantum numbers refer only to the gluon fields, not the transformation properties of $\bm{\hat{r}}$, which is $P$ and $C$ odd. The $\Sigma_g^{+}$ is not shown since it corresponds to the ground state. This table is taken from~\cite{Brambilla:1999xf}.}
\label{tab3}
\end{table}

The large time correlators are then given by
\begin{equation}
 \langle X_n,T/2|X_n,-T/2\rangle=\mathcal{N}e^{-iV_o(r)T}\langle0|P_n^a(T/2)\phi_{\mathrm{adj}}^{ab}(T/2,-T/2)P_n^b(-T/2)|0\rangle+\mathcal{O}\left(r^2\right)\,.
\end{equation}
The temporal Wilson line in the gluonic correlator ensures the gauge invariance of the expression. In $A_0^a=0$ gauges, which we assumed in the Hamiltonian, it can be replaced by a Kronecker delta, but in other gauges it is needed. The gluonic correlator can only be evaluated nonperturbatively, since it contains no physical scale except for $\Lambda_\mathrm{QCD}$, but on general grounds we can argue 
that
\begin{equation}
 \langle 0|P_n^a(T/2)\phi(T/2,-T/2)^{\mathrm{adj}}_{ab}P_n^b(-T/2)|0\rangle=|c_n|^2e^{-i\Lambda_HT}+|c_{n'}|^2e^{-i\Lambda_{H'}T}+\dots\,,
\end{equation}
so that we achieve the following matching condition between the static energy $E_n^{(0)}(r)$ in NRQCD and the static potential $V_o(r)$ in pNRQCD [cf.\ Eq.~\eqref{eelwl}]
\begin{equation}
 E_n^{(0)}(r)=\lim_{T\to\infty}\frac{i}{T}\log\langle X_n,T/2|X_n,-T/2\rangle=V_o(r)+\Lambda_H+\mathcal{O}\left(r^2\right)\,.
 \label{vH2}
\end{equation}
Again, for the singlet plus glueball states the calculation is analogous. The ground state corresponds to a singlet without a glueball operator, so
\begin{equation}
 E_{\Sigma_g^+}^{(0)}(r)=V_s(r)+\mathcal{O}\left(r^2\right)\,.
 \label{eqmy}
\end{equation}

At small distances, $r \ll 1/\Lambda_\mathrm{QCD}$, $V_s$ and $V_o$ can be calculated perturbatively. They are known at three loops with some partial results at four loops~\cite{Brambilla:2006wp,Brambilla:2009bi,Smirnov:2009fh,Anzai:2009tm,Pineda:2000gza,Pineda:2011db,Anzai:2013tja}. For a detailed comparison of $V_s$ to the lattice data in the short range, see~\cite{Brambilla:2010pp,Bazavov:2012ka,Bazavov:2014soa}.

Equations~\eqref{vH2} and~\eqref{eqmy} can be systematically improved by calculating higher orders in the multipole expansion. In particular, one can look at how the $\mathrm{O}(3)\otimes\mathrm{C}$ symmetry is softly broken to $D_{\infty h}$ in the short-distance limit. The leading correction coming from the multipole expansion to~\eqref{vH2} and~\eqref{eqmy} is at ${\cal O}(r^2)$ and can be calculated in pNRQCD in terms of nonperturbative correlators to be eventually evaluated on the lattice or in QCD vacuum models. Such a correction is necessary in order to form a bound state, since $V_o(r)$ itself is repulsive.

In this paper we consider only states of the lowest lying symmetry multiplet, i.e., the $\Sigma_u^-$ and $\Pi_u$ states. They are generated from a gluelump with quantum numbers $1^{+-}$. A good gluonic operator $P^a$ overlapping with this gluelump, which can be used in the large time correlator~\eqref{vH2}, is the chromomagnetic field $\bm{B}^a$, so we will call this gluelump operator $\bm{G}_B^a$.

For the projection on the $\Sigma_u^-$ state the unit vector $\bm{\hat{r}}=\left(\sin\theta\cos\varphi,\sin\theta\sin\varphi,\cos\theta\right)^T$ will be used, which gives the direction of the quark-antiquark axis. The other two projection vectors for the $\Pi_u$ states have to be orthogonal to $\bm{\hat{r}}$ and each other, but apart from that we are free to take any two convenient vectors. We will use $\bm{\hat{r}}^\pm=\left(\bm{\hat{\theta}}\pm i\bm{\hat{\varphi}}\right)/\sqrt{2}$, where $\bm{\hat{\theta}}=\left(\cos\theta\cos\varphi,\cos\theta\sin\varphi,-\sin\theta\right)^T$ and $\bm{\hat{\varphi}}=\left(-\sin\varphi,\cos\varphi,0\right)^T$ are the usual local unit vectors in a spherical coordinate system. The advantage of this choice is that with these complex vectors the projections of the gluelump operator transform as $\bm{\hat{r}}^\pm\cdot\bm{G}_B^a\to e^{\pm i\alpha}\,\bm{\hat{r}}^\pm\cdot\bm{G}_B^a$ under rotations by an angle $\alpha$ around the quark-antiquark axis. 

The leading order matching condition is then given by
\begin{align}
\left|1\Sigma_u^-;\bm{x}_1,\bm{x}_2\right\rangle^{(0)}&\stackrel{\sim}{=}O^{a\,\dagger}(\bm{r},\bm{R})\,\bm{\hat{r}}\cdot\bm{G}_B^a(\bm{R})\left|0\right\rangle+\mathcal{O}(r)\,,\\
\left|1\Pi^\pm_u;\bm{x}_1,\bm{x}_2\right\rangle^{(0)}&\stackrel{\sim}{=}O^{a\,\dagger}(\bm{r},\bm{R})\,\bm{\hat{r}}^\pm\cdot\bm{G}_B^a(\bm{R})\left|0\right\rangle+\mathcal{O}(r)\,.
\end{align}
Note that by this definition the index $\pm$ on the $\Pi_u$ states refers to the sign under rotations, while the index $-$ of the $\Sigma_u^-$ state refers to the sign under reflections.

\section{NRQCD Lattice determination of the gluonic static energies} \label{latdat}

The gluonic NRQCD static energies are calculated on the lattice through the logarithm of large time generalized static Wilson loops introduced in Eq.~\eqref{eelwl} divided by the interaction time. The generalized static Wilson loops are constructed using for the initial and final states NRQCD operators with the quantum numbers needed to select the desired static energy [see, for instance, Eq.~\eqref{latstat}].

The static energies for heavy quark-antiquark pairs have been computed in lattice QCD by several authors~\cite{Griffiths:1983ah,Campbell:1984fe,Ford:1988ki,Perantonis:1990dy,Lacock:1996ny,Juge:1997nc,Juge:2002br,Bali:2000vr}. In this section we review the latest available data sets obtained by Juge, Kuti, and Morningstar in~\cite{Juge:1997nc,Juge:2002br} and by Bali and Pineda in~\cite{Bali:2003jq}, which have been used in this paper.

Static energies were obtained in quenched lattice QCD by Juge, Kuti, and Morningstar on anisotropic lattices using an improved gauge action introduced in~\cite{Morningstar:1997ff}. They extracted the static energies from Monte Carlo estimates of generalized large Wilson loops for a large set of operators projected onto the different representations of the $D_{\infty h}$ group. The distance $r$ between the heavy quark-antiquark pair is fixed in the starting time slice. The use of anisotropic lattices with the temporal spacing much smaller than the spatial spacing is crucial to resolve the gluon excitation spectrum. The static energies for the $\Sigma$, $\Pi$ and $\Delta$ gluonic excitations were first computed in~\cite{Juge:1997nc} and then in larger lattice volumes in~\cite{Juge:2002br}. The lattice data from the latter reference consists of four different runs with lattice volumes: $(18^2\times 24)\times 54$, $(16^2\times 20)\times 80$, $14^3\times 56$, and the final one is a finite volume check. The corresponding lattice spacings for these runs are $\sim 0.12$~fm, $\sim 0.19$~fm, $\sim 0.22$~fm, and~$\sim 0.27$~fm. 

\begin{figure}[t]
 \centering
 \includegraphics[width=0.7\linewidth]{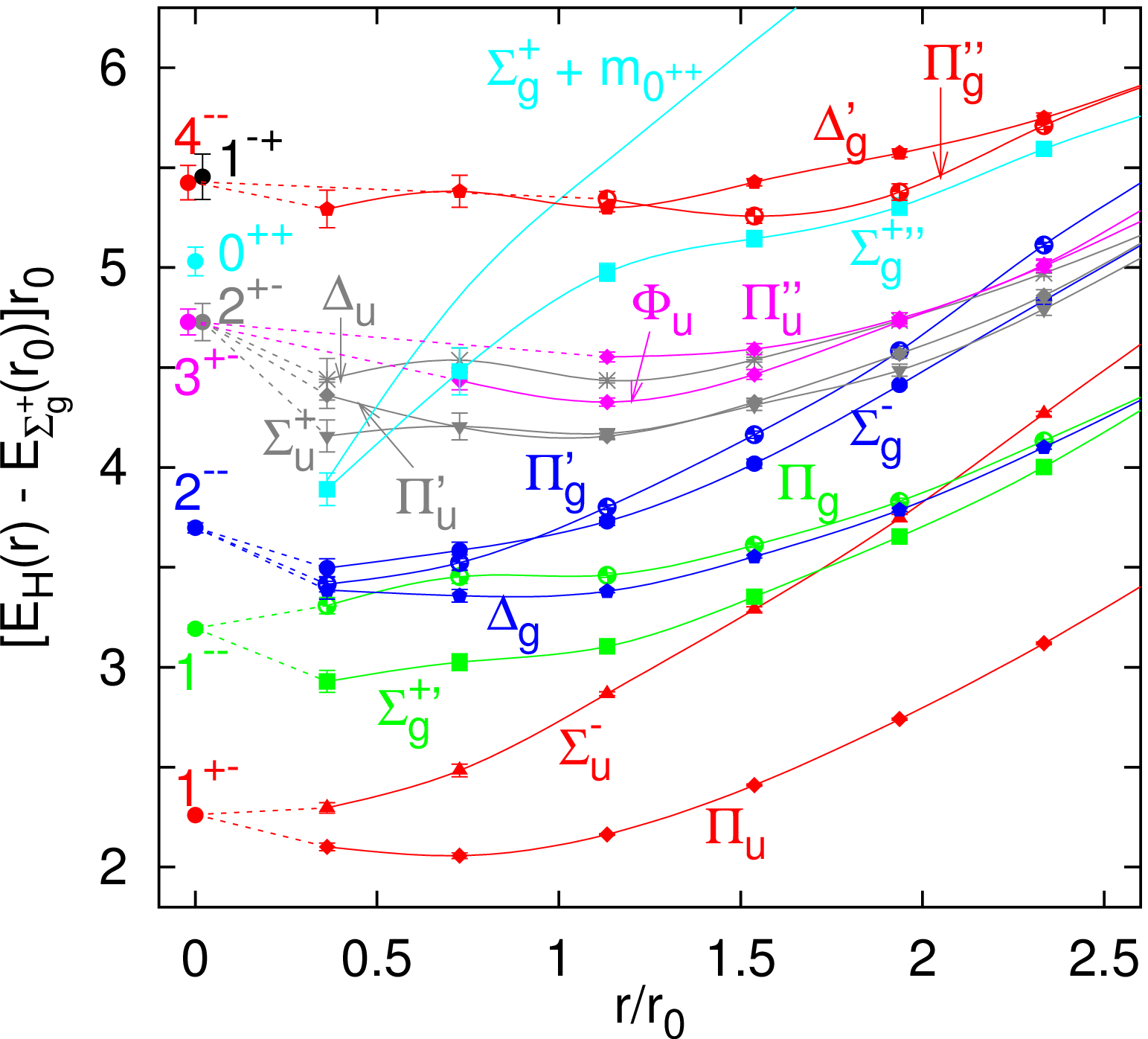}
 \caption{The lowest hybrid static energies~\cite{Juge:2002br} and gluelump masses~\cite{Foster:1998wu} in units of $r_0\approx 0.5$~fm. The absolute values have been fixed such that the ground state $\Sigma_g^+$ static energy (not displayed) is zero at $r_0$. The behavior of the static energies at short distances becomes rather unreliable for some hybrids, especially the higher exited ones. This is largely due to the difficulty in lattice calculations to distinguish between states with the same quantum numbers, which mix. For example, the $\Sigma_g^{+\prime\prime}$ static energy approaches the shape corresponding to a singlet plus $0^{++}$ glueball state (also displayed) instead of the $0^{++}$ gluelump. This picture is taken from~\cite{Bali:2003jq}.}
 \label{hybstat}
\end{figure}

Lattice simulations were carried out by Bali and Pineda in~\cite{Bali:2003jq} focusing on the short range static energies for the $\Pi_u$ and $\Sigma^-_u$ potentials. They performed two sets of computations using a Wilson gauge action in the quenched approximation. The first set was performed on an isotropic lattice with volume $24^3\times 48$ at $\beta=6.2$ and lattice spacing $\approx 0.07$~fm, the second set on three anisotropic lattices with spatial spacings $\approx 0.16,\,0.11,\,0.08$~fm and temporal spacing of one fourth of the spatial spacing, with $\beta=5.8,\,6.0,\,6.2$, respectively. The isotropic data was used as a consistency check and the anisotropic data was extrapolated to the continuum limit. 

The static energies computed on the lattice using generalized static Wilson loops contain divergent self-energy contributions in the temporal lines, in principle associated to the heavy quark mass. These self-energy contributions have to be removed in order to obtain the absolute value of the static energies. They could be removed by comparing the ground state static energy, $\Sigma^+_g$, with the Coulombic potential computed in perturbation theory at very short distances. In practice, however, lattice data is not available for such short distances in which the perturbative regime is valid. Instead, to remove the divergence, Juge, Kuti, and Morningstar fitted the $\Sigma^+_g$ static energy to $\Lambda_0+e_c/r+\kappa r$ and subtracted the value $\Lambda_0$, while Bali and Pineda chose to give the values of the static energies relative to the value of the $\Sigma^+_g$ static energy at $r=r_0\approx 0.5$~fm.

The ground state static energy, $\Sigma^+_g$, and the first gluonic excitation, $\Pi_u$, have been computed in unquenched lattice simulations in~\cite{Bali:2000vr}. The light quarks have unphysically large masses which are equivalent to a pion mass of $650$~MeV. Two lattice volumes were used, $16^3\times32$ and $24^3\times40$ with $\beta=5.6$ and a lattice spacing of $\approx0.076$~fm. Two quenched calculations were carried out in the same work and the results were found to agree within errors with the unquenched $\Sigma^+_g$ and $\Pi_u$ static energies below the quark-antiquark string breaking distance.

As explained in the previous section, in the short distance limit the heavy quark-antiquark pair gives origin to a local octet source, and the spectrum of gluonic static energies is related to the gluelump spectrum. In Fig.~\ref{hybstat} the lattice data from Ref.~\cite{Juge:2002br} is plotted and compared with the gluelump spectrum, computed also on the lattice, of Ref.~\cite{Foster:1998wu}. We can see that the two lowest-lying hybrid static energies are the $\Pi_u$ and $\Sigma^-_u$ states and they clearly tend to form a degenerate multiplet in the short range. The $\Pi_g-\Sigma_g^{+\prime}$, $\Delta_g-\Sigma^-_g-\Pi_g'$ and $\Delta_u-\Pi_u'-\Sigma^+_u$ multiplets are also expected to be degenerate in the short range~\cite{Brambilla:1999xf}, cf. Table~\ref{tab3}. 

\section{The Schr\"odinger equation: matching at order \texorpdfstring{$\bm{1/m}$}{1/m}}\label{sch}

\subsection{Beyond the static limit}

In this section we go beyond the static limit to obtain the bound state equation that gives the hybrid masses. Therefore, we consider the $1/m$ corrections to the NRQCD static Hamiltonian, see Eqs.~\eqref{H10} and~\eqref{H01}. We then match the NRQCD states and Hamiltonian to pNRQCD, obtaining the Schr\"odinger equation that describes the hybrids and the corresponding eigenstates.

The spectrum of the static Hamiltonian $H^{(0)}$, as of any Hermitian operator, provides a full basis of the corresponding Fock space. Therefore, we can express any state, in particular also the eigenstates $|N\rangle$ of the full Hamiltonian $H$, as a superposition of static states:
\begin{equation}
 |N\rangle=\sum_n\int d^3x_1\,d^3x_2\,|n;\bm{x}_1,\bm{x}_2\rangle^{(0)}\psi_n^{(N)}(\bm{x}_1,\bm{x}_2)\,.
 \label{|N>}
\end{equation}
In this notation, $N$ is a shorthand for all quantum numbers of the system described by the full Hamiltonian, which are generally different from the static quantum numbers $n$. The relation in Eq.~\eqref{|N>} is written in the most general way, but quantum numbers that are incompatible with $N$ do not, in fact, appear in the sum over $n$. For example, if a certain static quantum number is also a good quantum number in the non-static system, then the sum in Eq.~\eqref{|N>} can only contain one value for it. By writing the integrations over $\bm{x}_1$ and $\bm{x}_2$ explicitly, we already anticipate that the heavy quark and antiquark positions are not good quantum numbers, which is natural in the non-static system of the full Hamiltonian.

We want to use quantum mechanical perturbation theory in order to determine the leading coefficients in~\eqref{|N>} in the $1/m$ expansion. An important distinction to make here is whether to use degenerate or non-degenerate perturbation theory. In any quantum mechanical system with a Hamiltonian $H^{(0)}+\Delta H$ and a full set of unperturbed eigenstates satisfying $H^{(0)}|n\rangle^{(0)}=E_n^{(0)}|n\rangle^{(0)}$, the first two perturbative corrections to a non-degenerate energy eigenvalue of $H^{(0)}$ are given by
\begin{equation}
 E_n=E_n^{(0)}+{}^{(0)}\langle n|\Delta H|n\rangle^{(0)}+\sum_{n'\neq n}\frac{\left|{}^{(0)}\langle n'|\Delta H|n\rangle^{(0)}\right|^2}{E_n^{(0)}-E_{n'}^{(0)}}+\dots.
\end{equation}
The first correction to the leading term is usually small for a suitably chosen $\Delta H$, but the second correction term can only be considered small if $\langle\Delta H\rangle/\Delta E^{(0)}\ll1$, otherwise the second correction would be of the same order as the first, and the perturbative series would break down. If there is no degeneracy between the energies, i.e., $\Delta E^{(0)}\sim E_n^{(0)}$, then this condition is satisfied. The corresponding full eigenstate is given at leading order by exactly one unperturbed state.

However, if some of the energies are close enough or even identical, then because of the vanishing denominator in the second order term this expansion cannot be valid. Instead, one has to calculate the matrix elements of $H^{(0)}+\Delta H$ between all degenerate states and diagonalize the result. The full eigenstates at leading order are then no longer a single unperturbed state but a superposition of the degenerate states, and the coefficients of this superposition form the eigenvectors that diagonalize $H^{(0)}+\Delta H$ in the degenerate sector. The next correction to the energy is given by a term similar to the second order in the non-degenerate case, but the sum over $n'$ now contains none of the degenerate states (so there is no vanishing denominator), and the single state $|n\rangle$ and the energy $E_n^{(0)}$ have to be replaced by the superposition of degenerate states and the corresponding energy eigenvalue, respectively.

In our case the static states are clearly degenerate regarding the quark and antiquark positions $\bm{x}_1$ and $\bm{x}_2$. The question whether there are
degeneracies related to the other quantum numbers $n$ of the static states is harder to answer. We know that in the short distance limit the states belonging to the same gluelump multiplet are degenerate, and we can assume a mass gap of order $\Lambda_\mathrm{QCD}$ between the lowest gluelump and higher excited multiplets as well as the ground state (cf.\ Fig.~\ref{hybstat} and Ref.~\cite{Marsh:2013xsa}). Neglecting pion contributions is crucial for this assumption. At larger distances $r\sim \Lambda_\mathrm{QCD}^{-1}$ it is also reasonable to assume a mass gap of order $\Lambda_\mathrm{QCD}$ between the $\Pi_u$ and $\Sigma_u^-$ states, while at even larger distances the $\Sigma_u^-$ static energy starts to cross with higher excited states, although we do not expect those crossover regions to be of importance to the low lying hybrids. In any case at very large distance open flavor channels that we neglect will also play a role. So depending on the value of $r$ the static energies may or may not be degenerate, but since the lowest lying hybrids are expected to be located around the minimum of the potential, which is close to the short distance part, we will use degenerate perturbation theory with respect to the $\Pi_u$ and $\Sigma_u^-$ states.

The leading term for the energy in degenerate perturbation theory is obtained by diagonalizing the matrix elements between the degenerate states. For the static plus $1/m$ Hamiltonian, this can be done in two steps. We can write the matrix elements as
\begin{equation}
 ^{(0)}\langle\underline{n}';\bm{x}'_1,\bm{x}'_2|H^{(0)}+H^{(1)}|\underline{n};\bm{x}_1,\bm{x}_2\rangle^{(0)}=\left(\delta_{n'n}E_n^{(0)}+E_{n'n}^{(1)}\right)\delta^{(3)}(\bm{x}'_1-\bm{x}_1)\delta^{(3)}(\bm{x}'_2-\bm{x}_2)\,,
\end{equation}
where we use the abbreviation $H^{(1)}=H^{(1,0)}/m_Q+H^{(0,1)}/m_{\bar{Q}}$. The new energy term $E_{n'n}^{(1)}$ in this expression is a matrix-valued differential operator acting on the delta functions. Diagonalizing the matrix elements corresponds to finding the sets of eigenfunctions $\psi_n^{(N)}$ of $E^{(0)}+E^{(1)}$ satisfying
\begin{equation}
 \sum_{n}\left(\delta_{n'n}E_n^{(0)}+E_{n'n}^{(1)}\right)\psi_n^{(N)}=\mathcal{E}_N\,\psi_{n'}^{(N)}\,,
 \label{degpert}
\end{equation}
where the eigenvalue $\mathcal{E}_N$ gives the mass of the hybrid state as $m_H=m_Q+m_{\bar{Q}}+\mathcal{E}_N$ up to corrections of order $1/m^2$. So the first step corresponds to determining this differential operator, the second to solving the resulting eigenvalue problem.

We will first determine $E_{n'n}^{(1)}$ in the short distance limit, since it is in this regime where we have a strong degeneracy between the $\Sigma_u^-$ and $\Pi_u$ states. Accordingly, we will not calculate the matrix elements for the full $1/m$ Hamiltonian, but only for the leading order in the multipole expansion. The importance of each term can be determined by the standard power counting of weakly-coupled pNRQCD. All powers of $1/r$ including derivatives in $r$ scale as $mv$ with $v\ll1$, while all other dynamical fields scale as the next lower energy scale, which can either be $\Lambda_\mathrm{QCD}$ or $mv^2$, which is the scale of the potential terms. In this case the hierarchy $mv\gg\Lambda_\mathrm{QCD}\gg mv^2$ seems more appropriate.

In the octet sector the $1/m$ pNRQCD Hamiltonian is given by
\begin{equation}
 \mathcal{H}^{(1)}=\int d^3R\,d^3r\,O^{a\,\dagger}(\bm{r},\bm{R})\left[-\frac{\bm{\nabla}_r^2\delta^{ab}}{m}-\frac{\left(\bm{D}_R^2\right)^{ab}}{4m}+\frac{V^{(1)}(r)\delta^{ab}}{m}+\dots\right]O^b(\bm{r},\bm{R})\,.
 \label{1/mpNRQCD}
\end{equation}
Here we have assumed for simplicity that the quark and the antiquark have the same mass $m$, otherwise we would have to distinguish between reduced and total mass, i.e., replace the first denominator by $2m_Qm_{\bar{Q}}/(m_Q+m_{\bar{Q}})$ and the second by $2(m_Q+m_{\bar{Q}})$ etc. These are not all $1/m$ operators, the dots contain other terms that involve the gauge fields $\bm{E}$ and $\bm{B}$ at the same or higher orders in the multipole expansion, including spin interactions.

According to the power counting, the first term of $\mathcal{H}^{(1)}$, which is the kinetic term for the relative distance, scales as $mv^2$, while all other terms scale at most as $\Lambda_\mathrm{QCD}^2/m$ (in the weak coupling regime $V^{(1)}$ is of order $m^2v^4$~\cite{Brambilla:2000gk,*Pineda:2000sz}). We will include only the kinetic term, which means that our calculation will be valid up to corrections of order $\Lambda_\mathrm{QCD}^2/m$. The static Hamiltonian $\mathcal{H}^{(0)}$ itself is of order $mv^2$ in the heavy quark part, which contains the singlet and octet potentials, and of order $\Lambda_\mathrm{QCD}$ in the Yang-Mills part, which gives rise to the gluelump mass. 
So we see that at least the potential term of $\mathcal{H}^{(0)}$ and the kinetic term of $\mathcal{H}^{(1)}$ are of the same order, which is in accordance with the virial theorem of standard quantum mechanics.

In the long-distance limit, we cannot rely on the multipole expansion. Both $E_n^{(0)}$ and $E_{n'n}^{(1)}$ may be expressed as the expectation value of some generalized Wilson loop acting on quark-antiquark color singlet states. These generalized Wilson loops, involving the insertion of gauge fields in a static Wilson loop, can in principle be determined from lattice calculations. They have been in the case of $E_n^{(0)}$, see section~\ref{latdat}, but they have not been in the case of $E_{n'n}^{(1)}$. Hence we will be able to use the full nonperturbative information only for the static energies, while we will have to rely on short distance approximations, and in particular on the leading order term in the multipole expansion, in the case of the $1/m$ terms. This is a reasonable approximation for the lowest hybrid states that are expected to lie near the minimum of the potential, which is sufficiently close to the origin (a quantitative analysis can be found in section~\ref{htmp}).

In summary, we will use nearly degenerate perturbation theory for the static states $\Pi_u$ and $\Sigma_u^-$ belonging to the same $1^{+-}$ gluelump multiplet at short distances. We will use both perturbative and nonperturbative information for the static energies, $E_n^{(0)}$, while we will evaluate $E_{n'n}^{(1)}$ at short distances at leading order in the multipole expansion.

We turn to the evaluation of the matrix elements of the kinetic term in the short distance limit, which will lead to a coupled Schr\"odinger equation. The kinetic term acts on the static states corresponding to the lowest gluelump in the following way:
\begin{align}
 H_{kin}\left|\underline{n};\bm{x}_1,\bm{x}_2\right\rangle^{(0)}&\stackrel{\sim}{=}-\left[\int d^3R'\,d^3r'\,O^{a'\,\dagger}(\bm{r}',\bm{R}')\frac{\bm{\nabla}_{r'}^2}{m}O^{a'}(\bm{r}',\bm{R}'),O^{a\,\dagger}(\bm{r},\bm{R})\right]\bm{\hat{n}}\cdot\bm{G}_B^a(\bm{R})\left|0\right\rangle\notag\\
 &=-\left(\frac{\bm{\nabla}_r^2}{m}O^{a\,\dagger}(\bm{r},\bm{R})\right)\bm{\hat{n}}\cdot\bm{G}_B^a(\bm{R})\left|0\right\rangle\,,
\end{align}
where $\bm{\hat{n}}$ can be either $\bm{\hat{r}}$ or $\bm{\hat{r}}^\pm$ for $\Sigma_u^-$ or $\Pi_u$, respectively. The matrix elements are then given by
\begin{align}
 &^{(0)}\hspace{-2pt}\left\langle\underline{n}';\bm{x}'_1,\bm{x}'_2\right|H_{kin}\left|\underline{n};\bm{x}_1,\bm{x}_2\right\rangle^{(0)}\notag\\
 &=-\left\langle0\right|\bm{\hat{n}}^{\prime\,*}\cdot\bm{G}_B^{a'}(\bm{R'})\left[O^{a'}(\bm{r}',\bm{R}'),\left(\frac{\bm{\nabla}_r^2}{m}O^{a\,\dagger}(\bm{r},\bm{R})\right)\right]\bm{\hat{n}}\cdot\bm{G}_B^a(\bm{R})\left|0\right\rangle\notag\\
 &=-\left\langle0\right|\bm{\hat{n}}^{\prime\,*}\cdot\bm{G}_B^a(\bm{R})\,\bm{\hat{n}}\cdot\bm{G}_B^a(\bm{R})\left|0\right\rangle\frac{\bm{\nabla}_r^2}{m}\delta^{(3)}(\bm{r}-\bm{r}')\delta^{(3)}(\bm{R}-\bm{R}')\notag\\
 &=-\bm{\hat{n}}^{\prime\,*}(\theta',\varphi')\cdot\bm{\hat{n}}(\theta,\varphi)\frac{\bm{\nabla}_r^2}{m}\delta^{(3)}(\bm{r}-\bm{r}')\delta^{(3)}(\bm{R}-\bm{R}')\,.
 \label{me}
\end{align}
To evaluate the expectation value of the gluonic operators we have used the fact that the gluelump operators create orthonormal states. The dependence on the coordinates of the projection vectors has been made explicit in the last line.

If we now let the differential operator corresponding to these matrix elements act on the wave functions, which is equivalent to a convolution of Eq.~\eqref{me} with $\psi_n^{(N)}(\bm{r})$, then we obtain the following differential equation (replacing $\bm{r}'$ with $\bm{r}$)
\begin{equation}
 \sum_{n=\Sigma,\,\Pi^\pm}\bm{\hat{n}}^{\prime\,*}(\theta,\varphi)\cdot\left(-\frac{\bm{\nabla}_r^2}{m}+E_n^{(0)}(r)\right)\,\bm{\hat{n}}(\theta,\varphi)\,\Psi^{(N)}_n(\bm{r})=\mathcal{E}_N\,\Psi^{(N)}_{n'}(\bm{r})\,.
 \label{Diffeq}
\end{equation}
Comparing this result with Eq.~\eqref{degpert}, the scalar product of $\bm{\hat{n}}'$ and $\bm{\hat{n}}$ gives the $\delta_{n'n}$ in front of the static energy, and the first term gives the differential operator $E_{n'n}^{(1)}$, which will have a more complicated expression, because the derivatives act not only on the wave function, but also on $\bm{\hat{n}}$. The wave functions only need to depend on $\bm{r}$, because we have neglected the kinetic term for $\bm{R}$, so the center-of-mass coordinate is still a good quantum number. This corresponds to a hybrid at rest without any recoil effects between heavy quarks and gluons.

\subsection{The radial Schr\"odinger equation}

The Laplace operator $\bm{\nabla}_r^2$ can be split into a radial and an angular part, such that
\begin{equation}
 -\frac{\bm{\nabla}_r^2}{m}=-\frac{1}{m\,r^2}\left(\partial_r\,r^2\,\partial_r+\partial_x\left(1-x^2\right)\partial_x+\frac{1}{1-x^2}\,\partial^2_\varphi\right)\,,
\end{equation}
where we have replaced the angle $\theta$ by $x=\cos\theta$. The radial part $\partial_r\,r^2\,\partial_r$ acts only on the wave function $\Psi_n(\bm{r})$, and the scalar product of the projection vectors just gives a Kronecker delta: $\bm{\hat{n}}'\cdot\bm{\hat{n}}=\delta_{n'n}$.

The angular part usually has eigenfunctions in the spherical harmonics, however, the presence of the projection vectors modifies the defining differential equations in the diagonal entries $n'=n$ to
\begin{equation}
 -\left[\partial_x\left(1-x^2\right)\partial_x+\frac{1}{1-x^2}\left(\partial^2_\varphi-2i\lambda x\partial_\varphi-\lambda^2\right)\right]v_{l,\,m}^\lambda(x,\varphi)=l(l+1)v_{l,\,m}^\lambda(x,\varphi)\,,
\end{equation}
where $\lambda$ labels the different projection vectors, $\lambda=0$ for $\bm{\hat{r}}$ and $\lambda=\pm1$ for $\bm{\hat{r}}^\pm$. An explicit solution for these orbital wave functions can be given as
\begin{align}
v_{l,\,m}^\lambda(x,\varphi)&=\frac{(-1)^{m+\lambda}}{2^l}\sqrt{\frac{2l+1}{4\pi}\frac{(l-m)!}{(l+m)!(l-\lambda)!(l+\lambda)!}}P_{l,\,m}^\lambda(x)e^{im\varphi}\,,\\
P_{l,\,m}^\lambda(x)&=(1-x)^{\frac{m-\lambda}{2}}(1+x)^{\frac{m+\lambda}{2}}\partial_x^{l+m}(x-1)^{l+\lambda}(x+1)^{l-\lambda}\,.
\end{align}
A derivation of these functions can be found in textbooks such as~\cite{LandauLifshitz}. They are defined for $|m|\leq l$ and $|\lambda|\leq l$, and for $\lambda=0$ they are identical to the spherical harmonics.

The quantum numbers $l$ and $m$ correspond to the eigenvalues of the angular momentum $\bm{L}=\bm{L}_{Q\bar{Q}}+\bm{K}$, where $\bm{L}_{Q\bar{Q}}$ is the angular momentum operator of the heavy quarks, and $\bm{K}$ is the gluon angular momentum operator. These eigenvalues appear when the operator acts on the state, not only the wave function:
\begin{align}
 \bm{L}^2\int d\Omega\,\left(v_{l,\,m}^\lambda\,\bm{\hat{r}}^\lambda\cdot\bm{G}_B^a\,O^{a\,\dagger}\right)|0\rangle&=l(l+1)\int d\Omega\,\left(v_{l,\,m}^\lambda\,\bm{\hat{r}}^\lambda\cdot\bm{G}_B^a\,O^{a\,\dagger}\right)|0\rangle\,,\\
 L_3\int d\Omega\,\left(v_{l,\,m}^\lambda\,\bm{\hat{r}}^\lambda\cdot\bm{G}_B^a\,O^{a\,\dagger}\right)|0\rangle&=m\int d\Omega\,\left(v_{l,\,m}^\lambda\,\bm{\hat{r}}^\lambda\cdot\bm{G}_B^a\,O^{a\,\dagger}\right)|0\rangle\,.
\end{align}

The states with the orbital wave functions $v_{l,\,m}^\lambda$ are eigenstates of the angular momentum, but not yet of parity and charge conjugation, because acting with $P$ or $C$ turns $\lambda$ into $-\lambda$. We list here the transformation properties of all elements of the states:
\begin{align}
 v_{l,\,m}^\lambda&\stackrel{P}{\to}(-1)^lv_{l,\,m}^{-\lambda}\,, & v_{l,m}^\lambda&\stackrel{C}{\to}(-1)^lv_{l,\,m}^{-\lambda}\,,\\
 \bm{\hat{r}}^\lambda&\stackrel{P}{\to}(-1)^{\lambda+1}\bm{\hat{r}}^{-\lambda}\,, & \bm{\hat{r}}^\lambda&\stackrel{C}{\to}(-1)^{\lambda+1}\bm{\hat{r}}^{-\lambda}\,,\\
 \bm{G}_B^a&\stackrel{P}{\to}\bm{G}_B^a\,, & \bm{G}_B^a&\stackrel{C}{\to}-(-)^a\bm{G}_B^a\,,\\
 O_s^a&\stackrel{P}{\to}-O_s^a\,, & O_s^a&\stackrel{C}{\to}(-1)^s(-)^aO_s^a\,.
\end{align}
The factor $(-)^a$ comes from $T^a=(-)^a(T^a)^T$, but since it appears in front of the octet field and of the gluelump operator, it cancels for the gluelump states. The quantum number $s$ labels the total spin of the quark and the antiquark and can have values $0$ or $1$.

For $\lambda=0$ we already have parity and charge conjugation eigenstates:
\begin{align}
 v_{l,\,m}^0\,\bm{\hat{r}}\cdot\bm{G}_B^a\,O^{a\,\dagger}|0\rangle&\stackrel{P}{\to}(-1)^{l\phantom{+s}}\,v_{l,\,m}^0\,\bm{\hat{r}}\cdot\bm{G}_B^a\,O^{a\,\dagger}|0\rangle\,,\\
 v_{l,\,m}^0\,\bm{\hat{r}}\cdot\bm{G}_B^a\,O^{a\,\dagger}|0\rangle&\stackrel{C}{\to}(-1)^{l+s}\,v_{l,\,m}^0\,\bm{\hat{r}}\cdot\bm{G}_B^a\,O^{a\,\dagger}|0\rangle\,.
\end{align}
For $|\lambda|=1$ we can define even and odd parity or charge conjugation states:
\begin{align}
 \frac{1}{\sqrt{2}}\left(v_{l,\,m}^1\,\bm{\hat{r}}^+\pm v_{l,\,m}^{-1}\,\bm{\hat{r}}^-\right)\cdot\bm{G}_B^a\,O^{a\,\dagger}|0\rangle&\stackrel{P}{\to}\mp(-1)^{l\phantom{+s}}\,\frac{1}{\sqrt{2}}\left(v_{l,\,m}^1\,\bm{\hat{r}}^+\pm v_{l,\,m}^{-1}\,\bm{\hat{r}}^-\right)\cdot\bm{G}_B^a\,O^{a\,\dagger}|0\rangle\,,\\*
 \frac{1}{\sqrt{2}}\left(v_{l,\,m}^1\,\bm{\hat{r}}^+\pm v_{l,\,m}^{-1}\,\bm{\hat{r}}^-\right)\cdot\bm{G}_B^a\,O^{a\,\dagger}|0\rangle&\stackrel{C}{\to}\mp(-1)^{l+s}\,\frac{1}{\sqrt{2}}\left(v_{l,\,m}^1\,\bm{\hat{r}}^+\pm v_{l,\,m}^{-1}\,\bm{\hat{r}}^-\right)\cdot\bm{G}_B^a\,O^{a\,\dagger}|0\rangle\,.
\end{align}
We see that the combination with a relative minus sign has the same $P$ and $C$ transformation properties as the $\lambda=0$ state, while the positive combination has opposite behavior.

Now the angular momentum $\bm{L}$ and the spin $\bm{S}$ can be combined with the usual Clebsch-Gordan coefficients to form eigenstates of the total angular momentum $\bm{J}=\bm{L}+\bm{S}$. Since at this level of the approximation nothing depends on the spin, all the different spin combinations have the same energy and appear as degenerate multiplets. The $J^{PC}$ quantum numbers are then $\left\{l^{\pm\pm};(l-1)^{\pm\mp},l^{\pm\mp},(l+1)^{\pm\mp}\right\}$, where the first entry corresponds to the spin $0$ combination and the next three entries to the spin $1$ combinations. For $l=0$ there is only one spin $1$ combination as well as only one parity or charge conjugation state (see below), so we have $\left\{0^{++},1^{+-}\right\}$. In Table~\ref{multiplets} the first five degenerate multiplets that can be obtained are shown, arranged according to their energy eigenvalues (see section~\ref{htmp}). 

\begin{table}[t]
 \begin{tabular}{|c|c|c|c|c|}
 \hline
        & $\,l\,$ & $J^{PC}\{s=0,s=1\}$       & $E_n^{(0)}$           \\  \hline
  $H_1$ & $1$     & $\{1^{--},(0,1,2)^{-+}\}$ & $\Sigma_u^-$, $\Pi_u$ \\
  $H_2$ & $1$     & $\{1^{++},(0,1,2)^{+-}\}$ & $\Pi_u$               \\
  $H_3$ & $0$     & $\{0^{++},1^{+-}\}$       & $\Sigma_u^-$          \\
  $H_4$ & $2$     & $\{2^{++},(1,2,3)^{+-}\}$ & $\Sigma_u^-$, $\Pi_u$ \\
  $H_5$ & $2$     & $\{2^{--},(1,2,3)^{-+}\}$ & $\Pi_u$               \\
  \hline
 \end{tabular}
 \caption{$J^{PC}$ multiplets with $l\leq2$ for the $\Sigma_u^-$ and $\Pi_u$ gluonic states. We follow the naming notation $H_i$ used in~\cite{Juge:1999ie,Braaten:2014qka,*Braaten:2014ita,*Braaten:2013boa}, which orders the multiplets from lower to higher mass. The last column shows the gluonic static energies that appear in the Schr\"odinger equation of the respective multiplet.}
 \label{multiplets}
\end{table}

The $\lambda=0$ state will be convoluted with the radial wave functions $\psi^{(N)}_\Sigma(r)$, while the radial wave functions $\psi^{(N)}_{\pm\Pi}(r)$ will be convoluted with the $|\lambda|=1$ states that have the relative $\pm$ sign between the two projection vectors and orbital wave functions. The differential term $\bm{\hat{n}}'\cdot\nabla_r^2\,\bm{\hat{n}}$ in the coupled Schr\"odinger equation not only changes the differential equations for the orbital wave functions, it also adds additional diagonal and offdiagonal terms. The offdiagonal terms change the radial $\Sigma$ wave function to $\Pi$ and vice versa, however, they can not change the parity of the states. This means that $\psi_\Sigma^{(N)}$ mixes only with $\psi_{-\Pi}^{(N)}$, and $\psi_{+\Pi}^{(N)}$ decouples. We then have the following coupled radial Schr\"odinger equation for one parity state,
\begin{equation}
 \hspace{-4pt}\left[-\frac{1}{mr^2}\,\partial_rr^2\partial_r+\frac{1}{mr^2}\begin{pmatrix} l(l+1)+2 & 2\sqrt{l(l+1)} \\ 2\sqrt{l(l+1)} & l(l+1) \end{pmatrix}+\begin{pmatrix} E_\Sigma^{(0)} & 0 \\ 0 & E_\Pi^{(0)} \end{pmatrix}\right]\hspace{-4pt}\begin{pmatrix} \psi_\Sigma^{(N)} \\ \psi_{-\Pi}^{(N)}\end{pmatrix}=\mathcal{E}_N\begin{pmatrix} \psi_\Sigma^{(N)} \\ \psi_{-\Pi}^{(N)}\end{pmatrix}\,,
\label{rsheqnp}
\end{equation}
and for the other we get the conventional radial Schr\"odinger equation
\begin{equation}
 \left[-\frac{1}{mr^2}\,\partial_r\,r^2\,\partial_r+\frac{l(l+1)}{mr^2}+E_\Pi^{(0)}\right]\psi_{+\Pi}^{(N)}=\mathcal{E}_N\,\psi_{+\Pi}^{(N)}\,.
\label{rsheqpp}
\end{equation}

There is a special case for $l=0$ in that the offdiagonal terms in the coupled equation vanish, so the radial Schr\"odinger equations for $\psi_\Sigma^{(N)}$ and $\psi_{-\Pi}^{(N)}$ also decouple. In fact, $\psi_{-\Pi}^{(N)}$ is irrelevant, since there are no orbital wave functions with $|\lambda|=1$ for $l=0$. The same applies to $\psi_{+\Pi}^{(N)}$. So for $l=0$ there exists only one parity state, and its radial wave function is given by an almost ordinary Schr\"odinger equation with the $E_\Sigma^{(0)}$ potential, the only unusual element is that the angular part is $2/mr^2$ even though $l=0$.

In Appendix~\ref{RSEQ} we describe the derivation of the radial Schr\"odinger equations in more detail. For the uncoupled radial Schr\"odinger equations there exist well established numerical methods to find the wave functions and eigenvalues. These can also be extended to the coupled case, more details on the specific approach that we chose to get the numerical results are given in Appendix~\ref{appx} and~\cite{program}.

\subsection{Comparison with other descriptions of hybrids}\label{comodh}

We now compare the pattern of hybrid spin-symmetry multiplets that we have obtained in our approach with the one obtained in different pictures. The BO approximation for hybrids, as it has been employed in Refs.~\cite{Hasenfratz:1980jv,Griffiths:1983ah,Juge:1997nc,Braaten:2014qka,*Braaten:2014ita,*Braaten:2013boa}, produces spin-symmetry multiplets with the same $J^{PC}$ constituents as our $H_i$ multiplets in Table~\ref{multiplets}, however, in all the existing BO papers the masses of opposite parity states are degenerate.

In Ref.~\cite{Braaten:2014qka,*Braaten:2014ita,*Braaten:2013boa} the underlying assumptions of the BO approximation are given in more detail. Two main points are identified, an adiabatic approximation and a single-channel approximation. The adiabatic approximation states that the time scales for heavy and light degrees of freedom are very different, such that the light degrees of freedom adapt instantaneously to changes in the quark and antiquark positions and therefore always form a static eigenstate. This is equivalent to the $1/m$ expansion we have used here, where the hybrid states are expressed in terms of static states. The single-channel approximation states that at leading order the light degrees of freedom remain always in the same static eigenstate, because transitions to other states are suppressed by a mass gap of order $\Lambda_\mathrm{QCD}$. We make the same assumption regarding transitions to static states corresponding to excited gluelumps, but for the lowest gluelump states we go beyond the single-channel approximation, since at short distances they are nearly degenerate.

Consequently, we obtain terms that mix the static states through a coupled Schr\"odinger equation, in a way that is firmly based on QCD. Taking into account these mixing terms, we find that the degeneracy between opposite parity states is broken. In the BO approximation in the context of atomic molecules this effect is also known as \textit{$\Lambda$-doubling}~\cite{LandauLifshitz}. In the context of hybrids, $\Lambda$-doubling and the modified orbital wave functions $v_{l,\,m}^\lambda$ have been discussed here for the first time.

In the constituent gluon picture~\cite{Horn:1977rq}, hybrids are assumed to be composed of a gluonic excitation bound to a heavy-quark antiquark pair. The gluons are assumed to appear in $J^{PC}$ representations unlike the case of pNRQCD or BO descriptions, in which the gluonic states appear in $\Lambda^{\sigma}_{\eta}$ representations. The quantum numbers of the resulting hybrid are obtained by adding those of the gluon and those of the heavy-quark antiquark pair using the standard rules for addition of angular momentum. In this way one gets the same $J^{PC}$ quantum numbers as we do, but they are arranged in larger multiplets.

If, in the constituent gluon picture, we couple a chromomagnetic (i.e., $1^{+-}$) gluonic excitation with an $S$-wave heavy-quark antiquark pair in a spin singlet $\left\{0^{-+}\right\}$ or spin triplet $\left\{1^{--}\right\}$ state, then we get exactly the quantum numbers of $H_1$. Similarly, for $P$-wave quarkonium with quantum numbers $\left\{1^{+-},(0,1,2)^{++}\right\}$ (corresponding to different spin states) we get $H_2\cup H_3\cup H_4$. $H_5$ would then be included in the combination with the next quarkonium quantum numbers. Since for pNRQCD in the limit $r\rightarrow 0$ we recover spherical symmetry, we can see the constituent gluon picture as the short distance limit of the pNRQCD or BO pictures. Furthermore, one can interpret the finer multiplet structure of pNRQCD with respect to the constituent gluon picture as the effect of the finite distance $r$ between the heavy-quark pair.

The flux tube model~\cite{Isgur:1983wj,*Isgur:1984bm} (for a more recent comparison of the flux tube model with the constituent gluon picture see, e.g., \cite{Buisseret:2006wc}) arises from the idea that for QCD in the strong-coupling regime one can think of the gluonic degrees of freedom as having condensed into a collective stringlike flux tube. In this picture the spectrum of gluonic static energies can be interpreted as the vibrational excitation levels of the string. The lowest excitations of such a string will correspond to nonrelativistic, small, transverse displacement oscillations and as such should be well described by the Hamiltonian of a continuous string. The eigenstates of such a Hamiltonian are characterized by the phonon occupation number and their polarizations, while the spectrum  corresponds to the different phonon occupation numbers.

The hybrid quantum numbers are constructed by specifying the gluonic states via phonon operators. The value of $\Lambda$ corresponds to the number of phonons with clockwise polarization minus the number of phonons with anticlockwise polarization. From here one can construct the $J^{PC}$ quantum numbers of the hybrid states in an analogous way to the BO picture. The first excited energy level is a one-phonon state, which necessarily corresponds to a $\Lambda=1$ state, unlike in the pNRQCD case, where the first excited energy level can be $\Lambda=0,\,1$. Thus, the pattern of the spin-symmetry multiplets emerging from the flux tube model in the case of the first excited static energy is the one in Table~\ref{multiplets} except for the nonexistence of $H_3$.

\section{Solving the Schr\"odinger equation: hybrid potentials and masses} \label{htmp}

In order to obtain the hybrid masses, we have to identify the specific form of the hybrid potentials $E_{\Sigma}(r)$ and $E_{\Pi}(r)$ to be used in the coupled Schr\"odinger equations in~\eqref{rsheqnp} and~\eqref{rsheqpp}. In section~\ref{statpnr} we have reviewed the EFT understanding of these potentials arriving at the expression for the short distance hybrid potential in Eq.~\eqref{hypot} and the matching condition with the static energies given in Eq.~\eqref{vH2}. 

It is well known that the quark mass depends on the renormalon subtraction scheme used. This dependence is canceled in standard quarkonium by the analogous dependence of the singlet potential $V_s$~\cite{Pineda:1998id,*Hoang:1998nz}, leaving the total static energy of the singlet, which corresponds to the physical observable, scheme invariant. Similarly, the hybrid static energies are scheme independent, but not $V_o$ and $\Lambda_H$, which depend on the renormalon subtraction scheme used. It has been shown that in the On-Shell (OS) scheme the perturbative expansion of the octet potential has a poor convergence. This bad behavior is due to the presence of singularities in the Borel transform of the perturbative series. These singularities are, however, artificial and cancel out in physical observables such as the static energies.

One of the several possible schemes to improve the convergence of the matching coefficients is the so-called Renormalon Subtracted (RS) scheme. In the RS scheme the singularities in the Borel plane (renormalons) are subtracted from the matching coefficients. In Ref.~\cite{Pineda:2001zq} this scheme has been worked out for the heavy quark mass and the static singlet potential, in Ref.~\cite{Bali:2003jq} the analogous work was done for the octet potential and the lowest gluelump mass. Note that, when working in the RS scheme for the octet potential and gluelump mass, the quark mass in the hybrid static energy also has to be taken in the RS scheme. We have used the RS octet potential $V^{RS}_o(r)$ up to order $\alpha^3_s$ in perturbation theory and $\Lambda^{RS}_H$ at the subtraction scale $\nu_f=1$~GeV. We have summarized the necessary formulas for the octet potential in the RS scheme in Appendix~\ref{rescheme}. 

The next-to-leading order corrections to the hybrid static energies at short distances are proportional to $r^2$. The specific proportionality constant depends on nonperturbative dynamics and can be expressed in terms of chromoelectric and chromomagnetic field correlators in the EFT. It could be calculated on the lattice, but no calculations of these objects exist at the moment, or in QCD vacuum models.\footnote{For a computation in the framework of the stochastic vacuum model see~\cite{SVMthesis}.} We choose to fix this coefficient through a fit to the lattice data for the static energies. We are going to consider that this term takes different values for hybrid static energies corresponding to different representations of $D_{\infty h}$, thus breaking the degeneracy of the short range pNRQCD description of the $\Pi_u$ and $\Sigma^-_u$ static energies at leading order in the multipole expansion. The final form for the short distance hybrid potential we are going to use is then [cf.\ Eq.~\eqref{vH2}]
\begin{equation}
E_n(r)=V^{RS}_o(\nu_f)+\Lambda^{RS}_H(\nu_f)+b_n r^2\,, \quad \nu_f=1\,\mathrm{GeV}\,,
\label{hybridpot}
\end{equation}
and the values of the heavy quark and the $1^{+-}$ gluelump masses in the RS scheme at $\nu_f=1$~GeV are: $m^{RS}_c=1.477(40)$~GeV, $m^{RS}_b=4.863(55)$~GeV, and $\Lambda_H^{RS}=0.87(15)$~GeV~\cite{Pineda:2001zq,Bali:2003jq}.

We have prepared two different fits for the hybrid potentials to be used in the Schr\"odinger equations. The first relies only on information from the short distance regime and fits the quadratic term to the lattice data only up to distances where weakly coupled pNRQCD no longer makes sense. Going to larger distances in this potential is inconsistent. The second fit uses the short distance expression for the potential only for distances where weakly coupled pNRQCD is expected to work well, and uses some generic fit function to describe the lattice data of the static energies for larger distances. Comparing the results obtained from both of these fits gives some idea of the importance of the long range regime for hybrids.

In order to obtain the short range quadratic coefficients $b_n$ of Eq.~\eqref{hybridpot} in either case, we use lattice data from Refs.~\cite{Juge:2002br,Bali:2003jq} described in section~\ref{latdat}. To do these fits, it must be taken into account that the two sources of lattice data have different energy offsets with respect to the theoretical hybrid potential due to the different methods for the subtractions of the mass divergence of the lattice calculations. We extract $b_n$ and the energy offsets from both sets of lattice data by fitting the function
\begin{equation}
\mathcal{V}(r)=V_o^{RS}+c+b_n r^2\,,
\label{potfit}
\end{equation}
with $c$ and $b_n$ as free parameters.

The RS scheme does not affect the coefficient of the quadratic term $b_n$. The constant term $c$ is affected both by the $RS$ scheme and by the subtraction scheme used in the lattice calculation, however, at leading order in the multipole expansion the $\Pi_u$ and $\Sigma_u^-$ potentials are degenerate. Therefore, we perform a fit of both potentials of the form~\eqref{potfit} to the lattice data of both groups, restricting the value of $c$ to be the same for both potentials but different for each group and, conversely, restricting the value of $b_n$ to be the same for both groups but different for each potential.

\begin{figure}[t]
 \centering
 \begin{minipage}{0.49\linewidth}
  \includegraphics[width=\linewidth]{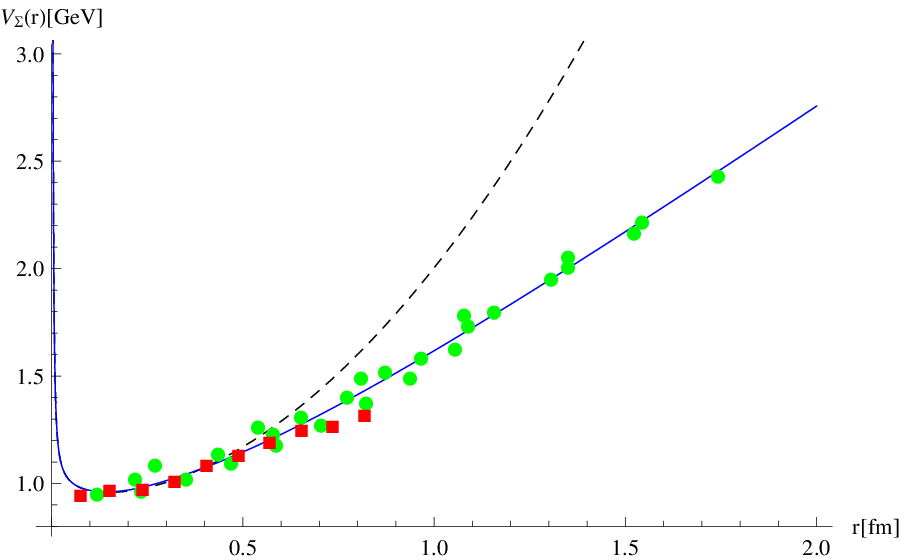}
 \end{minipage}
 \begin{minipage}{0.49\linewidth}
  \includegraphics[width=\linewidth]{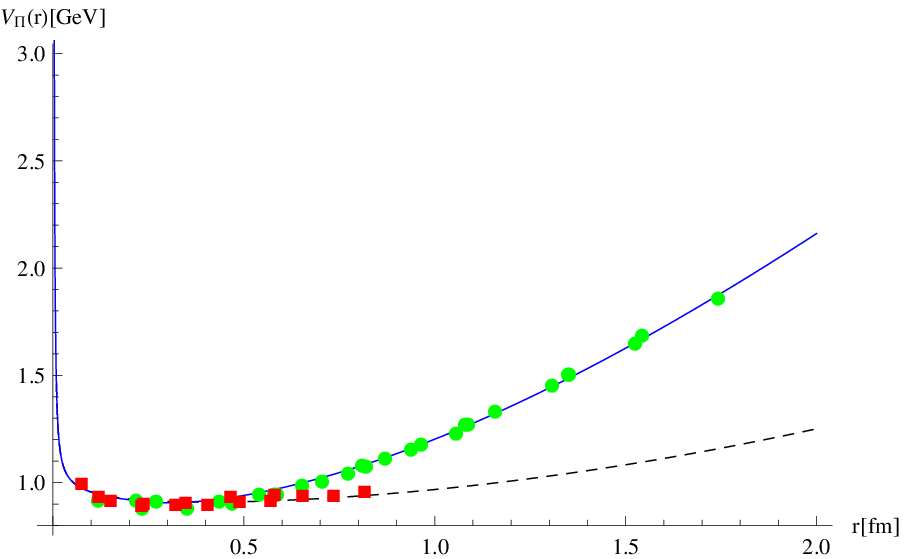}
 \end{minipage}
 \caption{Lattice data from Bali and Pineda~\cite{Bali:2003jq} is represented by red squares, the data from Juge, Kuti, and Morningstar~\cite{Juge:2002br} is represented by the green dots. In the left (right) figure we have plotted the data corresponding to $\Sigma^-_u$ ($\Pi_u$). The lattice data has been corrected by the offsets of~\eqref{offsets}. The black dashed line corresponds to $V_\Sigma^{(0.5)}$ ($V_\Pi^{(0.5)}$), the blue continuous line to $V_\Sigma^{(0.25)}$($V_\Pi^{(0.25)}$), see text.}
 \label{latdatpot}
\end{figure}

We first give the results for the short range fit. The weakly-coupled pNRQCD description of the hybrid static energy of~\eqref{hybridpot} is only valid up to $r\lesssim 1/\Lambda_\mathrm{QCD}$. Taking perturbation theory up to its limit of validity, we fit~\eqref{potfit} to lattice data in the range of $r=0-0.5$~fm. We obtain the following offsets for the two lattice data sources 
\begin{equation}
c_\mathrm{BP}=0.105\,\mathrm{GeV} ,\,\quad c_\mathrm{KJM}=-0.471\,\mathrm{GeV}\,,
\label{offsets}
\end{equation}
and the values for the coefficient of the quadratic term are
\begin{equation}
b^{(0.5)}_{\Sigma}=1.112\,\mathrm{GeV/fm^2},\,\quad b^{(0.5)}_{\Pi}=0.110\,\mathrm{GeV/fm^2}\,.
\label{bfit05}
\end{equation}
The potentials obtained from using the coefficients of the quadratic terms of~\eqref{bfit05} in Eq.~\eqref{hybridpot} will be called $V_\Pi^{(0.5)}$ and $V_\Sigma^{(0.5)}$ 
respectively (corresponding to the $\Pi_u$ and $\Sigma^-_u$ configurations). We have plotted $V_\Pi^{(0.5)}$ and $V_\Sigma^{(0.5)}$ in Fig.~\ref{latdatpot} with the lattice data corrected for the different offsets using the values from~\eqref{offsets}. 

For the second potential fit, which includes as much information as possible from the long range lattice data, we proceed as follows. For $r\leq 0.25$~fm we use the potential from~\eqref{hybridpot} with different $b_n$ factors for each of the low lying hybrid static energies $\Pi_u$ and $\Sigma^-_u$. Accordingly, we will call the potentials from this fit $V^{(0.25)}$. The $b_n$ factors are obtained through a fit of the function~\eqref{potfit} for each potential to lattice data up to $r=0.25$~fm from both sources with the offsets of~\eqref{offsets}. The quadratic term factors resulting from this fit are
\begin{equation}
b^{(0.25)}_{\Sigma}=1.246\,\mathrm{GeV/fm^2},\,\hspace{30pt} b^{(0.25)}_{\Pi}=0.000\,\mathrm{GeV/fm^2}\,.
\label{bfit025}
\end{equation}

For $r\geq 0.25$~fm we use a fit of the function
\begin{equation}
\mathcal{V}'(r)=\frac{a_1}{r}+\sqrt{a_2\,r^2+a_3}+a_4\,,
\label{lattfit}
\end{equation}
to all the lattice data with $r\geq 0.25$~fm using the offsets of~\eqref{offsets}. The particular form of~\eqref{lattfit} is not related to a specific model, but approaches the generally expected behavior at short and large distances. Indeed, in the long distance a linear behavior in $r$ is expected as a
string picture emerges~\cite{Luscher:1980ac,Arvis:1983fp,Juge:2002br,Andreev:2012mc}. The parameters have been left unconstrained (e.g., no universal string tension or short-range coupling have been imposed) to better reproduce the lattice data in the distance region where it is available. To ensure a smooth transition between the two pieces of the potential, we impose continuity up to first derivatives. The parameters obtained are
\begin{align}
a^{\Sigma}_1&=0.000\,\mathrm{GeV fm}\,,&a^{\Sigma}_2&=1.543\,\mathrm{GeV^2/fm^2},&a^{\Sigma}_3&=0.599\,\mathrm{GeV^2},&a^{\Sigma}_4&=0.154\,\mathrm{GeV}\,,\notag\\
a^\Pi_1&=0.023\,\mathrm{GeV fm}\,,&a^\Pi_2&=2.716\,\mathrm{GeV^2/fm^2},&a^\Pi_3&=11.091\,\mathrm{GeV^2},&a^\Pi_4&=-2.536\,\mathrm{GeV}\,.
\label{lattfitpar}
\end{align}

In Fig.~\ref{latdatpot} we can see both potential fits together with the lattice data. The $V^{(0.25)}$ potentials do a good job reproducing the whole range of lattice data, in fact, fitting with a potential of the form~\eqref{lattfit} also for $r<0.25$~fm does not change the results significantly. The $V^{(0.5)}$ potentials describe the lattice data well up to $r\lesssim 0.55-0.65$~fm, which corresponds to $1/r \gtrsim 0.36-0.30$~GeV.

\begin{table}[t]
 \begin{tabular}{||c|c||c|c|c|c||c|c|c|c||c|c|c|c||}
 \hline\hline
 \multirow{2}{*}{\text{multiplet}} & \multirow{2}{*}{$J^{PC}$} & \multicolumn{4}{c||}{$c\bar{c}$} & \multicolumn{4}{c||}{$b\bar{c}$} & \multicolumn{4}{c||}{$b\bar{b}$}\\
 \cline{3-6}\cline{7-10}\cline{11-14}
 & & $m_H$ & $\langle1/r\rangle$ & $E_{kin}$ & $P_\Pi$ & $m_H$ & $\langle1/r\rangle$ & $E_{kin}$ & $P_\Pi$ & $m_H$ & $\langle1/r\rangle$ & $E_{kin}$ & $P_\Pi$ \\
 \hline\hline
 \multicolumn{14}{c}{\vspace{-16pt}}\\
 \hline\hline
 $H_1$ & \multirow{2}{*}{$\{1^{--},(0,1,2)^{-+}\}$} & 4.05 & 0.29 & 0.11 & 0.94 & 7.40 & 0.31 & 0.08 & 0.94 & 10.73 & 0.36 & 0.06 & 0.95 \\
 \cline{3-14}
 $H_1'$ & & 4.23 & 0.27 & 0.20 & 0.91 & 7.54 & 0.30 & 0.16 & 0.91 & 10.83 & 0.36 & 0.11 & 0.92 \\
 \hline
 $H_2$ & \multirow{2}{*}{$\{1^{++},(0,1,2)^{+-}\}$} & 4.09 & 0.21 & 0.13 & 1.00 & 7.43 & 0.23 & 0.10 & 1.00 & 10.75 & 0.27 & 0.07 & 1.00 \\
 \cline{3-14}
 $H_2'$ & & 4.30 & 0.19 & 0.24 & 1.00 & 7.60 & 0.21 & 0.19 & 1.00 & 10.87 & 0.25 & 0.13 & 1.00\\
 \hline
 $H_3$ & $\{0^{++},1^{+-}\}$ & 4.69 & 0.37 & 0.42 & 0.00 & 7.92 & 0.42 & 0.34 & 0.00 & 11.09 & 0.50 & 0.23 & 0.00 \\
 \hline
 $H_4$ & $\{2^{++},(1,2,3)^{+-}\}$ & 4.17 & 0.19 & 0.17 & 0.97 & 7.49 & 0.25 & 0.14 & 0.97 & 10.79 & 0.29 & 0.09 & 0.98 \\
 \hline
 $H_5$ & $\{2^{--},(1,2,3)^{-+}\}$ & 4.20 & 0.17 & 0.18 & 1.00 & 7.51 & 0.19 & 0.15 & 1.00 & 10.80 & 0.22 & 0.10 & 1.00 \\
 \hline\hline
 \multicolumn{14}{c}{\vspace{-16pt}}\\
 \hline\hline
 $H_1$ & \multirow{2}{*}{$\{1^{--},(0,1,2)^{-+}\}$} & 4.15 & 0.42 & 0.16 & 0.82 & 7.48 & 0.46 & 0.13 & 0.83 & 10.79 & 0.53 & 0.09 & 0.86 \\ \cline{3-14}
 $H_1'$ & & 4.51 & 0.34 & 0.34 & 0.87 & 7.76 & 0.38 & 0.27 & 0.87 & 10.98 & 0.47 & 0.19 & 0.87 \\ \hline
 $H_2$ & \multirow{2}{*}{$\{1^{++},(0,1,2)^{+-}\}$} & 4.28 & 0.28 & 0.24 & 1.00 & 7.58 & 0.31 & 0.19 & 1.00 & 10.84 & 0.37 & 0.13 & 1.00 \\ \cline{3-14}
 $H_2'$ & & 4.67 & 0.25 & 0.42 & 1.00 & 7.89 & 0.28 & 0.34 & 1.00 & 11.06 & 0.34 & 0.23 & 1.00 \\ \hline
 $H_3$ & $\{0^{++},1^{+-}\}$ & 4.59 & 0.32 & 0.32 & 0.00 & 7.85 & 0.37 & 0.27 & 0.00 & 11.06 & 0.46 & 0.19 & 0.00\\
 \hline
 $H_4$ & $\{2^{++},(1,2,3)^{+-}\}$ & 4.37 & 0.28 & 0.27 & 0.83 & 7.65 & 0.31 & 0.22 & 0.84 & 10.90 & 0.37 & 0.15 & 0.87 \\
 \hline
 $H_5$ & $\{2^{--},(1,2,3)^{-+}\}$ & 4.48 & 0.23 & 0.33 & 1.00 & 7.73 & 0.25 & 0.27 & 1.00 & 10.95 & 0.30 & 0.18 & 1.00 \\
 \hline
 $H_6$ & $\{3^{--},(2,3,4)^{-+}\}$ & 4.57 & 0.22 & 0.37 & 0.85 & 7.82 & 0.25 & 0.30 & 0.87 & 11.01 & 0.30 & 0.20 & 0.89 \\
 \hline
 $H_7$ & $\{3^{++},(2,3,4)^{+-}\}$ & 4.67 & 0.19 & 0.43 & 1.00 & 7.89 & 0.22 & 0.35 & 1.00 & 11.05 & 0.26 & 0.24 & 1.00 \\
 \hline\hline
 \end{tabular}
 \caption{Hybrid energies obtained from solving the Schr\"odinger equation with the RS heavy quark masses for the $V^{(0.5)}$ potentials (upper table) and for the $V^{(0.25)}$ potentials (lower table). All values are given in units of GeV. The values of the heavy quark and the $1^{+-}$ gluelump masses in the RS scheme at $\nu_f=1$~GeV are: $m^{RS}_c=1.477(40)$~GeV, $m^{RS}_b=4.863(55)$~GeV, and $\Lambda_H^{RS}=0.87(15)$~GeV (see~\cite{Pineda:2001zq,Bali:2003jq}). For the $b\bar{c}$ systems we have used the corresponding reduced mass in the Schr\"odinger equation. The first row for each multiplet corresponds to the ground state, the second row corresponds to the first excited state. $P_{\Pi}$ is the integral over the square of the wave function associated with the $\Pi_u$ potential. It can be interpreted as the probability to find the hybrid in a $\Pi_u$ configuration, thus it gives a measure of the mixing effects.}
\label{meth12}
\end{table}

We have solved the coupled Schr\"odinger equations with both $V^{(0.5)}$ and $V^{(0.25)}$ potentials using the RS heavy quark masses. The results are displayed in Table~\ref{meth12}. The states obtained with $V^{(0.25)}$ lie above the ones obtained using $V^{(0.5)}$. The masses of the states with smaller sizes have a better agreement, since both potentials agree in the short range. The largest source of uncertainties for the hybrid masses lies in the RS gluelump mass, which is known with an uncertainty of $\pm0.15$~GeV.

If we look at the results obtained with $V^{(0.5)}$ for the average of the inverse distance $\langle 1/r\rangle$, which are displayed in Table~\ref{meth12}, we see that for the lowest states the condition that $\langle 1/r\rangle$ falls inside the region where the lattice data is well described by the fit is only marginally fulfilled. The condition that $\langle 1/r\rangle\gtrsim E_{kin}$, which is at the base of the multipole expansion, is instead fulfilled by almost all the states. Interestingly, adding a long range tail to the potential, as we do for $V^{(0.25)}$, pushes the heavy quarks closer together, in this way better justifying the short distance expansion of the matrix element of $H_{kin}$ that we performed in~\eqref{Diffeq}. For this reason we will use the $V^{(0.25)}$ potential in the following section as our reference potential for the comparison with data and other approaches.

\section{Comparison with experimental data and other determinations of the hybrid masses}\label{compare}

We compare our results for the hybrid masses with experimental observations in section~\ref{expid}, predictions obtained using the leading Born--Oppenheimer approximation in section~\ref{lbobraa}, and direct lattice results and sum rule calculations in sections~\ref{comdlat} and ~\ref{comsumrules}, respectively.

\subsection{Identification of hybrids with experimental states} \label{expid}

\begin{figure}
\centerline{
\includegraphics[width=0.9\linewidth]{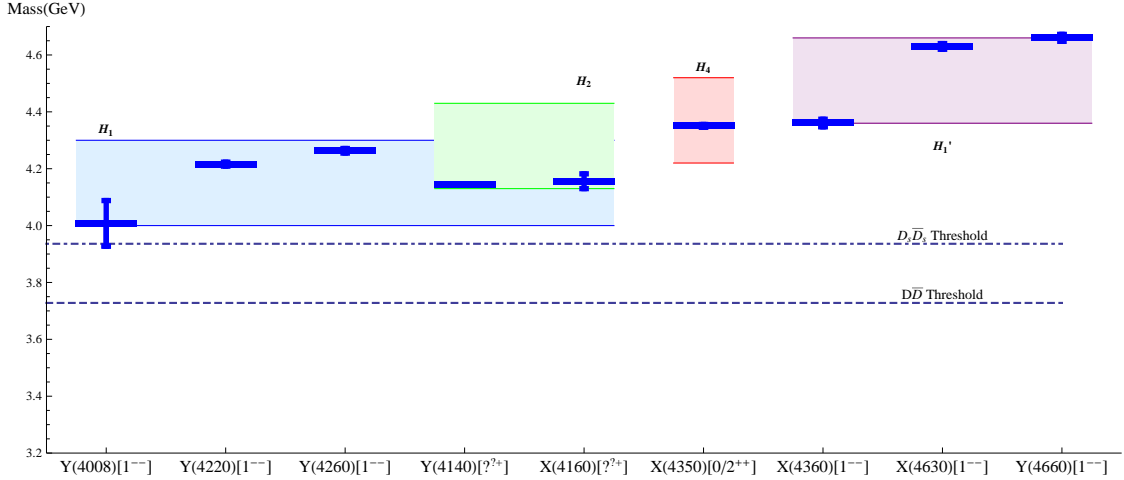}}
\caption{Comparison of the experimental candidate masses for the charmonium sector with our results using the $V^{(0.25)}$ potential. The experimental states are plotted in solid blue lines with error bars corresponding to the average of the lower and upper mass uncertainties (see Table~\ref{nmhc}). Our results for the $H_1$, $H_2$, $H_4$ and $H_1'$ multiplets have been plotted in error bands corresponding to the gluelump mass uncertainty of $\pm0.15$~GeV.}
\label{exp}
\end{figure}

The list of candidates for heavy quark hybrids consists of the neutral heavy quark mesons above open flavor threshold. An updated list~\cite{Brambilla:2014jmp} of the states fulfilling these conditions can be found in Table~\ref{nmhc}. Most of the candidates have $1^{--}$ or $0^{++}/2^{++}$, since the main observation channels are production by $e^+e^-$ or $\gamma\gamma$ annihilation, respectively, which constrains the $J^{PC}$ quantum numbers. It is important to keep in mind that the main source of uncertainty of our results in section~\ref{htmp} is the uncertainty of the gluelump mass $\Lambda_H^{RS}=0.87\pm0.15$~GeV. We have plotted the candidate experimental states in Fig.~\ref{exp}, except for the single one corresponding to the bottomonium sector, overlaid onto our results using the $V^{(0.25)}$ potential with error bands corresponding to the uncertainty of the gluelump mass.

Three $1^{--}$ states fall close to our mass for the charmonium hybrid from the $H_1$ multiplet,\footnote{Note that our hybrid multiplets are spin degenerate, i.e., do not include corrections to the mass due to spin effects.} the $Y(4008)$, $Y(4230)$, and $Y(4260)$. The $1^{--}$ hybrid from the $H_1$ multiplet is a spin singlet state, and as such the decays to spin triplet products are suppressed by one power of the heavy quark mass due heavy quark spin symmetry. All these three candidate states decay to spin triplet charmonium, which in principle disfavors the hybrid interpretation. Nevertheless, there might be enough heavy quark spin symmetry violation to explain those decays~\cite{Wang:2013kra}. On the other hand, the interpretation of these states as charmonium hybrids would make the decay into two $S$-wave open charm mesons forbidden~\cite{Kou:2005gt}, which would explain why such decays have not been observed for the $Y(4260)$. Nevertheless, the recent observation of the transition $Y(4260)\rightarrow X(3872)\gamma$~\cite{Ablikim:2013dyn} makes the identification of $Y(4260)$ as a hybrid highly unlikely.

The $Y(4220)$ is a narrow structure proposed in~\cite{Yuan:2013ffw} to fit the line shape of the annihilation processes $e^+e^-\rightarrow h_c\pi^+\pi^-$ observed by BESIII and CLEO-c experiments. Its mass is quite close to the one of the $H_1$ multiplet. Like the previous states, it is a $1^{--}$ state that would be identified as a spin singlet hybrid. However, unlike the previous states, the $Y(4220)$ has been observed decaying to spin singlet quarkonium, which makes it a very good candidate for a charmonium hybrid. However, the $Y(4220)$ falls very close to the $Y(4230)$~\cite{Ablikim:2014qwy} and it is possible that they are the same structure observed in different decay channels.

The $J^{PC}$ quantum numbers of the $Y(4140)$ and $Y(4160)$ have not yet been fully determined, however, their charge conjugation and mass suggest that they can be candidates for the spin triplet $1^{-+}$ member of the $H_1$ multiplet. Nevertheless, their mass is also compatible within uncertainties with the spin singlet $1^{++}$ member of the $H_2$ multiplet. In the case of the $Y(4160)$, it decays into $D^* \bar D^*$ which favors a molecular interpretation of this state.

If the $X(4350)$ turns out to be a $2^{++}$ state it can be a candidate for the spin singlet charmonium state of the $H_4$ multiplet, although its decay violates heavy quark spin symmetry. 

The three higher mass $1^{--}$ charmonia, the $X(4360)$, $X(4630)$, and $Y(4660)$,\footnote{It has been suggested that $X(4630)$ and $X(4660)$ might actually be the same particle~\cite{Cotugno:2009ys}.} have a mass that is compatible with the excited spin singlet member of the $H_1$ multiplet within uncertainties, although none of them falls very close to the central value. The $X(4360)$ and $Y(4660)$ decay into a spin triplet product, which violates heavy quark spin symmetry.

There is so far only one bottomonium candidate for a hybrid state, the $Y_b(10890)$, which can be identified with the spin singlet $1^{--}$ state of the $H_1$ bottomonium hybrid multiplet. However, its decay to the $\Upsilon$ violates heavy quark spin symmetry, which is expected to be a good symmetry for bottomonium states.

\begin{table}[tb]
\begin{center}
\begin{tabular}{||cllcll||}\hline\hline
   State      &   $M$~(MeV)       & $\Gamma$~(MeV) &  $J^{PC}$    &  Decay modes                                        & 1$^\mathrm{st}$~observation  \\
   \hline
   $X(3823)$  & $3823.1\pm1.9$    & $<24$          &  $?^{?-}$    & $\chi_{c1}\gamma$                                   &  Belle~2013               \\
   $X(3872)$  & $3871.68\pm0.17$  & $<1.2$         &  $1^{++}$    & $J/\psi\,\pi^+\pi^-$, $J/\psi\,\pi^+\pi^-\pi^0$,    &  Belle~2003               \\
              &                   &                &              & $D^0\bar{D}^0\pi^0$, $D^0\bar{D}^0\gamma$,          &                           \\
              &                   &                &              & $J/\psi\,\gamma$, $\psi(2S)\,\gamma$                &                           \\
   $X(3915)$  & $3917.5\pm1.9$    & $20\pm5$       & $0^{++}$     & $J/\psi\,\omega$                                    &  Belle~2004               \\ 
   $\chi_{c2}(2P)$ & $3927.2\pm2.6$  & $24\pm6$    & $2^{++}$     & $D\bar{D}$                                          &  Belle~2005               \\ 
   $X(3940)$    & $3942^{+9}_{-8}$  & $37^{+27}_{-17}$ & $?^{?+}$ & $D^*\bar{D}$, $D\bar{D}^*$                          &  Belle~2007               \\
   $G(3900)$    & $3943\pm21$       & $52\pm11$      & $1^{--}$   & $D\bar{D}$                                          &  Babar~2007               \\ 
   $Y(4008)$    & $4008^{+121}_{-\ 49}$ & $226\pm97$ & $1^{--}$   & $J/\psi\,\pi^+\pi^-$                                &  Belle~2007               \\
   $Y(4140)$    & $4144.5\pm2.6$     & $15^{+11}_{-\ 7}$ & $?^{?+}$ & $J/\psi\,\phi$                                    &  CDF~2009                 \\
   $X(4160)$    & $4156^{+29}_{-25}$ & $139^{+113}_{-65}$ & $?^{?+}$ & $D^*\bar{D}^*$                                   &  Belle~2007               \\
   $Y(4220)$    & $4216\pm7$        & $39\pm17$      & $1^{--}$   & $h_c(1P)\,\pi^+\pi^-$                               &  BESIII~2013              \\ 
   $Y(4230)$    & $4230\pm14$        & $38\pm14$     & $1^{--}$   & $\chi_{c0}\,\omega $                                &  BESIII~2014              \\ 
   $Y(4260)$    & $4263^{+8}_{-9}$  & $95\pm14$      & $1^{--}$   & $J/\psi\,\pi^+\pi^-$, $J/\psi\,\pi^0\pi^0$,         &  Babar~2005               \\ 
                &                   &                &            & $Z_c(3900)\,\pi$                                    &                           \\
   $Y(4274)$    & $4293\pm20$ & $35\pm16$ &          $?^{?+}$     & $J/\psi\,\phi$                                      &  CDF~2010                 \\
   $X(4350)$    & $4350.6^{+4.6}_{-5.1}$ & $13.3^{+18.4}_{-10.0}$ & $0/2^{++}$ & $J/\psi\,\phi$                         &  Belle~2009               \\ 
   $Y(4360)$    & $4354\pm11$       & $78\pm16$      & $1^{--}$   & $\psi(2S)\,\pi^+\pi^-$                              &  Babar~2007               \\ 
   $X(4630)$    & $4634^{+\ 9}_{-11}$ & $92^{+41}_{-32}$ & $1^{--}$ & $\Lambda_c^+\Lambda_c^-$                          &  Belle~2007               \\
   $Y(4660)$    & $4665\pm10$       & $53\pm14$      & $1^{--}$   & $\psi(2S)\,\pi^+\pi^-$                              &  Belle~2007               \\
   \hline
   $Y_b(10890)$ & $10888.4\pm3.0$ & $30.7^{+8.9}_{-7.7}$ & $1^{--}$   & $\Upsilon(nS)\pi^+\pi^-$                        &  Belle~2010               \\
\hline\hline
\end{tabular}
\end{center}
\caption{Neutral mesons above open flavor threshold excluding isospin partners of charged states.}
\label{nmhc}
\end{table}

\subsection{Comparison with the leading Born--Oppenheimer approximation}\label{lbobraa}

In a recently published paper Braaten, Langmack, and Smith~\cite{Braaten:2014qka,*Braaten:2014ita,*Braaten:2013boa} used the BO approximation to obtain the hybrid masses from the gluonic static energies computed on the lattice. They did not consider the hybrid potential mixing in the Schr\"odinger equation, which leads to the $\Lambda$-doubling effect, cf.\ section~\ref{comodh}. Considering the mixing terms results in the breaking of the degeneracy between the $H_1$ and $H_2$ multiplets as well as the $H_4$ and $H_5$ multiplets. In their approach they account for the breaking of this degeneracy by using different energy offsets for positive and negative parity potentials. These offsets were set in the charmonium sector to reproduce the spin averages of the hybrids from the direct lattice calculations of Ref.~\cite{Liu:2012ze} and in the bottomonium sector to reproduce the mass splittings between the $1^{--}$, $1^{++}$, and $0^{++}$ states from the NRQCD lattice computations of Ref.~\cite{Juge:1999ie}.

\begin{figure}
\centering
\includegraphics[width=0.8\linewidth]{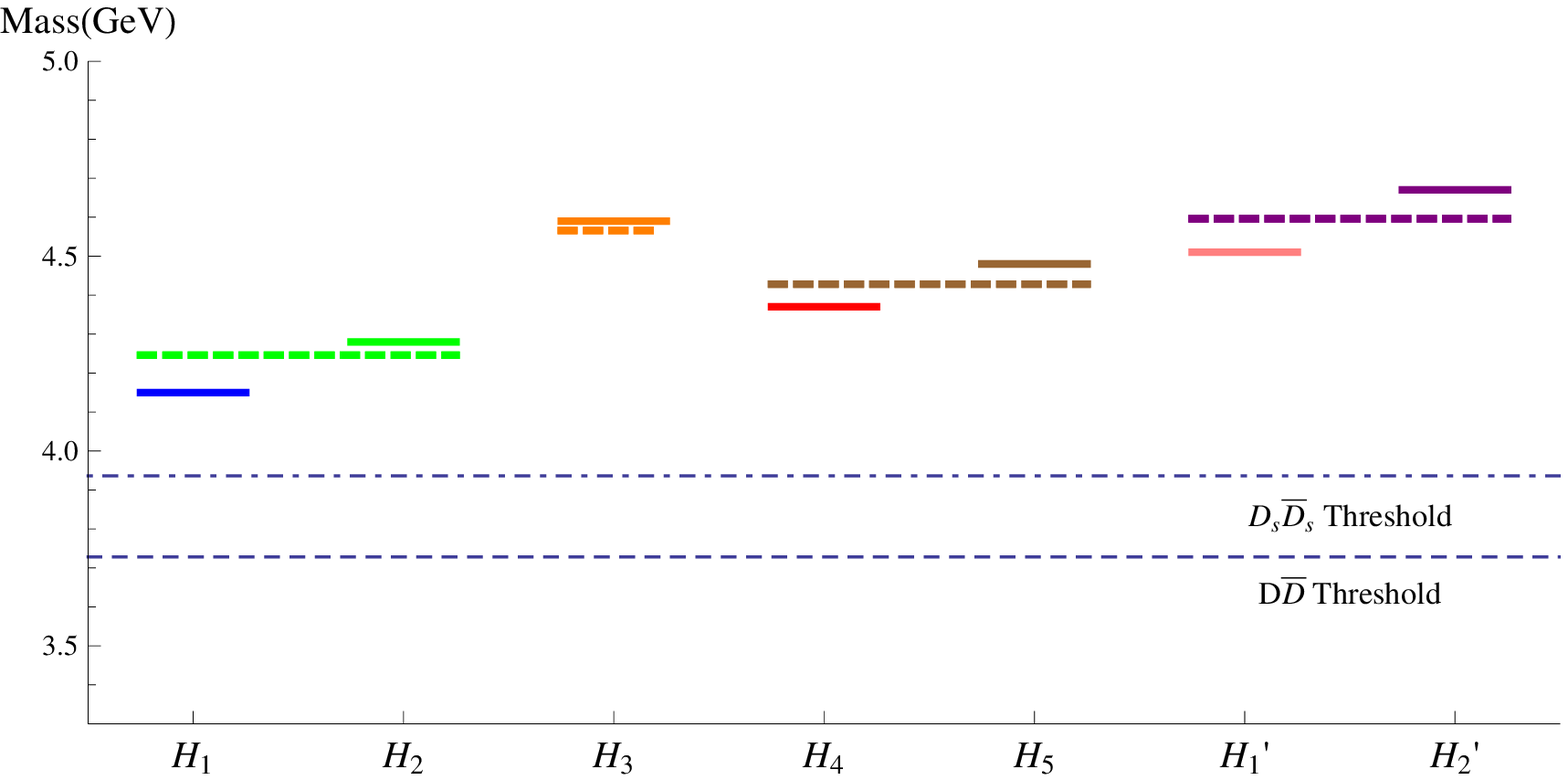}\\
\vspace{20pt}
\includegraphics[width=0.8\linewidth]{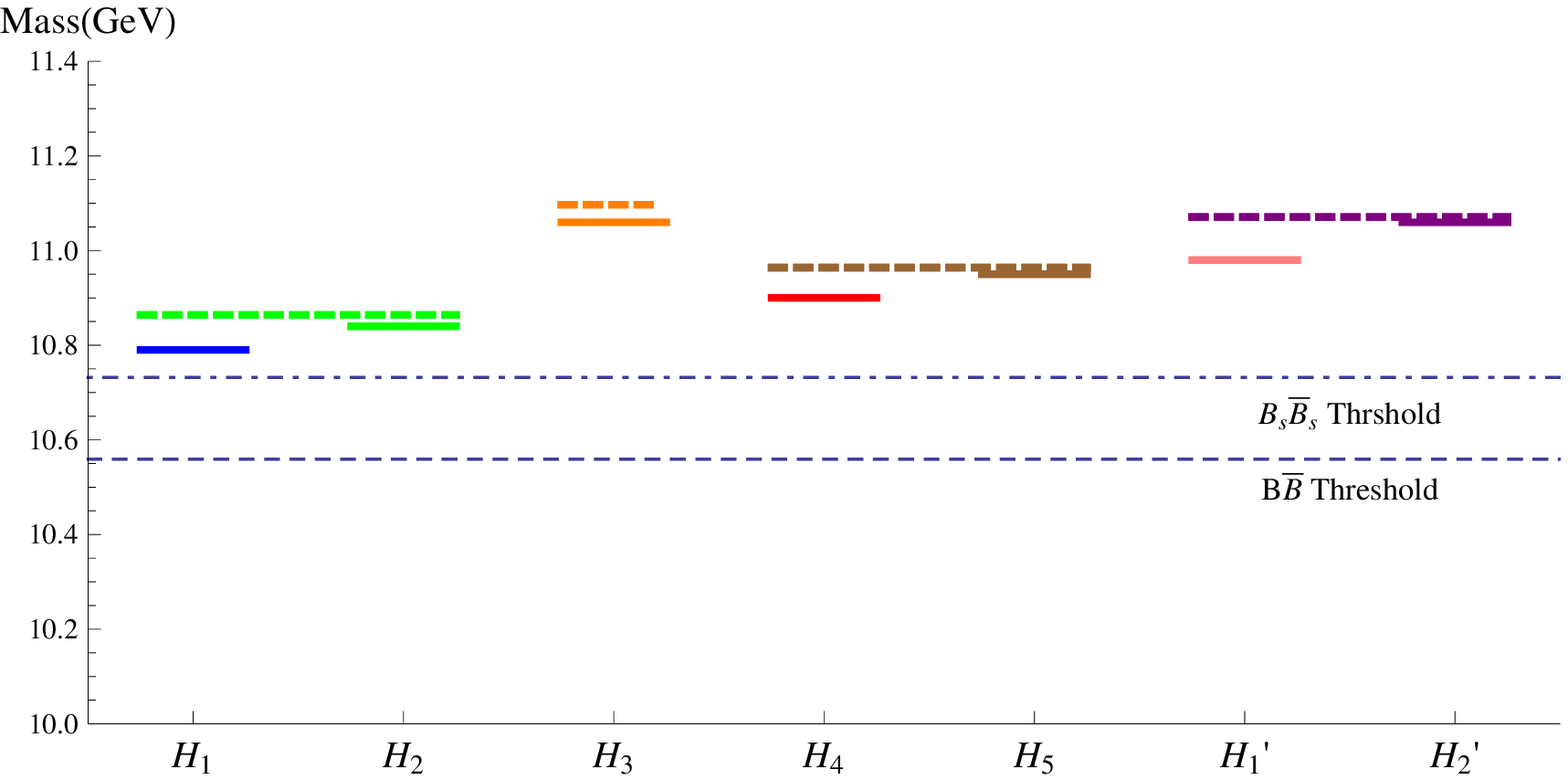}
\caption{Comparison of the hybrid multiplet masses in the charmonium (upper figure) and bottomonium (lower figure) sectors obtained by Braaten \emph{et al.}~\cite{Braaten:2014qka,*Braaten:2014ita,*Braaten:2013boa} (before adjusting to lattice data) with the results obtained using the $V^{(0.25)}$ potential. The Braaten \emph{et al.}\ results correspond to the dashed lines, while the solid lines correspond to the results obtained using $V^{(0.25)}$. The degeneracy of the masses of the $H_{1/2}$ and $H_{4/5}$ multiplets in Braaten \emph{et al.}\ is broken by the introduction of the mixing terms in our approach.}
\label{bccb}
\end{figure}

We have listed the results from~\cite{Braaten:2014qka,*Braaten:2014ita,*Braaten:2013boa} suitable for comparison with our results in Table~\ref{bracom}, and we have plotted them together with our results obtained using the $V^{(0.25)}$ potential in Fig.~\ref{bccb} for both charmonium and bottomonium hybrids. The predicted $H_{1/2}$ mass from Braaten \emph{et al.}\ (before adjusting to lattice data) should be compared with our $H_2$ mass, since this multiplet is a pure $\Pi_u$ potential state. Similarly, their $H_{4/5}$ mass should be compared with our $H_5$ mass. The $H_3$ multiplet is a pure $\Sigma^-_u$ potential state in both approaches and can also be compared. We can see that there is a good agreement with our results from Table~\ref{meth12}. If we shift the masses by the difference in the $H_{1/2}$ state $\sim30$~MeV, then the other states agree within $40$~MeV. The mass shift of $30$~MeV should be accounted for through the uncertainty of the gluelump mass and other systematic errors, so we can take the $40$~MeV discrepancy between our results and those of~\cite{Braaten:2014qka,*Braaten:2014ita,*Braaten:2013boa} to be the uncertainty coming from the fitting of the potentials and the solution of the Schr\"odinger equation. Overall, comparing with the results from~\cite{Braaten:2014qka,*Braaten:2014ita,*Braaten:2013boa}, we can see that the effect of introducing the $\Lambda$-doubling terms lowers the masses of the multiplets that have mixed contributions from the two hybrid static energies.

\begin{table}[t]
 \begin{tabular}{||c|c|c||}
 \hline\hline
             & $cg\bar{c}$ & $bg\bar{b}$ \\ \hline
  $H_{1/2}$  & $4.246$     & $10.864$    \\
  $H_3$      & $4.566$     & $11.097$    \\
  $H_{4/5}$  & $4.428$     & $10.964$    \\
  $H_{1/2}'$ & $4.596$     & $11.071$    \\
  \hline\hline
 \end{tabular}
 \caption{Predicted multiplet masses from~\cite{Braaten:2014qka,*Braaten:2014ita,*Braaten:2013boa} before adjusting to lattice data. The prime on a multiplet stands for the first excited state of that multiplet. All values are given in units of GeV.}
 \label{bracom}
\end{table}

\subsection{Comparison with direct lattice computations} \label{comdlat}

The spectrum of hybrids in the charmonium sector has recently been calculated by the Hadron Spectrum Collaboration~\cite{Liu:2012ze} using unquenched lattice QCD. The calculations were done using an anisotropic lattice with a Shekholeslami-Wohlert fermion action with tree-level tadpole improvements and three-dimensional stout-link smearing of the gauge fields. The calculations were performed on two lattice volumes $16^3\times128$ and $24^3\times128$ with a spatial spacing of $\sim 0.12$~fm. The light quarks were given unphysically heavy masses equivalent to a pion mass of $\approx 400$~MeV.

To interpret their results, the Hadron Spectrum Collaboration organizes the hybrid states into spin-symmetry multiplets. They generate these spin-symmetry multiplets in the constituent gluon picture. The spin-symmetry multiplet resulting from combining a $1^{+-}$ gluonic constituent with an $S$-wave heavy-quark pair generates the $J^{PC}$ quantum numbers corresponding to our $H_1$ multiplet. The $P$-wave heavy-quark pair generates a multiplet with the $J^{PC}$ quantum numbers corresponding to the ones in our $H_2$, $H_3$ and $H_4$ multiplets. Then the lattice results can be assigned to the $S$-wave or $P$-wave multiplets according to their $J^{PC}$ quantum numbers. The Hadron Spectrum Collaboration then argues that the closeness in the masses of the states of each multiplet validates the constituent gluon picture.

\begin{figure}
\centerline{
\includegraphics[width=0.9\linewidth]{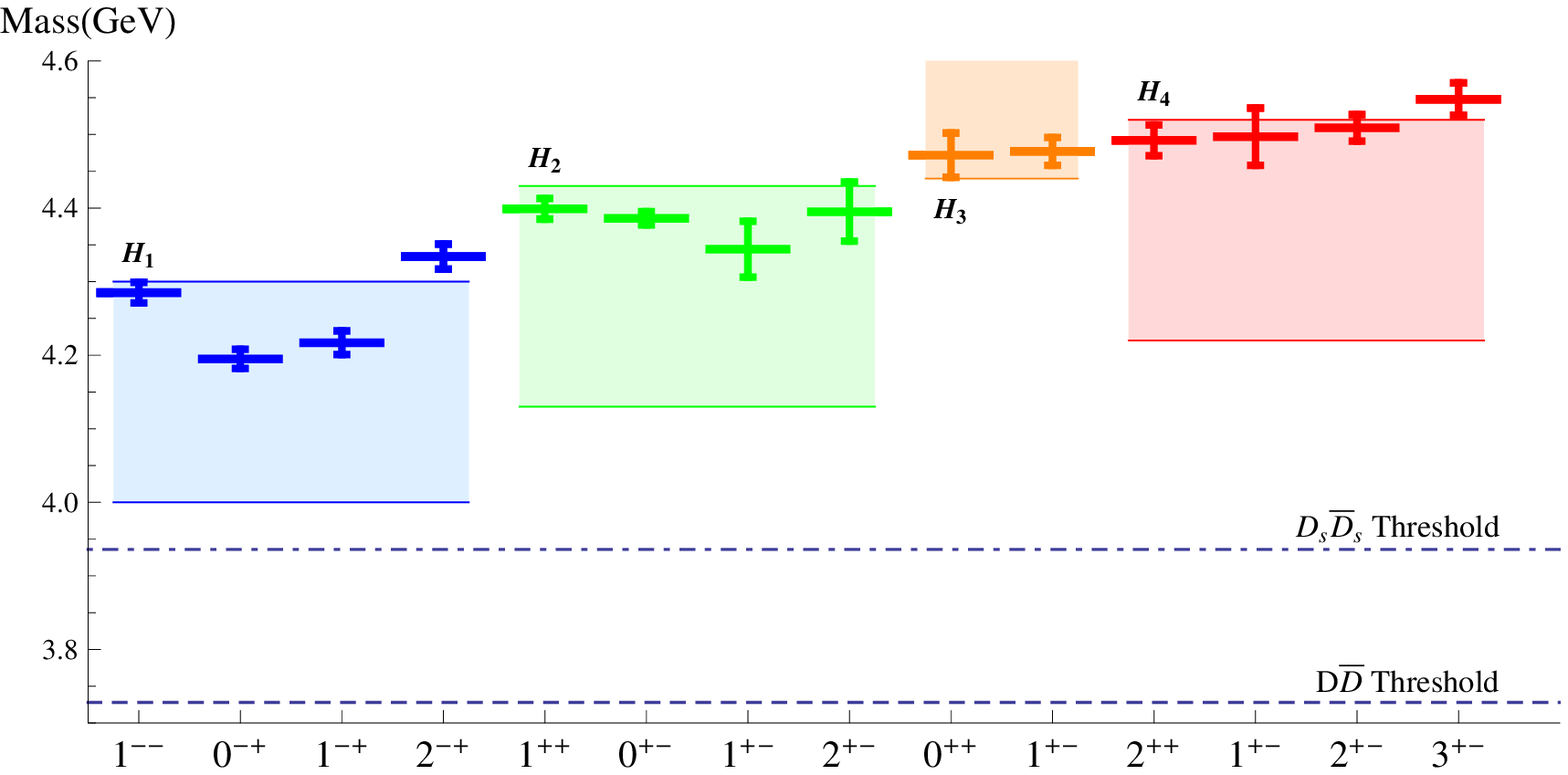}}
\caption{Comparison of the results from direct lattice computations of the masses for charmonium hybrids~\cite{Liu:2012ze} with our results using the $V^{(0.25)}$ potential. The direct lattice mass predictions are plotted in solid lines with error bars corresponding to the mass uncertainties. Our results for the $H_1$, $H_2$, $H_3$, and $H_4$ multiplets have been plotted in error bands corresponding to the gluelump mass uncertainty of $\pm0.15$~GeV.}
\label{charmDLC}
\end{figure}

Similarly, the direct lattice results can be assigned to the pNRQCD (or BO) multiplets of Table~{\ref{multiplets}}, however, this assignment is ambiguous because some $J^{PC}$ quantum numbers appear more than once in the $H_2$, $H_3$, and $H_4$ multiplets. We choose to work with the same assignment as was used in~\cite{Braaten:2014qka,*Braaten:2014ita,*Braaten:2013boa} (see Table~\ref{dudekcom}), which assigns states to a specific multiplet based on the closeness in mass. Looking at Fig.~\ref{charmDLC}, the direct lattice calculation seems to support the result of the pNRQCD and BO approaches that the hybrid states appear in three distinct multiplets ($H_2$, $H_3$, and $H_4$) as compared to the constituent gluon picture, where they are assumed to form one supermultiplet together (cf.\ also the discussion in~\cite{Braaten:2014qka,*Braaten:2014ita,*Braaten:2013boa}).

The results from~\cite{Liu:2012ze} are given with the $\eta_c$ mass subtracted and are not extrapolated to the continuum limit. In Table~\ref{dudekcom} we list their results with the experimental value of $m_{\eta_c}=2.9837(7)$~GeV added. In Fig.~\ref{charmDLC} the results from~\cite{Liu:2012ze} have been plotted together with our results using the $V^{(0.25)}$ potential. We have also computed the spin averaged mass of each multiplet in order to compare with our results from Table~\ref{meth12}.

\begin{table}[t]
\begin{tabular}{||c|c|c|c||} \hline \hline
\text{multiplet}& $J^{PC}$ & $m$ & spin average \\ \hline
$H_1$  & $1^{--}$ & $4.285(14)$ & $4.281(16)$ \\
       & $0^{-+}$ & $4.195(13)$ &             \\
       & $1^{-+}$ & $4.217(16)$ &             \\
       & $2^{-+}$ & $4.334(17)$ &             \\ \hline

$H_2$  & $1^{++}$ & $4.399(14)$ & $4.383(30)$ \\
       & $0^{+-}$ & $4.386(09)$ &             \\
       & $1^{+-}$ & $4.344(38)$ &             \\
       & $2^{+-}$ & $4.395(40)$ &             \\ \hline

$H_3$  & $0^{++}$ & $4.472(30)$ & $4.476(22)$ \\
       & $1^{+-}$ & $4.477(19)$ &             \\ \hline

$H_4$  & $2^{++}$ & $4.492(21)$ & $4.517(23)$ \\
       & $1^{+-}$ & $4.497(39)$ &             \\
       & $2^{+-}$ & $4.509(18)$ &             \\
       & $3^{+-}$ & $4.548(22)$ &             \\ 
\hline\hline
\end{tabular}
\caption{Spectrum of charmonium hybrids calculated by the Hadron Spectrum Collaboration~\cite{Liu:2012ze}. We have added the experimental value $m_{\eta_c}=2.9837(7)$~GeV. All values are given in units of GeV.}
\label{dudekcom}
\end{table}

Comparing the spin averages of the masses of the hybrid states from~\cite{Liu:2012ze} to our results, we see that the masses obtained using the $V^{(0.25)}$ potentials are closer to the direct lattice calculations than the ones obtained using the $V^{(0.5)}$ potentials. For the states obtained from $V^{(0.25)}$ our masses are $0.1-0.14$~GeV lower except for the $H_3$ multiplet, which is $0.11$~GeV higher. It is interesting to note that the $H_3$ multiplet is the only one dominated by the $\Sigma^-_u$ potential. For the states obtained using $V^{(0.5)}$ the differences roughly double.

To further illustrate this comparison, we give the mass splittings between the different multiplets in Table~\ref{mscomp}. Again, we find a better agreement of the lattice data with our calculation with the $V^{(0.25)}$ potentials. In particular, the mass difference between $H_1$ and $H_2$, which in our calculation is directly related to the $\Lambda$-doubling effect, is very close to our mass difference. The worst agreement is again found for the $H_3$ multiplet.

\begin{table}[t]
\begin{tabular}{||c|c|c|c||} \hline \hline
\text{splitting}& Ref.~\cite{Liu:2012ze}   & $V^{(0.5)}$  & $V^{(0.25)}$ \\ \hline
$\delta m_{H_2-H_1}$  & $0.10$ & $0.04$ & $0.13$ \\
$\delta m_{H_4-H_1}$  & $0.24$ & $0.12$ & $0.22$ \\
$\delta m_{H_4-H_2}$  & $0.13$ & $0.08$ & $0.09$ \\
$\delta m_{H_3-H_1}$  & $0.20$ & $0.64$ & $0.44$ \\ 
$\delta m_{H_3-H_2}$  & $0.09$ & $0.60$ & $0.31$ \\ \hline\hline
\end{tabular}
\caption{Mass splittings between $H_1$, $H_2$, $H_3$, and $H_4$ charmonium hybrid multiplets for the potentials $V^{(0.5)}$ and $V^{(0.25)}$ compared with the spin averages from the direct lattice calculation of~\cite{Liu:2012ze}. All values are given in units of GeV.}
\label{mscomp}
\end{table}

In the bottomonium sector direct lattice calculations have been carried out by Juge, Kuti, and Morningstar~\cite{Juge:1999ie} and by Liao and Manke~\cite{Liao:2001yh}. Juge, Kuti, and Morningstar did quenched simulations using anisotropic lattices with improved gauge-field actions for the gluons. The heavy quarks were treated in NRQCD for anisotropic lattices containing just a covariant temporal derivative term. Since the hybrid masses were expected to be large, anisotropic lattices with the temporal lattice spacing much smaller than the spatial spacing were used to reduce the statistical fluctuations. Two lattice volumes were used, $15^3\times45$ with $\beta=3.0$ and $10^3\times30$ with $\beta=2.6$.

They studied the correlation functions of five operators on the lattice, three of them corresponding to hybrid operators. They identified three hybrid states corresponding to the ground states of the $H_1$, $H_2$, and $H_3$ multiplets and one excited state of the $H'_1$ multiplet. Since no spin (or any relativistic) effects were included, the results given by Juge, Kuti, and Morningstar are the masses of the degenerate multiplets, which correspond to the ones in Table~\ref{multiplets}.

In Ref.~\cite{Juge:1999ie} the values of the multiplet mass splitting are given in units of $r_0$ relative to the mass of the $1S$ bottomonium states. We have used the most up-to-date value for $r_0=0.486\pm0.004$~fm from~\cite{Bazavov:2011nk}. Using this value as well as the spin average of the $1S$ bottomonium mass states we have computed the values for the multiplet masses from their largest lattice volume in Table~\ref{bbdirelat}.

Liao and Manke~\cite{Liao:2001yh} calculated the bottomonium spectrum using quenched lattice QCD on an anisotropic lattice. They were able to go beyond the nonrelativistic approximation by using a very fine discretization in the temporal spacing, which also allowed them to extrapolate the results for the hyperfine splitting of the standard bottomonium to the continuum. They used a standard Wilson action for the gluons with various link smearing, while for the heavy quarks in the gluonic background they used an anisotropic clover action. They explored five different lattice spacings from $0.04$~fm to $0.17$~fm and two anisotropy ratios.

They determined the masses for three $b\bar{b}$ mesons with explicit exotic quantum numbers. The results for the level splittings are presented in an analogous way to the Juge, Kuti, and Morningstar paper, and we have used the same spin independent masses for the $1S$ and $1P$ bottomonium states in order to generate the values displayed in Table~\ref{bbdirelat}.

\begin{table}[t]
\begin{tabular}{||c|c||c|c||} \hline \hline
multiplet & Ref.~\cite{Juge:1999ie} & $J^{PC}$(multiplet) & Ref.~\cite{Liao:2001yh} \\ \hline
$H_1$            & $10.830(30)$            & $1^{-+}(H_1)$              & $11.39(15)$             \\
$H_2$            & $10.865(54)$            & $0^{+-}(H_2)$              & $10.99(33)$             \\
$H_3$            & $11.138(28)$            & $2^{+-}(H_2)$              & $12.16(14)$             \\
$H_1'$           & $11.216(37)$            &                            &                         \\ \hline\hline
\end{tabular}
\caption{Masses of the bottomonium hybrids from direct lattice calculations. We present the results of the runs with size $10^3\times30$, $\beta=3.0$, and spatial lattice spacing $a\approx1.13$~fm of Ref.~\cite{Juge:1999ie} and with size $16^3\times128$, $\beta=6.3$, and $a\approx 0.0521$~fm from Ref.~\cite{Liao:2001yh}. All values are given in units of GeV.}
\label{bbdirelat}
\end{table}

\begin{table}[t]
\begin{tabular}{||c|c|c|c||} \hline \hline
splitting             & Ref.~\cite{Juge:1999ie} & $V^{(0.5)}$  & $V^{(0.25)}$ \\ \hline
$\delta m_{H_2-H_1}$  & $0.04$                  & $0.02$       & $0.05$       \\
$\delta m_{H_3-H_1}$  & $0.31$                  & $0.36$       & $0.27$       \\
$\delta m_{H_3-H_2}$  & $0.27$                  & $0.34$       & $0.22$       \\
$\delta m_{H_1'-H_1}$ & $0.39$                  & $0.10$       & $0.19$       \\ \hline\hline
\end{tabular}
\caption{Mass splittings between the $H_1$, $H_2$, $H_3$, and $H_1'$ bottomonium hybrid multiplets for the potentials $V^{(0.5)}$ and $V^{(0.25)}$ compared with the values from Ref.~\cite{Juge:1999ie}. All values are given in units of GeV.}
\label{mscomp2}
\end{table}

\begin{figure}
\centerline{
\includegraphics[width=0.9\linewidth]{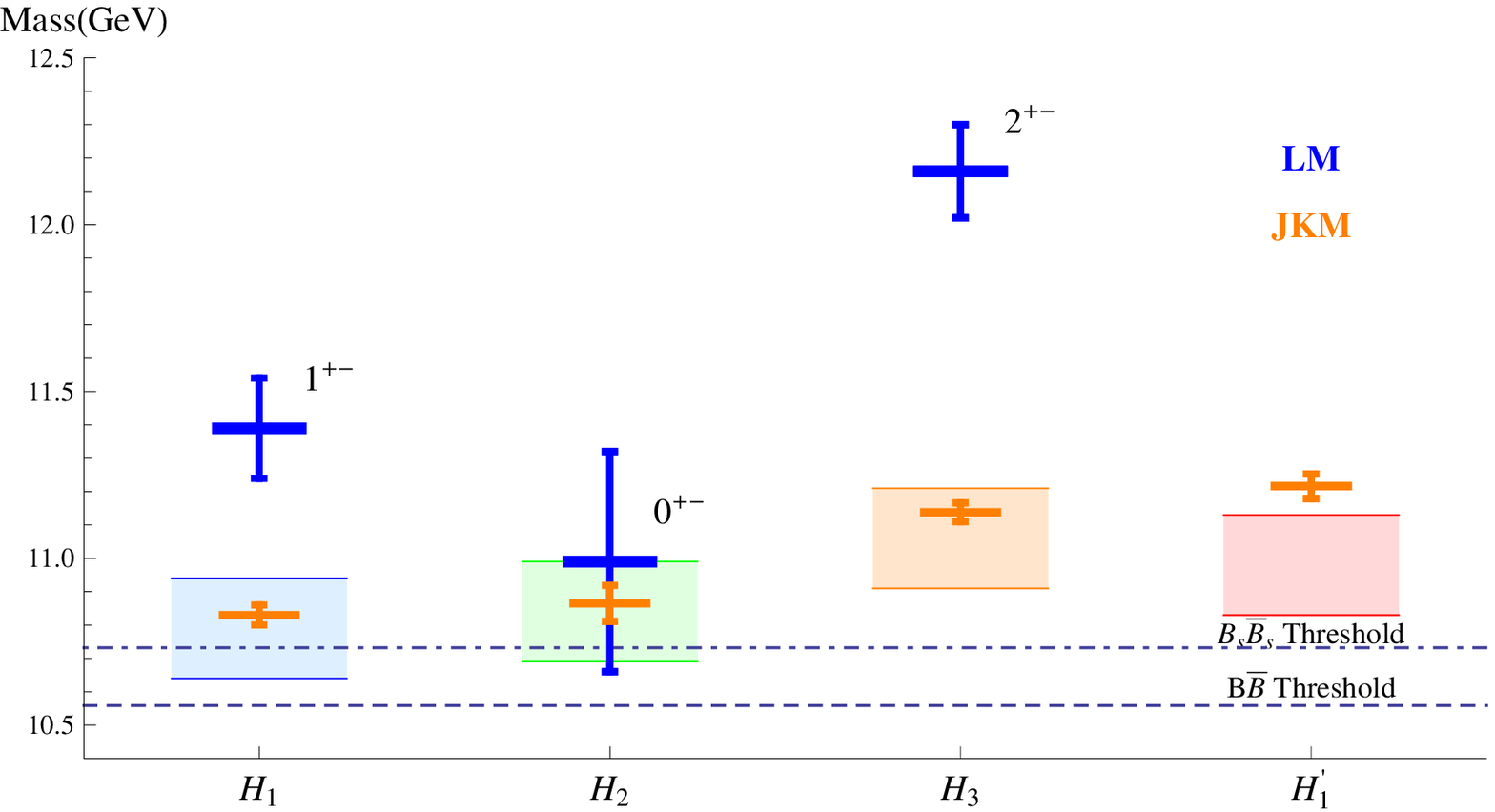}}
\caption{Comparison of the results from direct lattice computations of the masses for bottomonium hybrids from Juge, Kuti, and Morningstar (JKM)~\cite{Juge:1999ie} and Liao and Manke (LM)~\cite{Liao:2001yh} with our results using the $V^{(0.25)}$ potential. The direct lattice mass predictions are plotted in solid lines with error bars corresponding the mass uncertainties. Orange lines correspond to the results of JKM and blue lines to the ones of LM. The $J^{PC}$ quantum numbers in the figure correspond to the LM states. Our results for the $H_1$, $H_2$, $H_3$, and $H_1'$ multiplets have been plotted in error bands corresponding to the gluelump mass uncertainty of $\pm0.15$~GeV.}
\label{botomDLC}
\end{figure}

We have plotted the results from Juge, Kuti, and Morningstar and the ones from Liao and Manke together with our predictions for the masses of the bottomonium hybrid multiplets in Fig.~\ref{botomDLC}. If we compare our results from Table~\ref{meth12} with the values from direct lattice calculations from Table~\ref{bbdirelat}, we observe that our results are systematically lower by $0.05-0.15$~GeV except for the excited $H_1'$ state, for which the deviation is larger: $0.4$~GeV and $0.26$~GeV for the potentials $V^{(0.5)}$ and $V^{(0.25)}$, respectively. To eliminate possible systematic uncertainties we can look at the level splitting displayed in Table~\ref{mscomp2}. The values of the level splitting show considerable agreement, improving from using the $V^{(0.5)}$ potentials to using the $V^{(0.25)}$ potentials, with the only exception of the $H_1'$ state. In particular, the $\Lambda$-doubling effects seen in the mass splitting between $H_2$ and $H_1$ agree quite well with lattice predictions.

In general, the comparison of our results with direct lattice computations of hybrid masses shows a systematic energy offset but a reasonable agreement for the mass splittings between multiplets, particularly for the lower mass ones. The bottomonium sector results show more consistency with direct lattice computations than the charmonium sector, as expected. 

\subsection{Comparison with QCD sum rules} \label{comsumrules}

The method of QCD sum rules consists of a treatment in which hadrons are represented by their interpolating quark currents, taken at large virtualities, instead of in terms of constituent quarks. The correlation function of these currents is treated in the context of the operator product expansion, where the short and long distance physics are separated. The former is calculated using perturbation theory, whereas the latter is parametrized in terms of universal vacuum condensates or light-cone distribution amplitudes. The result of the calculation can then be related via dispersion relations to a sum over hadronic states.

A recent analysis of QCD sum rules for hybrid operators has been performed by Chen \emph{et al.}\ for $b\bar{b}$ and $c\bar{c}$ hybrids in~\cite{Chen:2013zia} and for $b\bar{c}$ hybrids in~\cite{Chen:2013eha}. Using hybrid operators and computing correlation functions and spectral functions up to dimension six condensates, they stabilized the sum rules and gave mass predictions for the heavy quark hybrids. 

\begin{figure}
\centerline{
\includegraphics[width=0.9\linewidth]{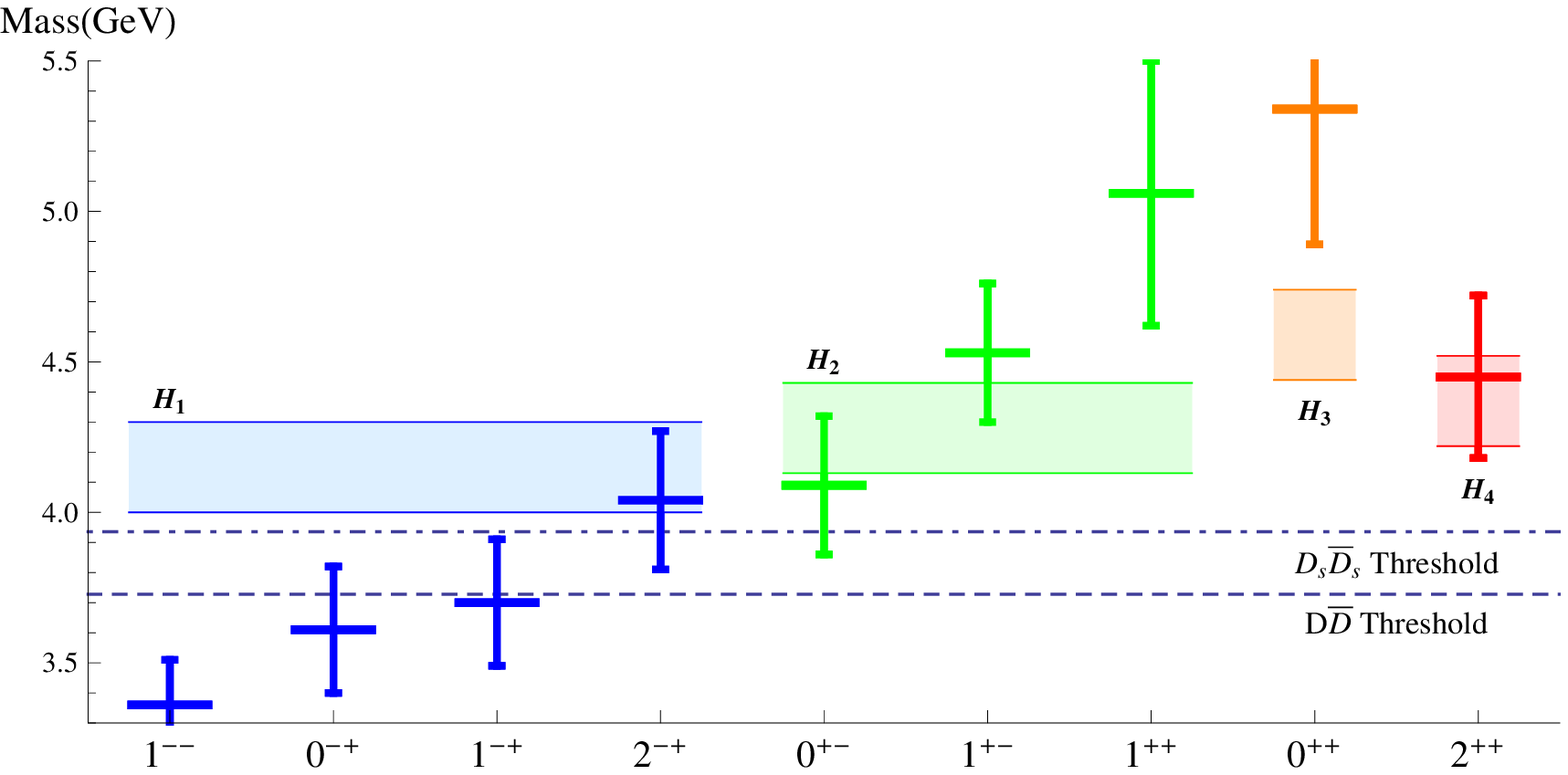}}
\vspace{0.5cm}
\centerline{
\includegraphics[width=0.9\linewidth]{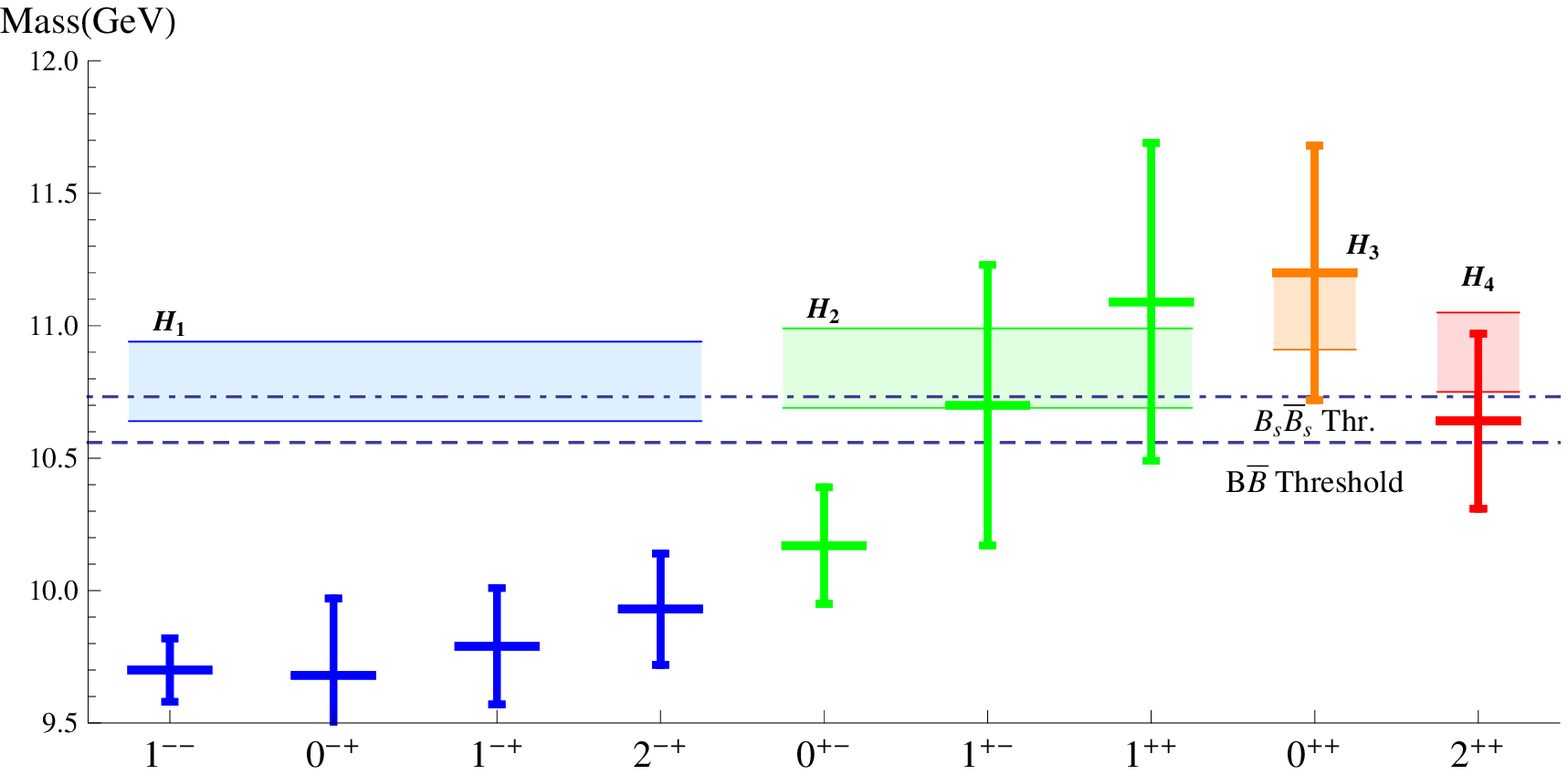}}
\caption{Comparison of the mass predictions for charmonium hybrids in the upper figure and for bottomonium hybrids in the lower figure, obtained using QCD sum rules~\cite{Chen:2013zia}, with our results using the $V^{(0.25)}$ potential. The solid lines correspond to the QCD sum rules masses with error bars corresponding to their uncertainties. Our results for the $H_1$, $H_2$, $H_3$, and $H_4$ multiplets have been plotted in error bands corresponding to the gluelump mass uncertainty of $\pm0.15$~GeV.}
\label{sumrulef}
\end{figure}

The pattern of hybrid states encountered by Chen \emph{et al.}\ in~\cite{Chen:2013zia}, which we show in Table~\ref{sumrult} and, plotted against our results using the $V^{(0.25)}$ potential, in Fig.~\ref{sumrulef}, is the same for $c\bar{c}$ and $b\bar{b}$ hybrid states. The lightest set of states they found corresponds to our $H_1$ multiplet. The next set of states consists of $0^{+-}$, $1^{+-}$, and $1^{++}$, which belong to the $H_2$ multiplet, $2^{++}$ and $0^{++}$, which are part of the multiplets $H_3$ and $H_4$, respectively, and $0^{--}$, which does not appear in any of the multiplets we have considered.

For charmonium the masses of the $H_1$ multiplet are between $3.36$~GeV and $4.04$~GeV with a spin average of $3.75(20)$~GeV, which is lower than our result for the $H_1$ multiplet (see Table~\ref{meth12}). The elements of $H_2$ show an important dispersion, but overall tend to be larger than our value for the mass of the $H_2$ multiplet, like in the case of the $2^{++}$ and $0^{++}$ masses when compared with our results for $H_3$ and $H_4$. A similar pattern emerges for $b\bar{b}$ hybrids. The $H_1$ multiplet ranges between $9.7$~GeV and $9.93$~GeV with a spin average of $9.81(19)$~GeV, which is about $1$~GeV below our estimates. Nevertheless, the $1^{+-}$, $1^{++}$, $2^{++}$, and $0^{++}$ states are within errors of our results.

\begin{table}[t]
\begin{tabular}{||c|ccc||cc||} \hline \hline
\text{multiplet} & $J^{PC}$   & $c\bar{c}$      & $b\bar{b}$      & $J^{P}$ & $b\bar{c}$  \\ \hline
$H_1$            & $1^{--}$   & $3.36(15)$      & $9.70(12)$      & $1^-$ & $6.83(16)$  \\ 
                 & $0^{-+}$   & $3.61(21)$      & $9.68(29)$      & $0^-$ & $6.90(22)$  \\ 
                 & $1^{-+}$   & $3.70(21)$      & $9.79(22)$      & $1^-$ & $6.95(22)$  \\ 
                 & $2^{-+}$   & $4.04(23)$      & $9.93(21)$      & $2^-$ & $7.15(23)$  \\ \hline
$H_2$            & $0^{+-}$   & $4.09(23)$      & $10.17(22)$     & $0^+$ & $7.37(31)$  \\ 
                 & $1^{+-}$   & $4.53(23)$      & $10.70(53)$     & $1^+$ & $7.77(24)$  \\ 
                 & $1^{++}$   & $5.06(44)$      & $11.09(60)$     & $1^+$ & $8.28(37)$  \\ \hline
$H_4$            & $2^{++}$   & $4.45(27)$      & $10.64(33)$     & $2^+$ & $7.67(18)$  \\ \hline
$H_3$            & $0^{++}$   & $5.34(45)$      & $11.20(48)$     & $0^+$ & $8.55(44)$  \\ \hline
                 & $0^{--}$   & $5.51(50)$      & $11.48(75)$     & $0^-$ & $8.48(67)$  \\ \hline \hline
\end{tabular}
\caption{Left panel: masses of the $c\bar{c}$ and $b\bar{b}$ hybrids obtained using QCD sum rules from~\cite{Chen:2013zia}. Right panel: masses of $b\bar{c}$ hybrids from~\cite{Chen:2013eha}. All values are given in units of GeV.}
\label{sumrult}
\end{table}

\begin{figure}
\centerline{
\includegraphics[width=0.9\linewidth]{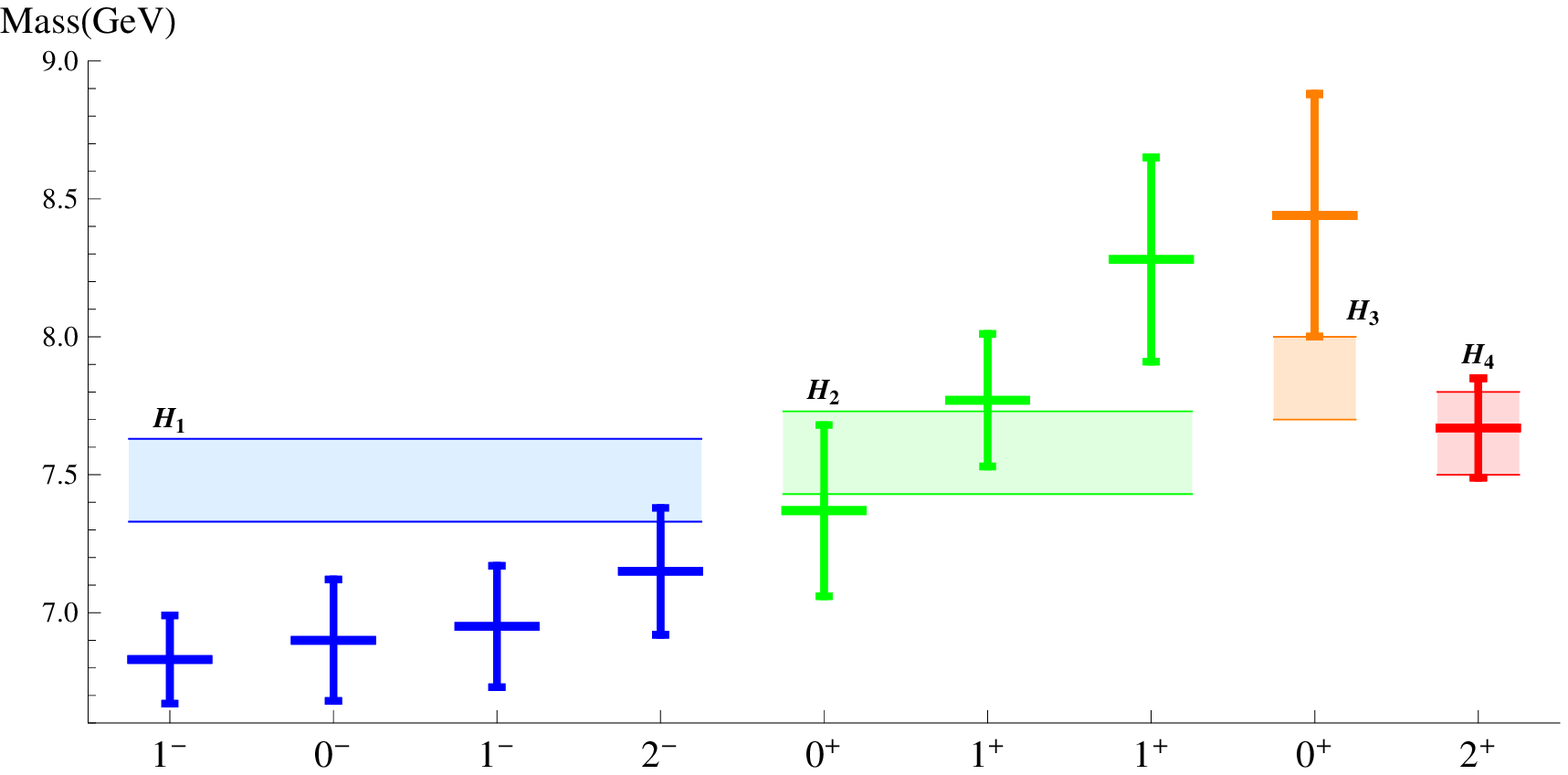}}
\caption{Comparison of the mass predictions for $b\bar{c}$ hybrids, obtained using QCD sum rules~\cite{Chen:2013eha}, with our results using the $V^{(0.25)}$ potential. The solid lines correspond to the QCD sum rules masses with error bars corresponding to their uncertainties. Our results for the $H_1$, $H_2$, $H_3$, and $H_4$ multiplets have been plotted in error bands corresponding to the gluelump mass uncertainty of $\pm0.15$~GeV.}
\label{bcsumrulef}
\end{figure}

The $b\bar{c}$ hybrids have also been studied with QCD sum rules by Chen \emph{et al.}\ in~\cite{Chen:2013eha}. In this case, since the heavy quark and antiquark are not the same, the interpolating currents that couple to the hybrids have no definite $C$-parity. The assignment of the $b\bar{c}$ states to each multiplet has been done by analogy of the interpolating currents that generate these states in $Q\bar{Q}$ and $b\bar{c}$. In Fig.~\ref{bcsumrulef} the results from Chen \emph{et al.}\ for $b\bar{c}$ hybrids are plotted alongside our results using the $V^{(0.25)}$ potential. The spin average for the $b\bar{c}$ $H_1$ multiplet is $7.00(16)$~GeV, which falls about $0.5$~GeV below our result.

\section{Conclusions} \label{conc}
In this paper we have constructed a nonrelativistic effective field theory description of heavy quarkonium hybrids. We started from QCD, excluding light quark degrees of freedom from our direct consideration, and aiming at describing exotic states at or above the strong decay threshold. Under these specifications we can restrict ourselves to a Fock space comprising heavy quarkonium, heavy quarkonium hybrids, and glueballs. We identify the symmetries of the system of a heavy quark, a heavy antiquark, and glue in the static limit. Corrections to this limit can be obtained order by order in the $1/m$ expansion as it is usually done in nonrelativistic effective field theories of QCD. At order $1/m$ in the expansion we obtain at the level of pNRQCD a system of coupled Schr\"odinger equations that describe the hybrid spin-symmetry multiplets constructed with the $\Sigma_u^-$ and $\Pi_u$ gluonic static energies. It is assumed that higher gluonic static energies do not mix with them. They would generate higher mass hybrid multiplets. The matching from NRQCD to pNRQCD allows us to identify the static interaction potentials entering the Schr\"odinger equation. In the short distance, the static potentials depend on two nonperturbative parameters. These are the gluelump mass and the quadratic slope. Both can be determined from lattice calculations. We adopt a renormalon subtraction scheme for the calculation of the perturbative part of the potential.

The Schr\"odinger equations couple the gluon to the heavy quark dynamics through the action of the angular part of the kinetic operator of the heavy quarks on the gluonic static states. The relevant matrix element could be computed on the lattice, but is at present unknown. We estimated it in the short-distance using a version of pNRQCD for which the multipole expansion holds. The matrix element generates terms that mix the contributions of different static energies into the hybrid states, an effect known as $\Lambda$-doubling in molecular physics. We have solved numerically the coupled Schr\"odinger equations for the heavy quarks and have obtained the masses for a large set of spin-symmetry multiplets for $c\bar{c}$, $b\bar{c}$, and $b\bar{b}$ hybrids. The $\Lambda$-doubling effect breaks the degeneracy between opposite parity spin-symmetry multiplets and has been found to lower the mass of the multiplets that get mixed contributions of different static energies.

We have compared our results with direct lattice computations in the charmonium and bottomonium sectors. We observe the same $\Lambda$-doubling pattern in direct lattice calculations, namely, the multiplets which receive mixed contributions from the $\Sigma_u^-$ and $\Pi_u$ have a lower mass than their parity partners that remain pure $\Pi_u$ states. On average, the direct lattice computations of hybrid masses lie above our values but within our uncertainty, which is dominated by the uncertainty of the gluelump mass of $\pm 0.15$~GeV. The mass shift remains fairly constant among the different multiplets, which could be an indication that it is due to systematic effects. Comparing the mass splits between multiplets, we obtain a good agreement in most of the cases. We have also compared our results with recent results from QCD sum rules. Sum rules predictions carry large uncertainties, particularly when compared to the direct lattice calculations. However, the same $\Lambda$-doubling pattern is also realized there, but the values of the multiplet masses have a large dispersion compared to our results. Up to now, works done in the BO approximation have not included the $\Lambda$-doubling terms.

To our knowledge this is the first attempt to develop from QCD a nonrelativistic EFT description of heavy quarkonium hybrids. Still there are some obvious limitations to what we have done so far. We have computed the relevant matrix element using information from weakly-coupled pNRQCD and taking advantage of the multipole expansion. This has allowed us to write and solve the coupled Schr\"odinger equations in a consistent setup. By fitting the nonperturbative parameters of the potential to the lattice static energy, we believe we have pushed our description to a sufficiently large value of the quark-antiquark distance to produce a realistic pattern of hybrid mass multiplets. This is confirmed also by the overall agreement that we see with the direct lattice calculations of the masses. The biggest uncertainty in our level predictions comes from the error of the lattice determination of the gluelump mass, which could be improved by new lattice calculations.

The next step will be to introduce in our framework spin contributions that will break the spin degeneracy and give a more detailed structure to the hybrid
multiplets. The long term goal is to introduce an EFT description of heavy quarkonium hybrids without using the multipole expansion. This would entail the
definition of the appropriate generalized Wilson loops that encode the dynamics of the nonperturbative matrix elements and obtaining in strongly-coupled pNRQCD the dynamical equations that couple them.

Neutral exotic quarkonia above open flavor thresholds are possible experimental candidates for quarkonium hybrids. Most of these candidates are $1^{--}$ states, due to these quantum numbers being the most easily accessible experimentally in electron-positron colliders. In Fig.~\ref{exp} we have overlaid the experimental candidates to our hybrid multiplet mass predictions. Most of these candidates decay into spin triplet quarkonium states, but their tentative hybrid identifications correspond to spin singlet states, which would mean that these decays violate the heavy quark spin symmetry. The most promising candidate, $Y(4220)$, is the only $1^{--}$ state that decays in a spin singlet quarkonium, however, this state is yet not well established. When comparing to the data, besides the prediction of the hybrid mass multiplets, it is hence important to develop ways to calculate transition and decay widths. Future EFT studies of heavy quarkonium hybrids will have to address the calculation of these quantities.
 
\acknowledgments

We thank Colin Morningstar and Gunnar Bali for giving us access to their lattice data for the static energies and Gunnar Bali and Antonio Pineda for giving us permission to use Fig.~\ref{hybstat}. N.B.\ acknowledges discussion with Estia Eichten. This work has been supported by the DFG and the NSFC through funds provided to the Sino-German CRC 110 ``Symmetries and the Emergence of Structure in QCD'', and by the DFG cluster of excellence ``Origin and structure of the universe'' (www.universe-cluster.de). 

\appendix
\section{Symmetries of the static system} \label{sss}

A system of two static opposite color sources (in our case the system formed by a heavy quark in position $\bm{x}_1$ and a heavy antiquark in position 
$\bm{x}_2$) remains invariant under the following symmetry transformations: rotations $R(\alpha)$ by an angle $\alpha\in]-\pi,\pi]$ around the axis defined by the two sources, space inversion $P$ in combination with charge conjugation $C$, reflections $M$ across a plane containing the two sources, and combinations thereof. These transformations form the group $D_{\infty h}$, which is the symmetry group of a cylinder. 

Since the static Hamiltonian is invariant under these transformations, we can use the quantum numbers of the representations of $D_{\infty h}$ to label its eigenstates. The conventional notation for the representations of $D_{\infty h}$ is $\Lambda_{\eta}^\sigma$. $\Lambda$ is the rotational quantum number, it can take non-negative integer values $0,1,2,3,\dots$, which are traditionally represented by capital Greek letters $\Sigma,\Pi,\Delta,\Phi,\dots$ corresponding to the atomic orbitals $s,p,d,f,\dots$, respectively. The eigenvalue of $CP$ is given as the index $\eta$. It can take the values $+1$ or $-1$, for which the labels $g$ (\textit{gerade}, i.e., even) and $u$ (\textit{ungerade}, i.e., odd) are used. The other index $\sigma$ gives the sign under reflections as $+$ or $-$, however, it is only written explicitly for the $\Sigma$-states, because for $\Lambda\geq1$ the states with opposite $\sigma$ are degenerate with respect to the static energy.

Physically, this can be understood in the following way. The static system itself has no preferred orientation for the plane across which the reflections are defined. In fact, through a combination of rotation and reflection operations one can define a new reflection operation $M'=R(-\alpha)MR(\alpha)$, where the reflection plane is rotated by an angle $\alpha$. The $\Sigma$-states are rotationally invariant, so $M$ and $M'$ give the same eigenvalue, but for $\Lambda\geq1$ they do not. If in the simplest case $\alpha$ is chosen to be $\pi/2$, then $M$ and $M'$ have opposite eigenvalues. However, the static Hamiltonian $H^{(0)}$ does not depend on the choice of $M$ or $M'$, so consequently its eigenvalues, the static energies, cannot depend on $\sigma$ unless $\Lambda=0$.

Mathematically, this can be explained by looking at the irreducible representations of $D_{\infty h}$. We can write $D_{\infty h}=O(2)\otimes Z_2$, where $Z_2$ corresponds to the sign $\eta$ under $CP$ transformations. There are two different one-dimensional irreducible representations of $O(2)$ and countably infinite two-dimensional ones. The two one-dimensional representations both map the rotations to unity and differ by the sign under reflections. These correspond to $\Lambda=0$ and positive or negative $\sigma$.

The two-dimensional representations are given by
\begin{equation}
 R(\alpha)=\begin{pmatrix} \cos\Lambda\alpha & \sin\Lambda\alpha \\ -\sin\Lambda\alpha & \cos\Lambda\alpha \end{pmatrix}\,,\hspace{20pt}M=\begin{pmatrix} 1 & 0 \\ 0 & -1 \end{pmatrix}\,.
\end{equation}
The basis for these representations was chosen such that $M$ is diagonal. It is possible to make a basis transformation that takes $M\to-M$ while $R(\alpha)$ remains the same. This means that the sign under reflections is irrelevant for the two-dimensional representations and $\sigma$ cannot label different representations. Since the static energies depend only on the representation, they must be independent of $\sigma$ for $\Lambda\geq1$. 

It is also possible to make a basis transformation that takes $R(\alpha)\to R(-\alpha)$ while $M$ remains the same. This means that negative values for $\Lambda$ do not correspond to a different representation but just to a different choice of basis, so by convention $\Lambda$ is defined to be non-negative. $\Lambda$ can only take integer values, because $R(2\pi)$ is required to be unity. Note that for $\Lambda=0$ the two-dimensional representation is diagonal and reduces to the two one-dimensional representations.

In the context of the spectrum of the static Hamiltonian, the two-dimensionality of the irreducible representations of $D_{\infty h}$ means that any eigenstate of $H^{(0)}$ with $\Lambda\geq1$ consists of two components, which correspond to $\sigma=\pm1$ in the basis given above. For the calculations in this paper it is advantageous to choose a different basis such that $R(\alpha)$ is diagonal:
\begin{equation}
 R(\alpha)=\begin{pmatrix} e^{i\Lambda\alpha} & 0 \\ 0 & e^{-i\Lambda\alpha} \end{pmatrix}\,,\hspace{20pt}M=\begin{pmatrix} 0 & \sigma_M^* \\ \sigma_M & 0 \end{pmatrix}\,.
\end{equation}
There are many ways in which one can make such a basis transformation, and this manifests itself in the phase $\sigma_M$ appearing in $M$, which is completely arbitrary. In this basis we can label the two components by $\lambda=\pm\Lambda$ such that they transform with $e^{i\lambda\alpha}$ under rotations. Because $M$ now is offdiagonal, irrespective of the choice of $\sigma_M$, the two components are exchanged under reflections, i.e., $\lambda\stackrel{M}{\to}-\lambda$.

The advantage of this choice of basis is that, if we introduce the angular momentum operator $\bm{K}$ of the light degrees of freedom, then $\lambda$ is the eigenvalue of $\bm{\hat{r}}\cdot\bm{K}$, where $\bm{\hat{r}}$ is the orientation of the quark-antiquark axis. $\Lambda$ is then given by the absolute value of $\bm{\hat{r}}\cdot\bm{K}$, which is also true for $\Lambda=0$. The operator $\bm{K}^2$ represents the fully three-dimensional rotations, i.e., the group $SO(3)$, so the static states are not eigenstates of $\bm{K}^2$ except for the limit of vanishing quark-antiquark distance, where this symmetry is restored.

\section{RS scheme}\label{rescheme}

The RS octet potential is defined as follows~\cite{Pineda:2001zq,Bali:2003jq}
\begin{equation}
V^{RS}_o(\nu_f)=V_o-\delta V^{RS}_o(\nu_f)\,,
\end{equation}
with
\begin{align}
V_o(r,\nu)&=\left(\frac{C_A}{2}-C_F\right)\frac{\alpha_{V_o}(\nu)}{r}\,, \\
\delta V^{RS}_o(\nu_f) &= \sum^{\infty}_{n=1}N_{V_o}\nu_f \left(\frac{\beta_0}{2\pi}\right)^n\alpha^{n+1}_s\left(\nu_f\right)\sum^{\infty}_{k=0}c_k\frac{\Gamma\left(n+1+b-k\right)}{\Gamma\left(1+b-k\right)}\,.
\end{align}
The value of $N_{V_o}=0.114001$ was computed in Ref.~\cite{Bali:2003jq}. The value of $\alpha_{V_o}$ up to order $\alpha^3_s$ is given by~\cite{Kniehl:2004rk}
\begin{equation}
\alpha_{V_o}\left(\nu\right)=\alpha_{V_s}\left(\nu\right)-\left(\frac{3}{4}-\frac{\pi^2}{16}\right)C^2_A \alpha^3_s\left(\nu\right)+\mathcal{O}\left(\alpha_\mathrm{s}^4\right)\,, 
\end{equation}
where $\alpha_{V_s}$ is
\begin{align}
\alpha_{V_s}\left(\nu\right)=\alpha_\mathrm{s}\left(\nu\right)&\left(1+(a_1+2\gamma_E\beta_0)\frac{\alpha_\mathrm{s}\left(\nu\right)}{4\pi}\right.\notag\\
&\left.+\left[\gamma_E(4a_1+\beta_0+2\beta_1)+\left(\frac{\pi^2}{3}+4\gamma^2_E\right)\beta^2_0+a_2\right]\frac{\alpha^2_s\left(\nu\right)}{16\pi^2}\right)\,.
\end{align}
The parameters $b$ and the first three $c_k$ appearing in $\delta V^{RS}_o$ are given by
\begin{align}
b&=\frac{\beta_1}{2\beta^2_0}\,,\hspace{60pt}c_0=1\,,\hspace{60pt}c_1=\frac{1}{4b\beta^3_0}\left(\frac{\beta^2_1}{\beta_0}-\beta_2\right)\,,\notag\\
c_2&=\frac{1}{32b(b-1)\beta^8_0}\left(\beta^4_1+4\beta^3_0\beta_1\beta_2-2\beta_0\beta^2_1\beta_2+\beta^2_0(\beta^2_2-2\beta^3_1)-2\beta^4_0\beta_3\right)\,.
\end{align}

\section{Detailed derivation of the radial Schr\"odinger equation} \label{RSEQ}

The Laplace operator $\bm{\nabla}_r^2$ can be split into a radial and an angular part, such that
\begin{equation}
 -\frac{\bm{\nabla}_r^2}{m}=-\frac{1}{m\,r^2}\left(\partial_r\,r^2\,\partial_r+\partial_x\left(1-x^2\right)\partial_x+\frac{1}{1-x^2}\partial^2_\varphi\right)\,,
\end{equation}
with the variable $x=\cos\theta$. The angular part of this acts on both the wave function and the projection vector in~\eqref{Diffeq}, and since we know $\bm{\hat{n}}$ explicitly for the $1^{+-}$ gluelump, we can work out the action of the angular part of $\bm{\hat{n}}'\cdot(-\bm{\nabla}_r^2/m)\,\bm{\hat{n}}$ in the form of a matrix acting on the three-component wave function $\Psi^{(N)}$. Then we get
\begin{equation}
 \left[-\frac{1}{mr^2}\partial_r\,r^2\,\partial_r+\frac{1}{mr^2}\left(\Delta_x+\Delta_\varphi\right)+V(r)\right]\Psi^{(N)}(\bm{r})=\mathcal{E}_N\Psi^{(N)}(\bm{r})\,,
\end{equation}
where we have defined $V(r)=\mathrm{diag}\left(E_\Sigma^{(0)}(r),E_\Pi^{(0)}(r),E_\Pi^{(0)}(r)\right)$ and
\begin{align}
 \Delta_x&=\begin{pmatrix} \displaystyle-\partial_x\left(1-x^2\right)\partial_x+2 & \displaystyle-\sqrt{2}\partial_x\sqrt{1-x^2} & \displaystyle-\sqrt{2}\partial_x\sqrt{1-x^2} \\ \displaystyle-\sqrt{2}\sqrt{1-x^2}\partial_x & \displaystyle-\partial_x\left(1-x^2\right)\partial_x+\frac{1}{1-x^2} & 0 \\ \displaystyle-\sqrt{2}\sqrt{1-x^2}\partial_x & 0 & \displaystyle-\partial_x\left(1-x^2\right)\partial_x+\frac{1}{1-x^2} \end{pmatrix}\,,\\
 \Delta_\varphi&=\begin{pmatrix} \displaystyle-\frac{1}{1-x^2}\partial^2_\varphi & \displaystyle\frac{\sqrt{2}}{\sqrt{1-x^2}}i\partial_\varphi & \displaystyle-\frac{\sqrt{2}}{\sqrt{1-x^2}}i\partial_\varphi \\ \displaystyle\frac{\sqrt{2}}{\sqrt{1-x^2}}i\partial_\varphi & \displaystyle-\frac{1}{1-x^2}\left(\partial^2_\varphi-2xi\partial_\varphi\right) & 0 \\ \displaystyle-\frac{\sqrt{2}}{\sqrt{1-x^2}}i\partial_\varphi & 0 & \displaystyle-\frac{1}{1-x^2}\left(\partial^2_\varphi+2xi\partial_\varphi\right) \end{pmatrix}\,.
\end{align}
The three columns correspond to $\bm{\hat{n}}=\bm{\hat{r}},\bm{\hat{r}}^+,\bm{\hat{r}}^-$ and the three rows to $\bm{\hat{n}}'=\bm{\hat{r}},\bm{\hat{r}}^+,\bm{\hat{r}}^-$ in that order.

This is a coupled Schr\"odinger equation, which differs from the standard example of the hydrogen atom by the appearance of different potentials for the different wave function components and the more complicated angular part. But like the hydrogen atom, it can be solved by a separation ansatz $\Psi^{(N)}(\bm{r})=\psi_m(\varphi)\,\psi_l(x)\,\psi^{(N)}(r)$. The angular wave functions $\psi_m(\varphi)$ and $\psi_l(x)$ are matrices acting on the vector $\psi^{(N)}(r)$. They are eigenfunctions of their respective differential operators $\Delta_\varphi$ and $\Delta_x$ in the following sense:
\begin{equation}
 \Delta_\varphi\psi_m(\varphi)=\psi_m(\varphi)M\,,\hspace{7pt}\mathrm{and}\hspace{10pt}(\Delta_x+M)\psi_l(x)=\psi_l(x)L\,,
 \label{eigen}
\end{equation}
where $M$ and $L$ are matrices. If we also require $\psi_m(\varphi)$ and $\psi_l(x)$ to commute with the potential matrix $V(r)$, and in addition $\psi_m(\varphi)$ to commute with $\Delta_x$, then the full Schr\"odinger equation reduces to a coupled radial Schr\"odinger equation for $\psi^{(N)}(r)$ with an effective potential $V_{eff}(r)=V(r)+L/mr^2$:
\begin{align}
 0&=\left[-\frac{1}{mr^2}\partial_r\,r^2\,\partial_r+\frac{1}{mr^2}\left(\Delta_x+\Delta_\varphi\right)+V(r)-\mathcal{E}_N\right]\psi_m(\varphi)\,\psi_l(x)\,\psi^{(N)}(r)\notag\\
 &=\psi_m(\varphi)\left[-\frac{1}{mr^2}\partial_r\,r^2\,\partial_r+\frac{1}{mr^2}\left(\Delta_x+M\right)+V(r)-\mathcal{E}_N\right]\psi_l(x)\,\psi^{(N)}(r)\\
 &=\psi_m(\varphi)\,\psi_l(x)\left[-\frac{1}{mr^2}\partial_r\,r^2\,\partial_r+\frac{1}{mr^2}L+V(r)-\mathcal{E}_N\right]\psi^{(N)}(r)\\
 &=\psi_m(\varphi)\,\psi_l(x)\left[-\frac{1}{mr^2}\partial_r\,r^2\,\partial_r+V_{eff}(r)-\mathcal{E}_N\right]\psi^{(N)}(r)\,.
\end{align}

We will now show that such matrices do indeed exist. A solution for $\psi_m(\varphi)$ can immediately be found by making the ansatz $\psi_m(\varphi)=e^{im\varphi}\,\openone$, where $\openone$ is the unit matrix. With this we have
\begin{equation}
 \Delta_\varphi\psi_m(\varphi)=
\psi_m(\varphi)\begin{pmatrix} \displaystyle\frac{m^2}{1-x^2} & \displaystyle-\frac{\sqrt{2}m}{\sqrt{1-x^2}} & \displaystyle\frac{\sqrt{2}m}{\sqrt{1-x^2}} \\ \displaystyle-\frac{\sqrt{2}m}{\sqrt{1-x^2}} & \displaystyle\frac{m^2-2mx}{1-x^2} & 0 \\ \displaystyle\frac{\sqrt{2}m}{\sqrt{1-x^2}} & 0 & \displaystyle\frac{m^2+2mx}{1-x^2} \end{pmatrix}\,.
\end{equation}

Also for the next wave function $\psi_l(x)$, a solution in the form of a diagonal matrix can be found, although now the diagonal entries differ from each other. The diagonal elements of $\Delta_x+M$ (without constant terms) all have the same form
\begin{equation}
 -\partial_x\left(1-x^2\right)\partial_x+\frac{m^2-2\lambda mx+\lambda^2}{1-x^2}\,,
\end{equation}
with $\lambda=0,1,-1$ for the first, second, and third entries, respectively. The eigenfunctions of this differential operator are generalizations of the associated Legendre polynomials, for $\lambda=0$ they even coincide, and their derivation can be found in textbooks such as~\cite{LandauLifshitz}.

Including the factor $e^{im\varphi}$ and proper normalization, they are given by
\begin{align}
 v_{l,\,m}^\lambda(x,\varphi)&=\frac{(-1)^{m+\lambda}}{2^l}\sqrt{\frac{2l+1}{4\pi}\frac{(l-m)!}{(l+m)!(l-\lambda)!(l+\lambda)!}}\,P_{l,\,m}^\lambda(x)e^{im\varphi}\,,\\
 P_{l,\,m}^\lambda(x)&=(1-x)^{\frac{m-\lambda}{2}}(1+x)^{\frac{m+\lambda}{2}}\partial_x^{l+m}(x-1)^{l+\lambda}(x+1)^{l-\lambda}\,.
\end{align}
The eigenvalue is $l(l+1)$, and just like for the spherical harmonics, solutions exist only for $l$ a non-negative integer, $|m|\leq l$, and $|\lambda|\leq l$. They are normalized such that
\begin{equation}
 \int d\Omega\,v_{l',\,m'}^{\lambda\,*}(x,\varphi)v_{l,\,m}^{\lambda}(x,\varphi)=\delta_{l'l}\delta_{m'm}\,,
\end{equation}
and they also satisfy the orthogonality relations
\begin{equation}
 \sum_{m=-l}^lv_{l,\,m}^{\lambda'\,*}(x,\varphi)v_{l,\,m}^{\lambda}(x,\varphi)=\frac{2l+1}{4\pi}\delta^{\lambda'\lambda}\,,
\end{equation}
\begin{equation}
 \sum_{\lambda=-l}^lv_{l,\,m'}^{\lambda\,*}(x,\varphi)v_{l,\,m}^{\lambda}(x,\varphi)=\frac{2l+1}{4\pi}\delta_{m'm}\,.
\end{equation}

The easiest way to construct these functions is to use ladder operators for $m$ and $\lambda$. These operators and their action on the $v_{l,\,m}^\lambda$ functions are given by
\begin{align}
 \left(\mp\sqrt{1-x^2}\,\partial_x-\frac{mx-\lambda}{\sqrt{1-x^2}}\right)e^{\pm i\varphi}\,v_{l,\,m}^\lambda(x,\varphi)&=\sqrt{l(l+1)-m(m\pm1)}\,v_{l,\,m\pm1}^\lambda(x,\varphi)\,,\\
 \left(\pm\sqrt{1-x^2}\,\partial_x-\frac{m-\lambda x}{\sqrt{1-x^2}}\right)v_{l,\,m}^\lambda(x,\varphi)&=\sqrt{l(l+1)-\lambda(\lambda\pm1)}\,v_{l,\,m}^{\lambda\pm1}(x,\varphi)\,.
\end{align}

If we now look at the offdiagonal elements of $\Delta_x+M$, we see that they are given exactly by the ladder operators for $\lambda$. So for $\psi_m(\varphi)\,\psi_l(x)=\mathrm{diag}\left(v_{l,\,m}^0(x,\varphi),v_{l,\,m}^1(x,\varphi),v_{l,\,m}^{-1}(x,\varphi)\right)$ Eq.~\eqref{eigen} becomes
\begin{equation}
 \left(\Delta_x+\Delta_\varphi\right)\psi_m(\varphi)\,\psi_l(x)=\psi_m(\varphi)\,\psi_l(x)\begin{pmatrix} l(l+1)+2 & \sqrt{2l(l+1)} & -\sqrt{2l(l+1)} \\ \sqrt{2l(l+1)} & l(l+1) & 0 \\ -\sqrt{2l(l+1)} & 0 & l(l+1) \end{pmatrix}\,.
\end{equation}

Before we write down the resulting radial Schr\"odinger equation, we will exploit the fact that we are free to multiply this expression by any constant matrix, which gives another solution to the angular differential equation with a modified but equivalent eigenvalue matrix $L$. If this constant matrix is $(1,2)$-block diagonal, then also $V(r)$ remains unchanged. In this way we will define a new orbital wave function matrix $\psi_{l,\,m}(x,\varphi)$ as
\begin{equation}
 \psi_{l,\,m}(x,\varphi)=\frac{1}{\sqrt{2}}\begin{pmatrix} \sqrt{2}\,v_{l,\,m}^0(x,\varphi) & 0 & 0 \\ 0 & v_{l,\,m}^1(x,\varphi) & v_{l,\,m}^1(x,\varphi) \\ 0 & -v_{l,\,m}^{-1}(x,\varphi) & v_{l,\,m}^{-1}(x,\varphi) \end{pmatrix}\,.
\end{equation}

The advantage of this redefinition is that now in the radial Schr\"odinger equation the effective potential is $(2,1)$-block diagonal.
\begin{align}
 &\left[-\frac{1}{mr^2}\partial_r\,r^2\,\partial_r+\frac{1}{mr^2}\left(\Delta_x+\Delta_\varphi\right)+V(r)\right]\psi_{l,\,m}(x,\varphi)\,\psi^{(N)}(r)\notag\\
 &=\psi_{l,\,m}(x,\varphi)\left[-\frac{1}{mr^2}\partial_r\,r^2\,\partial_r+\frac{1}{mr^2}\begin{pmatrix} l(l+1)+2 & 2\sqrt{l(l+1)} & 0 \\ 2\sqrt{l(l+1)} & l(l+1) & 0 \\ 0 & 0 & l(l+1) \end{pmatrix}+V(r)\right]\psi^{(N)}(r)\notag\\
 &=\mathcal{E}_N\,\psi_{l,\,m}(x,\varphi)\,\psi^{(N)}(r)\,.
 \label{radsheq}
\end{align}
We see here explicitly the decoupling of the opposite parity states described in the main part of this paper. One solution is of the form $\left(\psi_\Sigma^{(N)}(r),\psi_{-\Pi}^{(N)}(r),0\right)^T$, the other $\left(0,0,\psi_{+\Pi}^{(N)}(r)\right)^T$.

If those are multiplied by the orbital wave function matrix $\psi_{l,\,m}(x,\varphi)$, and spin and angular momentum indices are combined through Clebsch-Gordan coefficients, then we get the following expressions for the hybrid states:
\begin{equation}
 \sum_{m_l,\,m_s}\int d^3r\,C_{j,\,m;\,l,\,s}^{\,m_l,\,m_s}\left[v_{l,\,m_l}^0\bm{\hat{r}}\,\psi^{(N)}_\Sigma+\frac{1}{\sqrt{2}}\left(v_{l,\,m_l}^1\bm{\hat{r}}^+-v_{l,\,m_l}^{-1}\bm{\hat{r}}^-\right)\psi^{(N)}_{-\Pi}\right]\cdot\bm{G}_B^a\,O^{a\,\dagger}_{s,\,m_s}|0\rangle\,,
\end{equation}
\begin{equation}
 \sum_{m_l,\,m_s}\int d^3r\,C_{j,\,m;\,l,\,s}^{\,m_l,\,m_s}\,\frac{1}{\sqrt{2}}\left(v_{l,\,m_l}^1\bm{\hat{r}}^++v_{l,\,m_l}^{-1}\bm{\hat{r}}^-\right)\psi^{(N)}_{+\Pi}\cdot\bm{G}_B^a\,O_{s,\,m_s}^{a\,\dagger}|0\rangle\,.
\end{equation}
The first gives the hybrid multiplets $H_1$, $H_1'$, $H_3$, $H_4$, and $H_6$, the second gives $H_2$, $H_2'$, $H_5$, and $H_7$, for different values of $l$, $s$, and $N$. Note that the different $P$ and $C$ eigenstate combinations come out correct.

We will now show that the hybrid states we have constructed are in fact eigenstates of the total angular momentum operator $\bm{L}=\bm{L}_{Q\bar{Q}}+\bm{K}$, where $\bm{K}$ is the angular momentum operator of the gluons and $\bm{L}_{Q\bar{Q}}$ the one of the relative coordinate of the quark-antiquark system. The center-of-mass coordinate $\bm{R}$ is fixed in the current approximation, which corresponds to a hybrid at rest, so there is no contribution to the total angular momentum from this coordinate.

The $1^{+-}$ gluelump operator is a (pseudo) vector, so $\bm{K}$ acts on it as
\begin{equation}
 \left[K_i,G_{B\,j}^a\right]=i\,\epsilon_{ijk}G_{B\,k}^a\,.
\end{equation}
The relative angular momentum operator in the octet sector is given by
\begin{equation}
 \bm{L}_{Q\bar{Q}}=\int d^3r\,d^3R\,O^{a\,\dagger}(\bm{r},\bm{R})\begin{pmatrix} \displaystyle-i\sqrt{1-x^2}\sin\varphi\,\partial_x+\frac{ix\cos\varphi}{\sqrt{1-x^2}}\,\partial_\varphi \\ \displaystyle i\sqrt{1-x^2}\cos\varphi\,\partial_x+\frac{ix\sin\varphi}{\sqrt{1-x^2}}\,\partial_\varphi \\ \displaystyle-i\partial_\varphi \end{pmatrix} O^a(\bm{r},\bm{R})\,.
\end{equation}
Acting with $\bm{L}_{Q\bar{Q}}$ on the hybrid states is equivalent to acting with the differential operator between the two octet fields on the wave functions and projection vectors. In a slight abuse of notation, we will also use the symbol $\bm{L}_{Q\bar{Q}}$ for this differential operator. It should be clear which one is meant by whether it acts on a state or on a wave function.

It is straightforward to show that
\begin{equation}
 -i\partial_\varphi\,\bm{\hat{n}}^T(x,\varphi)=\bm{\hat{n}}^T(x,\varphi)\begin{pmatrix} -i\partial_\varphi & -i & 0 \\ i & -i\partial_\varphi & 0 \\ 0 & 0 & -i\partial_\varphi \end{pmatrix}\hspace{10pt}\mathrm{for\ all}\hspace{10pt}\bm{\hat{n}}=\bm{\hat{r}},\,\bm{\hat{r}}^\pm\,,
\end{equation}
and by construction the orbital wave functions satisfy $-i\partial_\varphi\,v_{l,\,m}^\lambda(x,\varphi)=m\,v_{l,\,m}^\lambda(x,\varphi)$. So acting with $L_3$ on the hybrid states (before combining spin and angular momentum indices) gives
\begin{align}
 &L_3\int d^3r\,O^{a\,\dagger}(\bm{r},\bm{R})\sum_{n,\,i}\hat{n}_i(x,\varphi)G_{B\,i}^a(\bm{R})\Psi_n^{(N)}(\bm{r})|0\rangle\notag\\
 &=\int d^3r\,O^{a\,\dagger}(\bm{r},\bm{R})\sum_{n,\,i,\,j}\hat{n}_i(x,\varphi)\left[\begin{pmatrix} -i\partial_\varphi & -i & 0 \\ i & -i\partial_\varphi & 0 \\ 0 & 0 & -i\partial_\varphi \end{pmatrix}+\begin{pmatrix} 0 & i & 0 \\ -i & 0 & 0 \\ 0 & 0 & 0 \end{pmatrix}\right]_{ij}G_{B\,j}^a(\bm{R})\Psi_n^{(N)}(\bm{r})|0\rangle\notag\\
 &=\int d^3r\,O^{a\,\dagger}(\bm{r},\bm{R})\sum_{n,\,i}\hat{n}_i(x,\varphi)G_{B\,i}^a(\bm{R})\left(-i\partial_\varphi\,\Psi_n^{(N)}(\bm{r})\right)|0\rangle\notag\\*
 &=m\int d^3r\,O^{a\,\dagger}(\bm{r},\bm{R})\sum_{n,\,i}\hat{n}_i(x,\varphi)G_{B\,i}^a(\bm{R})\Psi_n^{(N)}(\bm{r})|0\rangle\,.
\end{align}

For $L^2$ we can write
\begin{equation}
 L^2=L_{Q\bar{Q}}^2+2\bm{L}_{Q\bar{Q}}\cdot\bm{K}+K^2\,.
 \label{fullL2}
\end{equation}
We already know the effect of $L_{Q\bar{Q}}^2$ on $\bm{\hat{n}}$ from the previous section:
\begin{equation}
 L_{Q\bar{Q}}^2\,\bm{\hat{n}}=\left(-\partial_x\left(1-x^2\right)\partial_x-\frac{1}{1-x^2}\partial_\varphi^2\right)\bm{\hat{n}}=\sum_{n'}\bm{\hat{n}}'\left(\Delta_x+\Delta_\varphi\right)_{n'n}\,.
\end{equation}
Note that here and in the following we use the indices $n$ and $n'$ to denote matrices that are defined in the basis of the different static states, $\Sigma_u^-$ and $\Pi_u^\pm$, which correspond to the projection vectors $\bm{\hat{r}}$ and $\bm{\hat{r}}^\pm$, respectively. In contrast, the indices $i$ and $j$ will always be used for the components of vectors and matrices defined in three-dimensional position space.

The last term in Eq.~\eqref{fullL2} $K^2$ just gives a constant factor $k(k+1)$, which is equal to $2$ in our case. So there only remains to determine the effect of $\bm{L}_{Q\bar{Q}}\cdot\bm{K}$ on $\bm{\hat{n}}$. We can write it as a matrix of differential operators, where the matrix nature comes from the action of the $\bm{K}$ part on the gluelump. An explicit calculation gives
\begin{align}
 \bm{L}_{Q\bar{Q}}\cdot\bm{K}\,\hat{n}_i=&{}\sum_j\begin{pmatrix} 0 & 0 & \displaystyle-\sqrt{1-x^2}\cos\varphi\partial_x \\\vspace{-15pt}\\ 0 & 0 & \displaystyle-\sqrt{1-x^2}\sin\varphi\partial_x \\\vspace{-15pt}\\ \displaystyle\sqrt{1-x^2}\cos\varphi\partial_x & \displaystyle\sqrt{1-x^2}\sin\varphi\partial_x & 0 \end{pmatrix}_{ij}\hat{n}_j\notag\\
 &+\sum_j\begin{pmatrix} 0 & \displaystyle-\partial_\varphi & \displaystyle-\frac{x\sin\varphi}{\sqrt{1-x^2}}\partial_\varphi \\\vspace{-15pt}\\ \displaystyle\partial_\varphi & 0 & \displaystyle\frac{x\cos\varphi}{\sqrt{1-x^2}}\partial_\varphi \\\vspace{-15pt}\\ \displaystyle\frac{x\sin\varphi}{\sqrt{1-x^2}}\partial_\varphi & \displaystyle-\frac{x\cos\varphi}{\sqrt{1-x^2}}\partial_\varphi & 0 \end{pmatrix}_{ij}\hat{n}_j\notag\\
 =&{}\sum_{n'}\hat{n}'_i\begin{pmatrix} -2 & \displaystyle\frac{1}{\sqrt{2}}\partial_x\sqrt{1-x^2} & \displaystyle\frac{1}{\sqrt{2}}\partial_x\sqrt{1-x^2} \\\vspace{-15pt}\\ \displaystyle-\frac{1}{\sqrt{2}}\sqrt{1-x^2}\partial_x & -1 & 0 \\\vspace{-15pt}\\ \displaystyle-\frac{1}{\sqrt{2}}\sqrt{1-x^2}\partial_x & 0 & -1 \end{pmatrix}_{n'n}\notag\\
 &+\sum_{n'}\hat{n}'_i\begin{pmatrix} 0 & \displaystyle-\frac{i}{\sqrt{2}}\frac{1}{\sqrt{1-x^2}}\partial_\varphi & \displaystyle\frac{i}{\sqrt{2}}\frac{1}{\sqrt{1-x^2}}\partial_\varphi \\\vspace{-15pt}\\ \displaystyle-\frac{i}{\sqrt{2}}\frac{1}{\sqrt{1-x^2}}\partial_\varphi & 0 & 0 \\\vspace{-15pt}\\ \displaystyle\frac{i}{\sqrt{2}}\frac{1}{\sqrt{1-x^2}}\partial_\varphi & 0 & 0 \end{pmatrix}_{n'n}\,.
\end{align}

We now see that in $L_{Q\bar{Q}}^2+2\bm{L}_{Q\bar{Q}}\cdot\bm{K}+K^2$ all offdiagonal elements of $\Delta_x$ and $\Delta_\varphi$ cancel, as well as all constant terms in the diagonal elements. What remains is
\begin{align}
 &\left(L_{Q\bar{Q}}^2+2\bm{L}_{Q\bar{Q}}\cdot\bm{K}+K^2\right)\sum_{n}\bm{\hat{n}}(x,\varphi)\Psi_n^{(N)}(\bm{r})\notag\\
 &=\sum_{n,\,n'}\bm{\hat{n}}'(x,\varphi)\begin{pmatrix} L_{Q\bar{Q}}^2 & 0 & 0 \\ 0 & L_{Q\bar{Q}}^2+\dfrac{2ix\partial_\varphi+1}{1-x^2} & 0 \\ 0 & 0 & L_{Q\bar{Q}}^2+\dfrac{-2ix\partial_\varphi+1}{1-x^2}\end{pmatrix}_{n'n}\Psi_n^{(N)}(\bm{r})\notag\\
 &=l(l+1)\sum_{n}\bm{\hat{n}}(x,\varphi)\Psi_n^{(N)}(\bm{r})\,.
\end{align}
The last equality follows, because the diagonal entries are exactly the defining differential equations for the orbital wave functions.

\section{Numerical solution of the Schr\"odinger equations} \label{appx}

The Schr\"odinger equations in~\eqref{rsheqnp} and~\eqref{rsheqpp} can be solved numerically (see, e.g.,~\cite{Falkensteiner1985287,*Lucha:1998xc}). In the uncoupled case the nodal theorem can be used to determine the energy eigenvalues. Any value $\mathcal{E}$ one inserts in these equations in the place of $\mathcal{E}_N$ defines a linear differential equation of second order. These have in general two linearly independent solutions. Such a solution can only be interpreted as a wave function, if it is normalizable.

Two independent solutions can be distinguished by their behavior at the origin,
\begin{equation}
 \psi_{+\Pi}^{(N)}(r)\propto r^l+{\cal O}(r^{l+1})\hspace{30pt}\mathrm{or}\hspace{30pt}\psi_{+\Pi}^{(N)}(r)\propto r^{-l-1}+\mathcal{O}\left(r^{-l}\right)\,.
\end{equation}
The second expression is singular at the origin and therefore not normalizable. The first expression defines initial conditions for the wave function and its derivative, such that for any value of $\mathcal{E}$ the differential equation~\eqref{rsheqpp} has a unique solution. This solution generally diverges for large $r$, only for particular values of $\mathcal{E}=\mathcal{E}_N$ does it approach zero and is normalizable. These are the desired wave function solutions of the Schr\"odinger equation. The order $N$ of the eigenvalue is equal to the number of zeros in the wave function. For the special case of $l=0$ the initial conditions for $\psi_\Sigma^{(N)}$ are the same as for $\psi_{+\Pi}^{(N)}$ with $l=1$.

A similar approach can be used to determine the energy eigenvalues of the coupled Schr\"odinger equation~\eqref{rsheqnp} for $l\geq1$. A system of two linearly coupled differential equations of second order has in general four linearly independent solutions, of which now two are singular at the origin. The remaining two can also be distinguished by their behavior at the origin, which is given by
\begin{equation}
 \begin{pmatrix} \psi_\Sigma^{(N1)}(r) \\ \psi_{-\Pi}^{(N1)}(r) \end{pmatrix}\propto\begin{pmatrix} \sqrt{l}\,r^{l-1} \\ -\sqrt{l+1}\,r^{l-1} \end{pmatrix}+\mathcal{O}\left(r^l\right)\,,
\end{equation}
or
\begin{equation}
 \begin{pmatrix} \psi_\Sigma^{(N2)}(r) \\ \psi_{-\Pi}^{(N2)}(r) \end{pmatrix}\propto\begin{pmatrix} \sqrt{l+1}\,r^{l+1} \\ \sqrt{l}\,r^{l+1} \end{pmatrix}+\mathcal{O}\left(r^{l+2}\right)\,.
\end{equation}

Again, the solutions to the two coupled differential equations with these initial conditions diverge for general $\mathcal{E}$ at large $r$. For particular values of $\mathcal{E}=\mathcal{E}_N$ there exists one linear combination
\begin{equation}
 \begin{pmatrix} \psi_\Sigma^{(N)}(r) \\ \psi_{-\Pi}^{(N)}(r) \end{pmatrix}=\begin{pmatrix} \psi_\Sigma^{(N1)}(r) \\ \psi_{-\Pi}^{(N1)}(r) \end{pmatrix}+\nu\begin{pmatrix} \psi_\Sigma^{(N2)}(r) \\ \psi_{-\Pi}^{(N2)}(r) \end{pmatrix}\,,
\end{equation}
which approaches zero for large $r$, while any other combination with a different $\nu$ will still diverge. This gives the desired wave functions.

So now one has to tune two independent parameters in order to find the solutions, $\mathcal{E}$ and $\nu$. Fortunately, the two can be determined separately. Instead of counting zeros of the wave function in order to find the eigenvalues $\mathcal{E}_N$ like in the uncoupled case, one now has to look at the determinant of the two independent solutions~\cite{Amann:1995}
\begin{equation}
 U(r)=\det\begin{pmatrix} \psi_\Sigma^{(N1)}(r) & \psi_\Sigma^{(N2)}(r) \\ \psi_{-\Pi}^{(N1)}(r) & \psi_{-\Pi}^{(N2)}(r) \end{pmatrix}\,.
\end{equation}
This function diverges in the large $r$ limit for general $\mathcal{E}$ but converges for $\mathcal{E}=\mathcal{E}_N$ and then has exactly $N$ zeros. In this way $\mathcal{E}_N$ can be determined without knowledge of $\nu$.

Then in order to obtain the wave functions $\psi_\Sigma^{(N)}(r)$ and $\psi_{-\Pi}^{(N)}(r)$ one can determine $\nu$ through
\begin{equation}
 \nu=-\lim_{r\to\infty}\frac{\psi_\Sigma^{(N1)}(r)}{\psi_\Sigma^{(N2)}(r)}=-\lim_{r\to\infty}\frac{\psi_{-\Pi}^{(N1)}(r)}{\psi_{-\Pi}^{(N2)}(r)}\,,
\end{equation}
after $\mathcal{E}$ has been fixed to the eigenvalue $\mathcal{E}_N$ from the previous step. Alternatively, $(1,\nu)^T$ is the eigenvector of the wave function matrix (i.e., the matrix of which $U(r)$ is the determinant) at $r\to\infty$ with eigenvalue zero.

These properties of the solutions of the radial Schr\"odinger equations can be exploited in an algorithm to numerically find the eigenvalues and wave functions. The details of this will be described elsewhere~\cite{program}.

\bibliographystyle{apsrev4-1}
\bibliography{hyb}

\end{document}